\documentclass[aps,prb,twocolumn,floatfix,showpacs,superscriptaddress]{revtex4-1}
\usepackage{graphicx}
\usepackage[colorlinks=true]{hyperref}
\usepackage{xcolor}
\usepackage{amsmath}
\usepackage{amssymb}

\begin{document}

\title{Extended dynamical mean-field study of the Hubbard model with long range interactions}
\author{Li Huang}
\affiliation{Department of Physics, University of Fribourg, 1700 Fribourg, Switzerland}
\author{Thomas Ayral}
\affiliation{Centre de Physique Th\'{e}orique, Ecole Polytechnique, CNRS-UMR7644, 91128 Palaiseau, France}
\affiliation{Institut de Physique Th\'{e}orique (IPhT), CEA, CNRS, URA 2306, 91191 Gif-sur-Yvette, France}
\author{Silke Biermann}
\affiliation{Centre de Physique Th\'{e}orique, Ecole Polytechnique, CNRS-UMR7644, 91128 Palaiseau, France}
\author{Philipp Werner}
\affiliation{Department of Physics, University of Fribourg, 1700 Fribourg, Switzerland}
\date{\today}

\begin{abstract}
Using extended dynamical mean-field theory and its combination with the $GW$ approximation, we compute the phase diagrams and local spectral functions of the single-band extended Hubbard model on the square and simple cubic lattices, considering long range interactions up to the third nearest neighbors. The longer range interactions shift the boundaries between the metallic, charge-ordered insulating and Mott insulating phases, and lead to characteristic changes in the screening modes and local spectral functions. Momentum-dependent self-energy contributions enhance the correlation effects and thus compete with the additional screening effect from longer range Coulomb interactions. Our results suggest that the influence of longer range intersite interactions is significant, and that these effects deserve attention in realistic studies of correlated materials.   
\end{abstract}


\maketitle

\section{introduction\label{sec:intro}}
In condensed matter physics, electron-electron correlations give rise to many intriguing phenomena ranging from simple energy band renormalization to complex phase diagrams with charge-, spin-, or orbital ordering.\cite{RevModPhys.70.1039} The essential physics is the competition between electron localization and itinerancy. The Hubbard model is one of the simplest models which captures this competition, and it is therefore often used to investigate correlation effects in lattice systems.\cite{RevModPhys.68.13,RevModPhys.78.865} For instance, it is generally believed that the two-dimensional single-band Hubbard model with static onsite Coulomb interaction $U$ can be used to explain some underlying physics of cuprate high-temperature superconductors.\cite{RevModPhys.78.17} One widely accepted assumption in these studies is that the electron-electron interaction is local, i.e., that long range intersite interactions are fully screened or may be ignored. When additional intersite Coulomb interactions are considered, the model becomes an extended Hubbard model, which can be used for example to explore charge-ordering and Wigner-Mott transitions.\cite{PhysRevB.82.155102} This model also describes the screening of local interactions by the nonlocal interactions. Both the charge-ordering transition and the screening effect in the extended Hubbard model have been investigated in numerous theoretical studies.\cite{PhysRevLett.90.086402,PhysRevB.66.085120,PhysRevB.88.245110,PhysRevB.88.054504,PhysRevB.86.165136,PhysRevB.86.085113,PhysRevB.84.075161,PhysRevB.85.024516,PhysRevLett.109.226401,PhysRevB.87.125149}

The physical properties of the Hubbard model have been studied extensively using the dynamical mean-field theory (DMFT).\cite{RevModPhys.68.13,RevModPhys.78.865} This approximate scheme describes the generic behavior of high-dimensional lattice systems. In particular, at half-filling and low temperature, the DMFT solution for the hypercubic lattice will be an antiferromagnetically ordered insulator, whose character changes from a Slater-type antiferromagnet at weak interactions, to a Heisenberg-type antiferromagnet with local moments at large interaction. If the calculations are restricted to the paramagnetic phase, the DMFT method predicts a transition from a Fermi-liquid metal to a Mott insulator at a temperature-dependent critical value of the onsite interaction $U$ (comparable to the bandwidth). This paramagnetic Mott transition can be considered as the generic physical situation at sufficiently high temperatures, or in the magnetically frustrated case. The extended Hubbard model with strong non-local interactions (parametrized by $V$) exhibits a transition to a charge-ordered state characterized by a freezing of charge carriers and a spatial modulation of the charge density.\cite{PhysRevB.82.155102} To describe this transition one may resort to the extended dynamical mean-field theory (EDMFT) framework.\cite{PhysRevB.52.10295,PhysRevLett.77.3391,PhysRevB.59.5341,PhysRevB.63.115110,PhysRevLett.84.3678,PhysRevB.61.5184,PhysRevLett.92.196402,PhysRevB.66.085120} The basic idea of EDMFT was originally developed in studies of heavy-fermion systems and spin glasses with non-local Coulomb interactions.\cite{PhysRevB.52.10295,PhysRevLett.77.3391} The physical effects induced by the nonlocal interaction $V$, including a frequency dependence of the effective local interaction and a sizable reduction of the static value of $U$, are well captured by the EDMFT scheme. Since EDMFT takes into account the spatially nonlocal interactions beyond the Hartree level, it is a sophisticated numerical tool for studying the extended Hubbard model. However, EDMFT is still based on a local approximation, i.e., it assumes a $k$-independent self-energy function and polarization function. To further incorporate spatially nonlocal contributions into these functions, one can combine the EDMFT approach with the $GW$ approximation.\cite{PhysRevLett.90.086402,PhysRevB.66.085120,PhysRevLett.92.196402,PhysRevLett.109.226401,PhysRevB.87.125149}

While the EDMFT and $GW$ + EDMFT schemes have been developed more than ten years ago, there has been a recent revival in interest in these approaches, due to methodological improvements which enable an efficient and accurate solution of the self-consistency equations. In the previous studies, phase diagrams in the space of onsite interaction $U$ and the nearest neighbor interaction $V$, fully screened and retarded interactions, and local spectral functions have been calculated for the extended Hubbard model on square and simple cubic lattices.\cite{PhysRevLett.84.3678,PhysRevB.66.085120,PhysRevLett.92.196402,PhysRevLett.109.226401,PhysRevB.87.125149} It has been found that the critical charge-ordering lines $V_c(U)$ between the Mott insulator phase and the charge-ordered insulator phase obtained by the EDMFT and $GW$ + EDMFT approaches are substantially steeper than the naive mean-field estimate $V_c = U/z$, where $z$ is the number of the nearest neighbors.\cite{PhysRevB.87.125149} This may point to an overestimation of the local interactions in the EDMFT and $GW$ + EDMFT schemes or a non-trivial screening effect. Further issues left open in previous work concern the physical interpretation of the dominant screening processes, and their dependence on the parameters of the model. In Ref.~\onlinecite{PhysRevLett.111.036601}, it was proposed that the effective local interaction incorporating screening by neighboring lattice sites can be well approximated by simple estimates in terms of onsite and intersite interactions. The recent $GW$ + EDMFT study of Ref.~\onlinecite{PhysRevB.87.125149} was consistent with this simple picture in the correlated metallic case in two dimensions with nearest neighbor interactions. However, the usefulness and accuracy of these estimates in the higher dimensional case or with longer range interactions remains an open question.

The early studies of the three-dimensional extended Hubbard model \cite{PhysRevLett.92.196402,PhysRevB.66.085120} used a modified Hirsch-Fye algorithm to solve the effective impurity problem and could not reach low temperatures. In these calculations, the fermionic part of the impurity model was handled by a standard Hirsch-Fye algorithm,\cite{RevModPhys.68.13,RevModPhys.78.865} while the statistical weight due to the continuous bosonic fields was obtained directly by computing the corresponding Boltzmann factor.\cite{PhysRevB.62.12800} This algorithm is not as efficient and accurate as the recently developed continuous time quantum Monte Carlo (CT-QMC) solver\cite{PhysRevLett.99.146404,PhysRevLett.97.076405,PhysRevLett.104.146401,RevModPhys.83.349} which can treat systems with a frequency-dependent retarded interaction without any approximations. Thus, it is worthwhile to reinvestigate the model using the EDMFT and $GW$ + EDMFT approaches in combination with the state-of-the-art CT-QMC quantum impurity solver. This was done in Refs.~\onlinecite{PhysRevLett.109.226401,PhysRevB.87.125149} for the two-dimensional model with local and nearest neighbor interactions. Here, we extend the investigation to the three-dimensional model and to interactions of longer range. Indeed, recent constrained random phase approximation calculations \cite{Hansmann2013} and a recent $GW$ + EDMFT study \cite{PhysRevLett.110.166401} of adatom systems Si(111):$X$, with $X$ = Sn, Si, C, Pb, suggests that taking into account substantially longer range interactions is mandatory to understand experimentally observed trends from Mott physics toward charge-ordering physics along this series. In particular, it was shown that long-range interactions (for the surface systems, the full Coulomb tail was considered) can decrease the effective local interaction by up to a factor of two. Similar conclusions were drawn in Ref.~\onlinecite{PhysRevLett.111.036601} for other two-dimensional systems like graphene, silicene and benzene. Other studies suggest that the superconducting $T_c$ is generally suppressed in some pairing channels as the strength of longer range interactions increases.\cite{PhysRevB.85.024516} It thus appears that longer range intersite interactions beyond the nearest neighbors may be important, at least for low dimensional systems. So, it is worth investigating in a simple model context how longer range intersite interactions modify the phase diagrams and various local and nonlocal observables.  

The purpose of this paper is to gain qualitative and quantitative insights into the role of screening from non-local Coulomb interactions. For this, we study the extended Hubbard model on the square (2D) and simple cubic (3D) lattices using a modern EDMFT and $GW$ + EDMFT implementation with a numerically exact CT-QMC impurity solver. The calculations are restricted to repulsive interactions $ U > 0 $ and $V > 0$, and to the paramagnetic phase, so that we can investigate the particularly interesting screening effects in the correlated metal, close to the Mott or charge ordered insulator phase boundaries. In particular, we extract the dominant screening modes and analyze the effects of longer range intersite interactions on local, but energy dependent observables, such as spectral functions. At first, we will perform self-consistent EDMFT calculations to map out the entire $U-V$ phase diagram, and then compare to $GW$ + EDMFT results at some representative points to gain insights into the effects of nonlocal self-energy and polarization contributions.

The rest of this paper is organized as follows. Section~\ref{sec:method} defines the extended Hubbard model used in this study. The flowcharts for the EDMFT and $GW$ + EDMFT methods and the computational details are also briefly summarized in this section. Section~\ref{sub:edmft} shows the results obtained using the EDMFT approach. The phase diagrams, fully screened and retarded interactions induced by the $V$ term, and local spectral functions are presented and discussed in detail. Especially, doping-dependent phase diagrams and related bosonic spectral functions are also presented in this section. Some representative results obtained with the $GW$ + EDMFT approach are discussed in Sec.~\ref{sub:gw_edmft}. A brief summary and outlook are given in Sec.~\ref{sec:conclu}. Appendix \ref{app:vk} describes the long range intersite interactions considered in the 2D and 3D extended Hubbard models, while Appendix \ref{app:mem} details the maximum entropy based analytical continuation method used to extract the spectral functions for the frequency-dependent fully screened and retarded interactions.

\section{model and methods\label{sec:method}}
\subsection{Extended Hubbard model}
\begin{figure}[tp]
\centering
\includegraphics[width=\columnwidth]{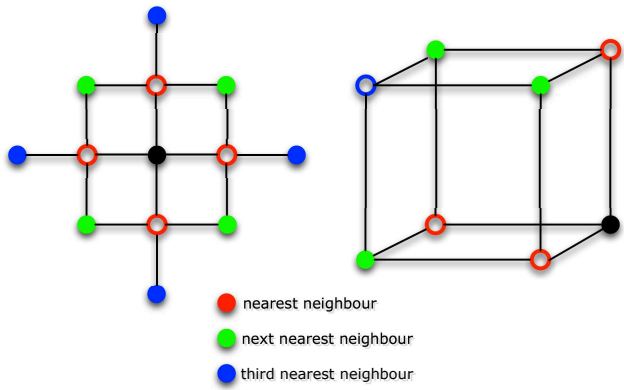}
\caption{(Color online) Schematic picture of the one-band half-filled extended Hubbard model in the charge-ordered state for the square lattice (left) and simple cubic lattice (right). The full dots represent doubly occupied sites and the open dots empty sites. The red, green, and purple dots denote the NN, NNN, and 3NN sites of the black dot, respectively. \label{fig:lattice}}
\end{figure}

In the present study, we consider the single-band extended Hubbard models on a two-dimensional square lattice and a three-dimensional simple cubic lattice, respectively (see schematic picture in Fig.~\ref{fig:lattice}). The grand-canonical Hamiltonian can be written as
\begin{align}
\label{eq:ham}
H = &-\sum_{(i,j),\sigma} t_{ij} (c^{\dagger}_{i\sigma}c_{j\sigma} + H.c) - \mu \sum_{i} n_{i} \nonumber \\
&+ U \sum_{i} n_{i\uparrow}n_{i\downarrow} + \sum_{(i,j)} V_{ij} n_{i} n_{j},
\end{align}
where $i$ and $j$ are site indices and $(i,j)$ denotes a pair of sites $i$ and $j$. $c_{i\sigma}$ and $c^{\dagger}_{i\sigma}$ are the annihilation and creation operators of an electron of spin $\sigma$ at the lattice site $i$. $n_{i\sigma}$ is the orbital occupation operator, and $n_i = n_{i\uparrow} + n_{i\downarrow}$. $t_{ij}$ is the hopping matrix element between two different sites, $\mu$ the chemical potential, $U$ the onsite interaction, and $V_{ij}$ the intersite interaction between sites $i$ and $j$. 

When $i = j$, both $t_{ij}$ and $V_{ij}$ must be zero. Only the hopping between the nearest neighbor (NN) sites is allowed in this study, namely, $t_{ij} = t_{\langle ij \rangle} = t > 0$. However, for the nonlocal repulsive interactions $V_{ij}$ we also consider the next nearest neighbor (NNN) and the third nearest neighbor (3NN) sites. Our definitions for the NN, NNN and 3NN sites are shown in Fig.~\ref{fig:lattice}. We further assume that $V_{ij}$ can be calculated by scaling $V$ with $a/|\vec{r}_i - \vec{r}_j|$, in other words, with the inverse distance in units of the NN distance $a$. In this sense, $V$ is not only the NN interaction, but also the parameter which determines the strength of all the long range Coulomb interactions. The detailed formulas of the Fourier-transformed $t_{ij}$ and $V_{ij}$ are given in Appendix~\ref{app:vk}.

\subsection{EDMFT and $GW$ + EDMFT}
We solve the single-band extended Hubbard model [see Eq.~(\ref{eq:ham})] with fully self-consistent EDMFT and $GW$ + EDMFT calculations. The EDMFT approach with the ``$UV$ decoupling" scheme\cite{PhysRevB.87.125149} formally treats the local interactions and nonlocal intersite interactions on the same footing. It can be used to describe the Mott transition and charge-ordering transition in the extended Hubbard model.\cite{PhysRevLett.77.3391,PhysRevB.61.5184,PhysRevB.52.10295,PhysRevLett.84.3678} The idea of the combined $GW$ + EDMFT\cite{PhysRevLett.90.086402} scheme is the following: One takes the local part of the self-energy (or polarization) from the EDMFT calculation and adds to it the nonlocal component of the $GW$ self-energy (or polarization). Thus, a momentum dependence is introduced into the self-energy (or polarization), and the scheme captures the interplay of screening and nonlocal correlations at least to some extent. While the accuracy of the scheme has not been systematically tested, self-consistent $GW$ + EDMFT calculations can be obtained in the whole interaction range from the weakly correlated region to the atomic limit. A detailed derivation of the $GW$ + EDMFT formulation for extended Hubbard model can be found in Ref.~\onlinecite{PhysRevB.87.125149}.

The typical $GW$ + EDMFT self-consistency loop involves the following steps.\cite{PhysRevB.87.125149,PhysRevB.66.085120} One starts with an initial guess for the $k$-dependent fermionic self-energy $\Sigma(k,i\omega_n)$ and the bosonic self-energy (or polarization) $\Pi(k,i\nu_n)$, with Matsubara frequencies $\omega_n = (2n + 1)\pi/\beta$ and $\nu_n = 2n \pi / \beta$ for integer $n$. The initial $\Sigma(k,i\omega_n)$ and $\Pi(k,i\nu_n)$ can be obtained from previously calculated results, or chosen to be zero. Then one calculates the lattice Green's function $G(k, i\omega_n)$ and fully screened interaction $W(k, i\nu_n)$ using the lattice Dyson equations 
\begin{equation}
\label{eq:lattice_dyson_g}
G(k, i\omega_n) = \frac{1}{i\omega_n + \mu  - \epsilon_k - \Sigma(k,i\omega_n)},
\end{equation}
and
\begin{equation}
\label{eq:lattice_dyson_w}
W(k, i\nu_n) = \frac{1}{v_k^{-1} - \Pi(k,i\nu_n)}.
\end{equation}
Here, $\epsilon_k$ is the band dispersion and $v_{k}$ is the bare interaction in reciprocal space (see Appendix \ref{app:vk} for more details). Then the local counterparts of $G$, $W$, $\Sigma$ and $\Pi$ are calculated by averaging over the whole Brillouin zone, for instance ($N_k$ is the number of $k$-points),
\begin{equation}
G(i\omega_n) = \frac{1}{N_k}\sum_k G(k,i\omega_n).
\end{equation}
Next, the local bath Green's function $\mathcal{G}(i\omega_n)$ and frequency dependent retarded interaction $\mathcal{U}(i\nu_n)$ are calculated through the impurity Dyson equations, namely,
\begin{equation}
\label{eq:local_dyson_g}
\mathcal{G}^{-1}(i\omega_n) = G^{-1}(i\omega) + \Sigma(i\omega_n),
\end{equation}
and
\begin{equation}
\label{eq:local_dyson_u}
\mathcal{U}^{-1}(i\nu_n) = W^{-1}(i\nu_n) + \Pi(i\nu_n).
\end{equation}
Then the quantum impurity model defined by $\mathcal{G}(i\omega_n)$ and $\mathcal{U}(i\nu_n)$ is solved numerically. The impurity solver directly yields the new $G(i\omega_n)$. On the other hand, the calculation of the new $W(i\nu_n)$ involves as an intermediate step, the calculation of the connected charge-charge correlation function $\chi(\tau) = \langle \mathcal{T} \bar{n}(\tau)\bar{n}(0) \rangle$ with $\bar{n} = n - \langle n \rangle$. From the Fourier-transformed $\chi(i\nu_n)$ and $\mathcal{U}(i\nu_n)$, we finally obtain the new $W(i\nu_n)$ via
\begin{equation}
\label{eq:W}
W(i\nu_n) = \mathcal{U}(i\nu_n) - \mathcal{U}(i\nu_n)\chi(i\nu_n)\mathcal{U}(i\nu_n).
\end{equation}
Using these $G(i\omega_n)$ and $W(i\nu_n)$ as inputs, the new local self-energy functions $\Sigma(i\omega_n)$ and $\Pi(i\nu_n)$ are determined by using Eqs.~(\ref{eq:local_dyson_g}) and (\ref{eq:local_dyson_u}) again. Within the $GW$ approximation, one evaluates the momentum-dependent $GW$ self-energy and polarization functions as $\Sigma^{\text{GW}} = -GW$ and $\Pi^{\text{GW}} = 2GG$.\cite{PhysRevLett.90.086402} Here the factor $2$ comes from the contribution of the spin degree of freedom. Finally, one has to separate the local and nonlocal parts of these $GW$ self-energies and polarizations,
\begin{equation}
\Sigma^{\text{GW}}_{\text{loc}}(i\omega_n) = \frac{1}{N_k}\sum_k \Sigma^{\text{GW}}(k,i\omega_n),
\end{equation}
\begin{equation}
\Pi^{\text{GW}}_{\text{loc}}(i\nu_n) = \frac{1}{N_k}\sum_k \Pi^{\text{GW}}(k,i\nu_n),
\end{equation}
\begin{equation}
\label{eq:nonlocal_s}
\Sigma^{\text{GW}}_{\text{nonloc}} (k,i\omega_n) = \Sigma^{\text{GW}} (k,i\omega_n) - \Sigma^{\text{GW}}_{\text{loc}}(i\omega_n),
\end{equation}
\begin{equation}
\label{eq:nonlocal_p}
\Pi^{\text{GW}}_{\text{nonloc}} (k,i\nu_n) = \Pi^{\text{GW}} (k,i\nu_n) - \Pi^{\text{GW}}_{\text{loc}}(i\nu_n),
\end{equation}
and then combine the nonlocal parts with the local contributions obtained from the impurity calculations, i.e.,
\begin{equation}
\Sigma(k,i\omega_n) = \Sigma^{\text{GW}}_{\text{nonloc}} (k,i\omega_n) + \Sigma(i\omega_n),
\end{equation}
and
\begin{equation}
\Pi(k,i\nu_n) = \Pi^{\text{GW}}_{\text{nonloc}} (k,i\nu_n) + \Pi(i\nu_n).
\end{equation}
The new self-energy and polarization functions, $\Sigma(k,i\omega_n)$ and $\Pi(k,i\nu_n)$, serve as the starting point of the next iteration. This completes the self-consistent loop. 

The EDMFT self-consistency loop can be viewed as a simplification of the full $GW$ + EDMFT iteration, where one ignores the calculations of the $GW$ self-energies $\Sigma^{\text{GW}} (k,i\omega_n)$ and polarizations $\Pi^{\text{GW}} (k,i\nu_n)$, and adopts the following local approximations 
\begin{equation}
\Sigma(k,i\omega_n) = \Sigma(i\omega_n),
\end{equation}
and
\begin{equation}
\Pi(k,i\nu_n) = \Pi(i\nu_n).
\end{equation}

In the following calculations, we consider half-filled single-band extended Hubbard models on the square lattice and simple cubic lattice (some results for the 2D model away from half-filling can be found in Sec.~\ref{sub:edmft}). The $k$-sums are discretized in the irreducible Brillouin zone on $81 \times 81$ and $19 \times 19 \times 19$ grid points, respectively. We used the hybridization expansion quantum impurity solver to solve the effective impurity problems.\cite{PhysRevLett.104.146401,RevModPhys.83.349} The imaginary time Green's function $G(\tau)$ and charge-charge correlation function $\chi(\tau)$ are measured on $N = 1024$ equally spaced time points. We used $4t$ as the unit of energy and performed calculations at inverse temperature $\beta = 100$, restricting our study to the paramagnetic phase. Up to 40 EDMFT and $GW$ + EDMFT iterations are required to reach convergence when the system is close to the Mott or charge-ordering transition.

\subsection{Analytical continuation\label{sub:maxent}}
Since the self-consistency loop is implemented fully on the imaginary time/frequency axis, we have to analytically continue the converged $G(\tau)$, $\mathcal{U}(i\nu)$, and $W(i\nu)$ to obtain meaningful information about single particle excitations and screening modes. 

The frequency dependence of the retarded interaction $\mathcal{U}(i\nu)$ affects the single particle spectral function $A(\omega)$, and in particular induces satellites at energies which are determined by the dominant screening frequencies.\cite{PhysRevLett.104.146401,PhysRevB.85.035115,Werner2012} However, the classical maximum entropy method,\cite{mem:1996} which is commonly used to perform analytical continuations of $G(\tau)$, tends to smooth out these high-energy features. To overcome this obstacle, we adopted the algorithm proposed by Casula \emph{et al}.\cite{PhysRevB.85.035115} and proceed as follows: From the spectral function $\text{Im}\mathcal{U}(\nu)$ we calculate the bosonic function 
\begin{equation}
B(\tau)=\exp[K(0)-K(\tau)],
\end{equation}
where\cite{PhysRevLett.109.126408}
\begin{equation}
K(\tau)=\int_0^\infty \text{d}\nu \frac{\text{Im}\mathcal{U}(\nu)}{\nu^2}\frac{\cosh[\nu(\beta/2-\tau)]}{\sinh(\nu\beta/2)},
\end{equation}
 and the corresponding spectral function $A_{\text{B}}(\nu)$. We then define the auxiliary fermionic Green's function $G_\text{aux}(\tau)=G(\tau)/B(\tau)$, which later is analytically continued using the conventional maximum entropy method to yield $A_{\text{aux}}(\omega)$. Finally, the spectral function for $G(\tau)$ is obtained from the convolution 
\begin{equation}
A(\omega) = \int \text{d}\epsilon 
\frac{A_{\text{B}}(\epsilon)A_\text{aux}(\omega-\epsilon)(1+e^{-\beta\omega})}
     {(1+e^{\beta(\epsilon-\omega)})(1-e^{-\beta\epsilon})}.
\end{equation}  

This procedure requires an accurate estimate of the spectral function $\text{Im}\mathcal{U}(\nu)$. In previous studies, the Pad\'{e} approximation was used.\cite{PhysRevB.87.125149} However, we found that the Pad\'{e} results are very sensitive to the data quality of $\mathcal{U}(i\nu)$. Small fluctuations in $\mathcal{U}(i\nu)$, which are almost unavoidable [see Eq.~(\ref{eq:local_dyson_u})], can lead to drastic modifications in the Pad\'{e} estimation of Im$\mathcal{U}(\nu)$. Thus, a robust procedure with respect to the typical level of numerical noise is crucial. The maximum entropy method is superior in this respect, and we have adapted it to the problem of analytically continuing the retarded interaction $\mathcal{U}(i\nu)$ and fully screened interaction $W(i\nu)$. The details of this procedure are explained in Appendix~\ref{app:mem}.
 
\section{results and discussion\label{sec:results}}
\subsection{EDMFT results\label{sub:edmft}}
In this subsection, we present self-consistent EDMFT results for the paramagnetic, half-filled single-band $U$-$V$ Hubbard model on the square lattice and simple cubic lattice. All results are for inverse temperature $\beta=100$.

\subsubsection{$U$-$V$ phase diagrams}
\begin{figure*}[tp]
\centering
\includegraphics[width=\columnwidth]{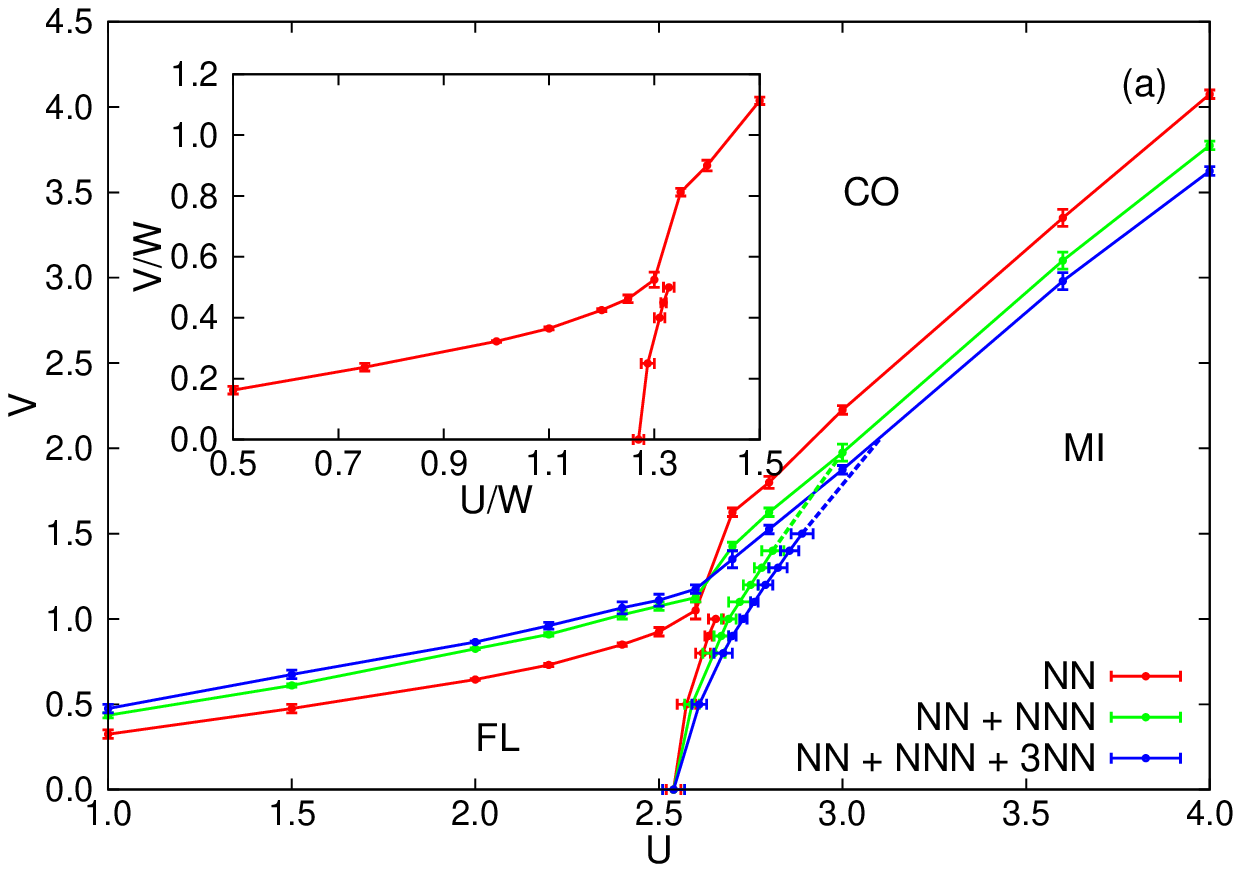}
\includegraphics[width=\columnwidth]{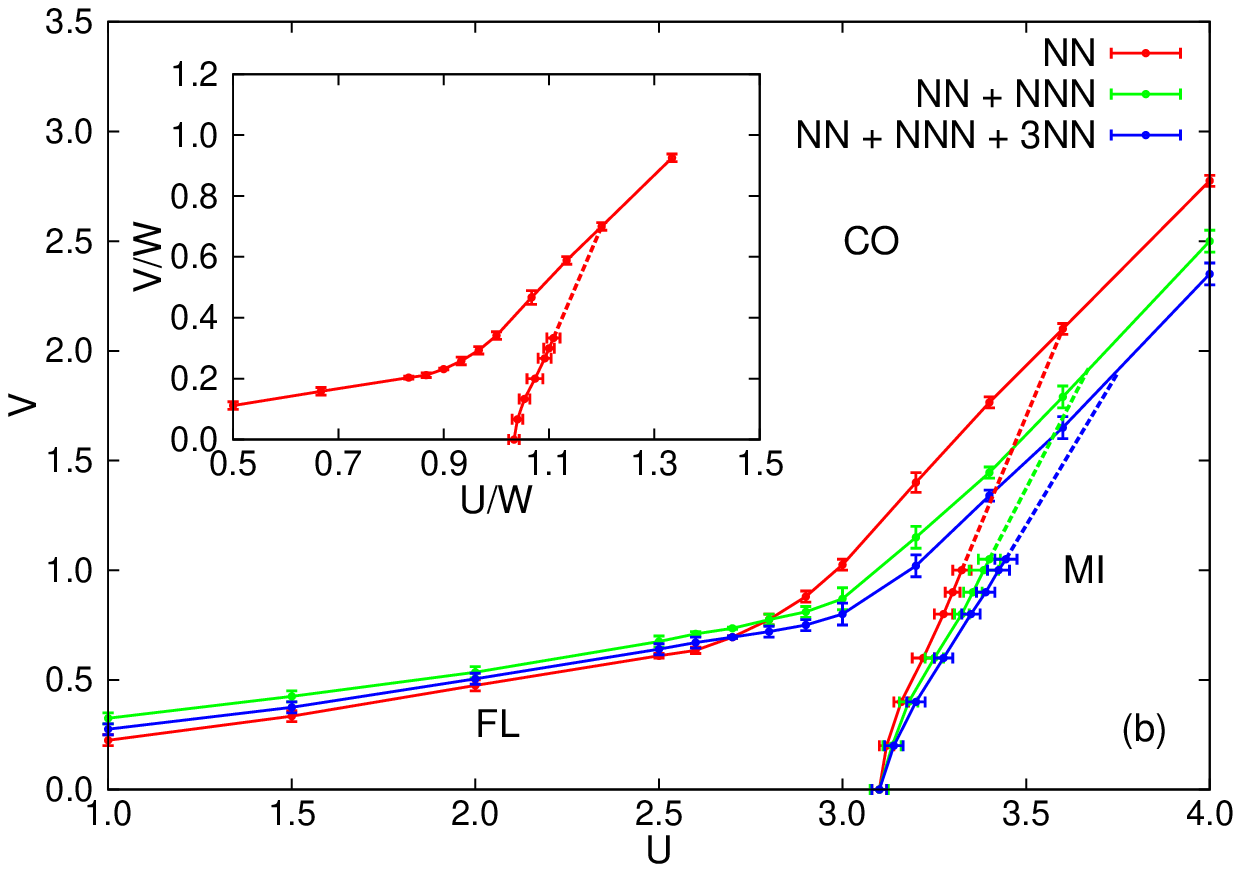}
\caption{(Color online) The paramagnetic $U$-$V$ phase diagrams for the single-band half-filled extended Hubbard model determined by EDMFT calculations. (a) Phase diagram for the 2D square lattice. (b) Phase diagram for the 3D simple cubic lattice. Here CO denotes a charge-ordered insulating phase, FL the metallic state, and MI the Mott insulator. The dashed lines are extrapolated FL-MI phase boundaries. The insets in (a) and (b) show the phase diagrams with axes rescaled by the bandwidth. \label{fig:phase}}
\end{figure*}

Figure~\ref{fig:phase} shows the phase diagrams in the space of the parameters $U$ and $V$. In this figure, the left panel shows the result for the square lattice, and the right panel corresponds to the simple cubic lattice. Both phase diagrams exhibit three phases: a metallic Fermi-liquid (FL) phase in which the kinetic energy dominates the interactions, the Mott insulating (MI) phase with one particle per site, where $U$ is dominant, and the charge-ordered (CO) insulator with a charge density wave (CDW) when $V$ prevails. The insets plot the phase diagrams with axes rescaled by the bandwidth ($8t$ for the square lattice and $12t$ for the simple cubic lattice), to emphasize the similarities and differences between the 2D and 3D cases. 

The paramagnetic phase diagram for the extended Hubbard model with the NN interactions on the square lattice is consistent with the result by Ayral \emph{et al.}\cite{PhysRevB.87.125149} The paramagnetic phase diagram for the simple cubic lattice with the NN interactions has been calculated in the pioneering paper by Sun \emph{et al.}\cite{PhysRevB.66.085120} Their calculations however were performed at a much higher temperature ($\beta = 5$), above the end-point of the FL-MI transition. Also, the quantum impurity solver used in that study was a modified Hirsch-Fye algorithm with Bose factor approximation,\cite{PhysRevB.62.12800} which is not as accurate as the numerically exact CT-QMC algorithm.\cite{PhysRevLett.104.146401} Taking into account these differences, the phase diagram presented in Fig.~\ref{fig:phase}(b) appears to be qualitatively consistent with the previous result by Sun \emph{et al.}\cite{PhysRevB.66.085120} When the temperature is increased, the $V_c(U)$ line shifts upwards, and the $U_c(V)$ line is shifted to the left. In contrast to the paramagnetic MI, the CO insulator does not have a large entropy of $\ln 2$ per site (the phase boundary is determined from the divergence in the charge susceptibility, see subsection \ref{sub:transition} for further details).

In the previous calculations, only the NN intersite interactions have been included. In the present work, we also consider the effects of longer range interactions, more specifically the NNN and 3NN intersite interactions, as depicted in Fig.~\ref{fig:lattice}. In a future study, it would be interesting to consider the effect of an infinite range Coulomb $1/r$-type tail. A proper treatment of it requires an Ewald lattice summation, as is shown by Hansmann \emph{et al.}\cite{PhysRevLett.110.166401}

The modifications in the phase diagram for the square lattice are shown in Fig.~\ref{fig:phase}(a). When $U$ is small, the $V_c(U)$ lines is shifted upward if the NNN and 3NN interactions are added, which means that these longer range intersite interactions destabilize the CO state. This is not surprising, since the left panel of Fig.~\ref{fig:lattice} shows that both the NN and 3NN interactions act between sites of the same sub-lattice, and hence penalize the CDW. In the strongly correlated region, the $V_c(U)$ line is shifted downward, which means that the MI state is suppressed by longer range intersite interactions, which can be interpreted as the result of the enhanced screening of the onsite interaction. For the same reason, the $U_c(V)$ line is slightly shifted to the right. Finally, if only the NN intersite interaction is considered, the $V_c(U)$ line ``jumps" in the region where the $V_c(U)$ and $U_c(V)$ lines intersect, and this jump is accompanied by a change of the slope. If longer range interactions are included, the metallic phase extends to larger values of $U$, so that the transition between MI and CO phases is no longer a direct one, at least for $2.5 \lesssim U \lesssim 3.0$. As a result of this intermediate metallic phase, the jump in the $V_c(U)$ line disappears. We note that the shape of the metallic phase with longer range interactions is qualitatively similar to the FL phase in the single-band Holstein-Hubbard model with large phonon frequency.\cite{PhysRevLett.104.146401} One difference is that the phase diagram for the Holstein-Hubbard model does not have a sudden slope change in the phase boundary to the CO phase in the vicinity of the Mott transition. This suggests that the slope change in the extended Hubbard model originates from changes in the screening processes near $U_c$. We will investigate this issue in more detail in subsection~\ref{sub:screening}.

Next, let's turn to the simple cubic lattice case [see Fig.~\ref{fig:phase}(b)]. Here, for small $U$, the $V_c(U)$ phase boundary is shifted upward when the NNN interaction is added, just as in the 2D case, but the 3NN interaction has the opposite effect. Therefore, the shift is not monotonous any more. This can be understood by looking at the right hand panel of Fig.~\ref{fig:lattice}. While the NNN interactions act between sites on the same sub-lattice, and hence frustrate the CDW, the 3NN interactions act between sites on different sub-lattices, and thus favor the CO phase. Another difference to the 2D case is that the metallic region between the MI and CO phases is larger, so that there is no obvious ``kink" or sudden ``jump" in the $V_c(U)$ line near the Mott transition. In fact, for the model with only the NN interactions, the slope change in the $V_c(U)$ line happens already quite a bit before the Mott transition ($U_c \sim 3.1$) at $V = 0$.

\subsubsection{Charge-ordering and Mott metal-insulator transitions\label{sub:transition}}
\begin{figure*}[tp]
\centering
\includegraphics[width=\columnwidth]{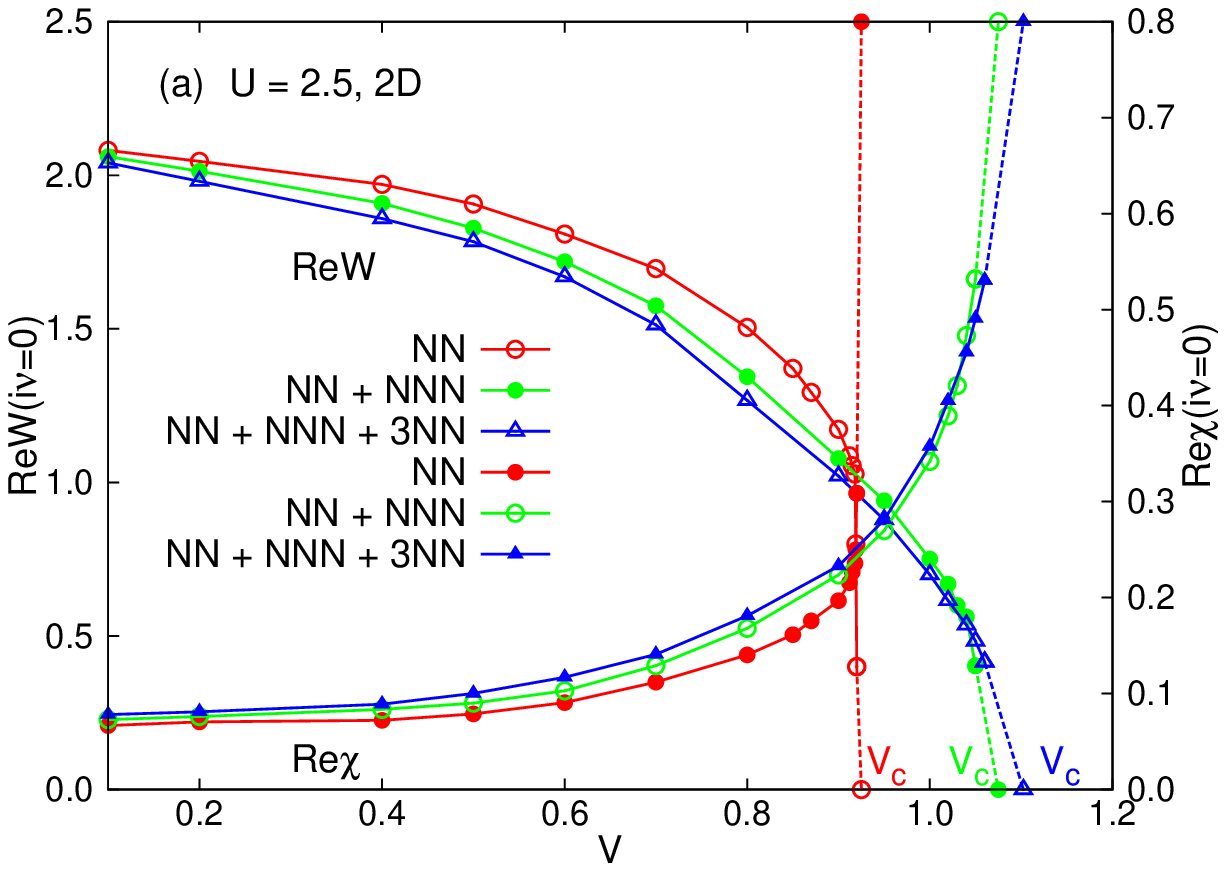}
\includegraphics[width=\columnwidth]{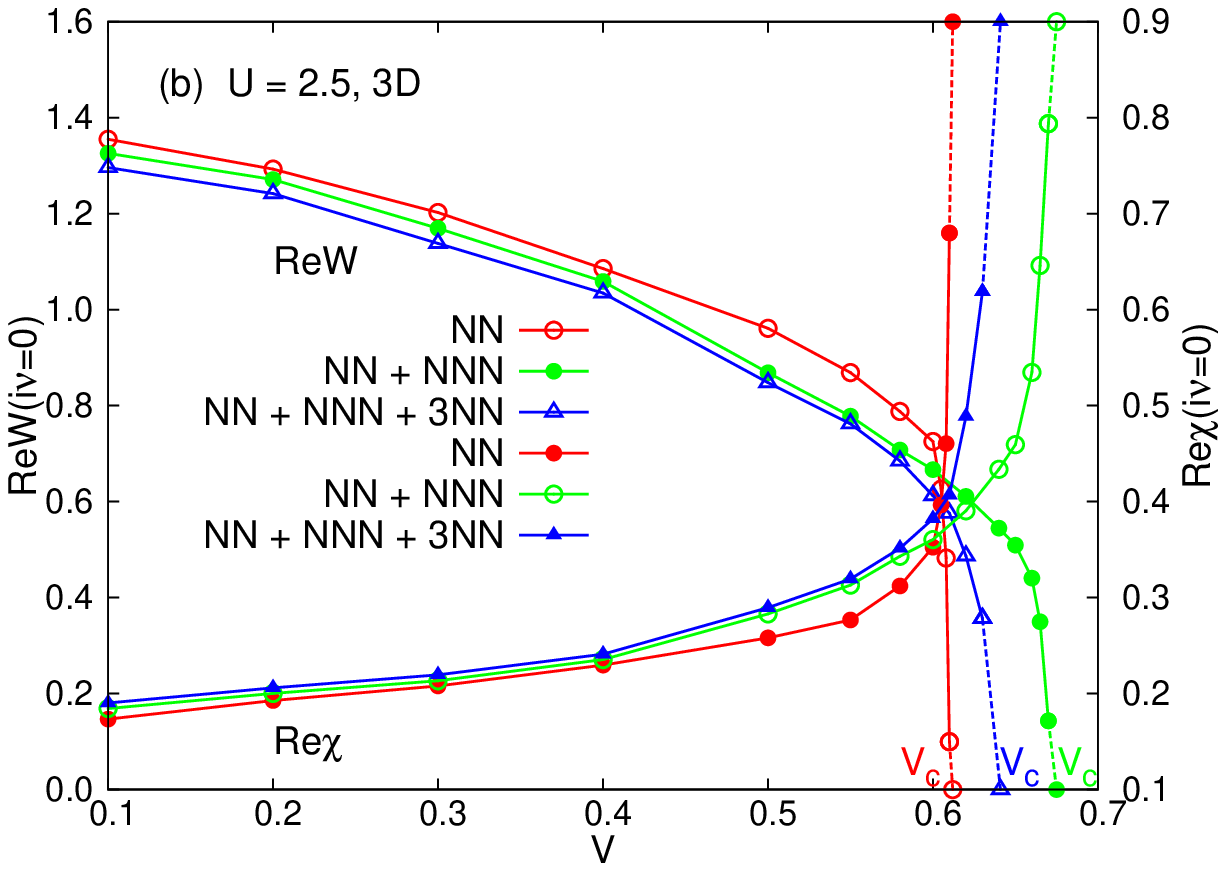}
\caption{(Color online) Re$W(i\nu = 0)$ (see left $x$-axis) and Re$\chi(i\nu = 0)$ (see right $y$-axis) as a function of $V$. $U = 2.5$. (a) Results for the square lattice. (b) Results for the simple cubic lattice. The dashed lines are used to determine $V_c$ for the charge-ordering transition. \label{fig:wzero}}
\end{figure*}

\begin{figure*}[tp]
\centering
\includegraphics[width=\columnwidth]{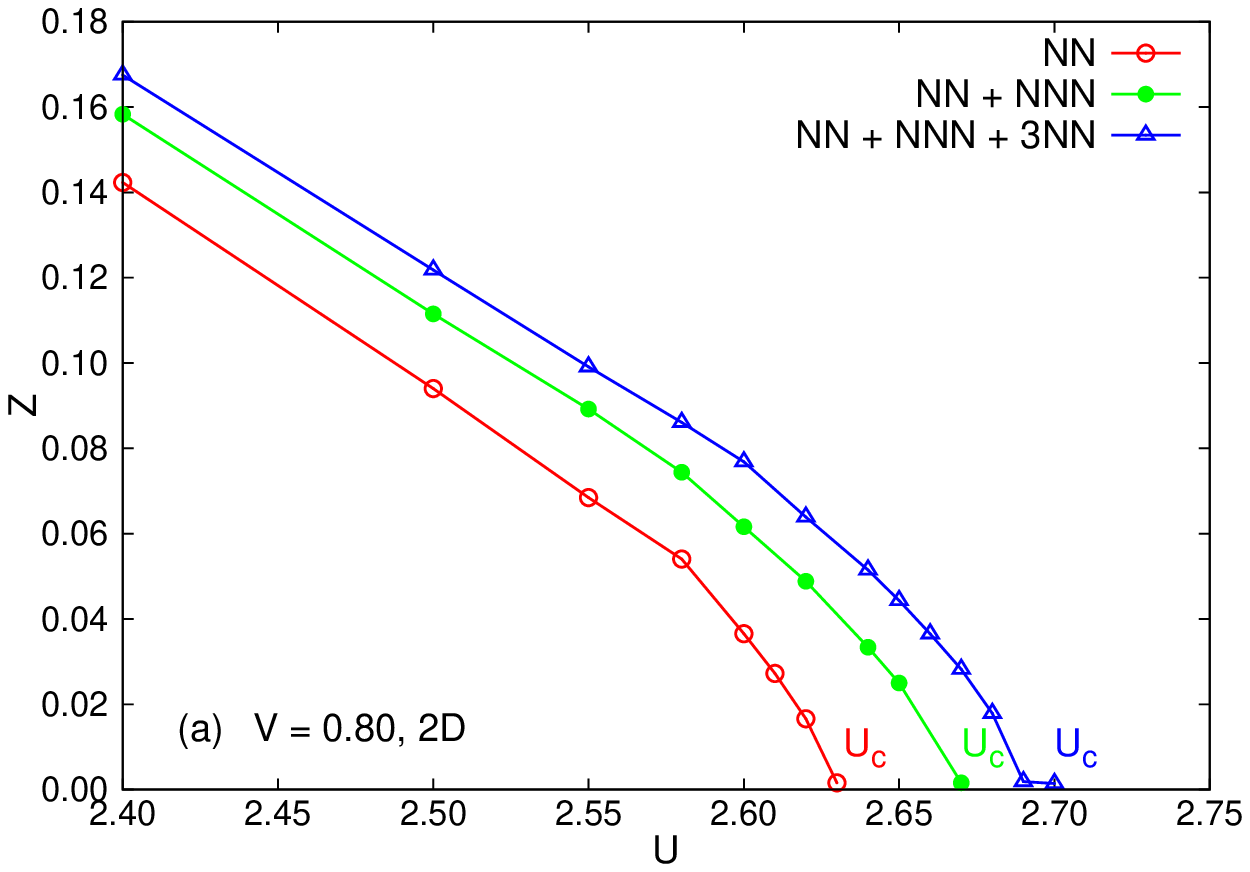}
\includegraphics[width=\columnwidth]{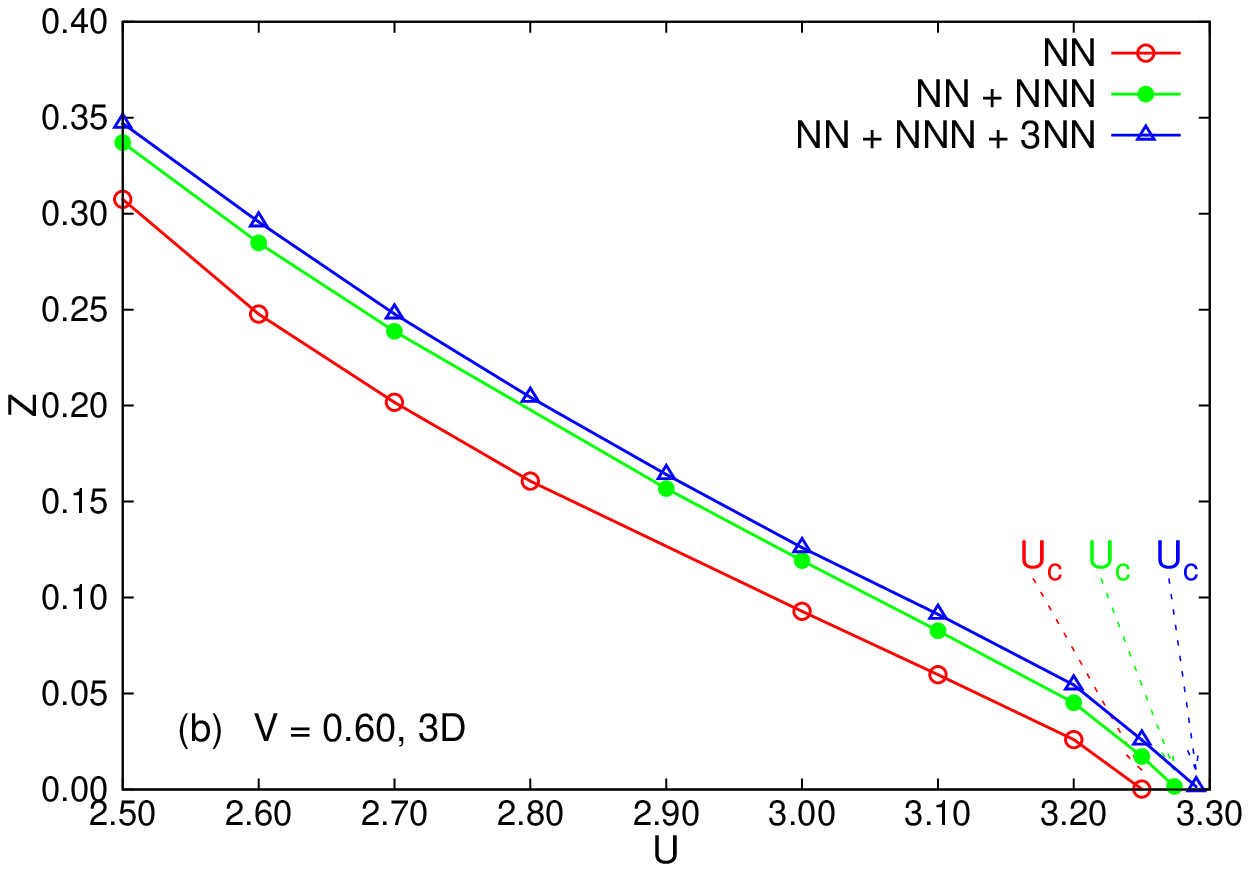}
\caption{(Color online) Quasiparticle weight $Z$ as a function of $U$. (a) Results for the square lattice, $V = 0.80$. (b) Results for the simple cubic lattice, $V = 0.60$. When $Z$ goes to zero, the Mott-Hubbard metal-insulator transition occurs. The corresponding $U$ is $U_{c}$. In panel (b) the dashed lines are used to guide the eyes. \label{fig:mott}}
\end{figure*}

The phase transition from the FL and MI phases to the CO phase is signaled by a diverging charge susceptibility $\chi(i\nu = 0)$.\cite{PhysRevB.66.085120} This divergence almost coincides with a sign change in the fully screened interaction Re$W(i\nu=0)$ [see Eq.~(\ref{eq:W})]. When $V$ increases, Re$W(i\nu = 0)$ gets smaller, and when it reaches zero, the cost for the formation of doublons vanishes.\cite{PhysRevB.87.125149} In Fig.~\ref{fig:wzero}, the real parts of $W(i\nu = 0)$ and $\chi(i\nu = 0)$ are plotted against $V$ for $U = 2.5$, which is still in the metallic state for the square and simple cubic lattices. The phase boundary to the CO state has been located by approaching the phase transition from below $V_c$. Actually, before Re$\chi(i\nu=0)$ diverges or Re$W(i\nu = 0)$ reaches zero, we already encounter a numerical instability which prevents the convergence of the EDMFT self-consistency loop. Thus, we extrapolate the curves using $(V - V_c)^{-1}$, as shown by the dashed lines in Fig.~\ref{fig:wzero}, to determine the critical $V_{c}$. While the extrapolation procedure is somewhat arbitrary, the trend is unambiguous: In the square lattice case, $V_c$ increases as we add longer range interactions, even though for $V \lesssim 0.9$, the trend is actually opposite (due to an increasing screening effect). For the simple cubic lattice, the screening effect leads to a reduction of Re$W(i\nu=0)$ with increasing range of the interaction for $V \lesssim 0.6$, but then the drop to zero occurs in a non-monotonic way, for reasons related to lattice geometry as discussed above. In the large-$U$ region, close to the Mott transition, the $V_c(U)$ phase boundary shifts down with increasing range of the interaction, both for the square and the simple cubic lattice. This indicates that the interaction induced changes in the screening function should play the dominant role there.

The phase boundary between metal and Mott insulator is signaled by a vanishing spectral weight at the Fermi level. We increased the onsite interaction $U$ step by step to approach the phase transition from the FL metallic side, so that our $U_c$ values indicate the stability region of the metallic phase ($U < U_c$). In our calculations, the Mott metal-insulator transition is determined by computing the quasiparticle weight $Z$\cite{RevModPhys.68.13}
\begin{equation}
Z = \Bigg[1 - \frac{\text{Im}\Sigma(i\omega_0)}{\omega_0}\Bigg]^{-1},
\end{equation}
where $\omega_0$ is the first Matsubara frequency $\omega_0 = \pi / \beta$. Strictly speaking, this equation is only valid at zero temperature, but our temperature is low enough ($\beta=100$) that it can be regarded as a good approximation. In Fig.~\ref{fig:mott}, the calculated quasiparticle weights $Z$ for the square and simple cubic lattices are plotted for selected $V$ parameters. This figure shows that longer range intersite interactions lead to a larger $Z$ and hence to a larger $U_c$. The reason is again a larger screening effect. 

\subsubsection{Screened and retarded interactions\label{sub:screening}}
\begin{figure*}[tp]
\centering
\includegraphics[width=0.32\textwidth]{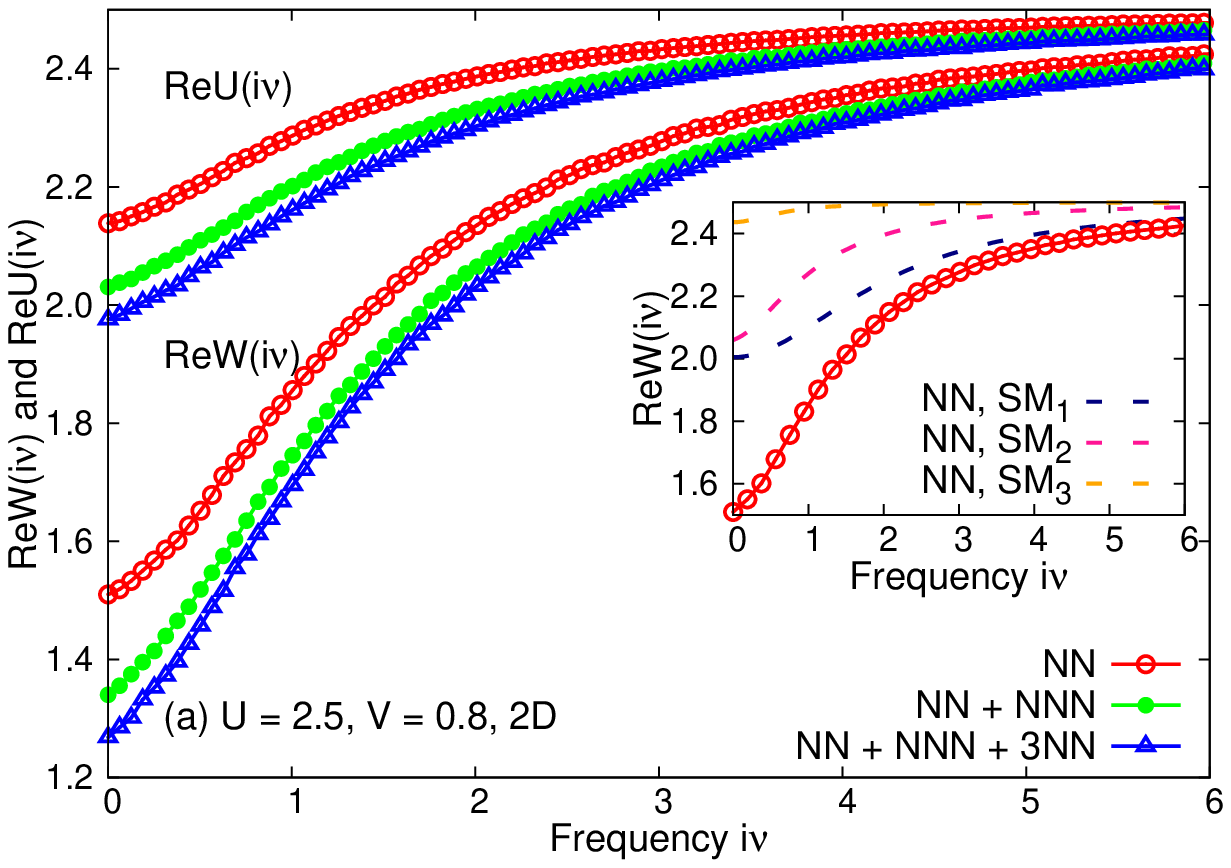}
\includegraphics[width=0.32\textwidth]{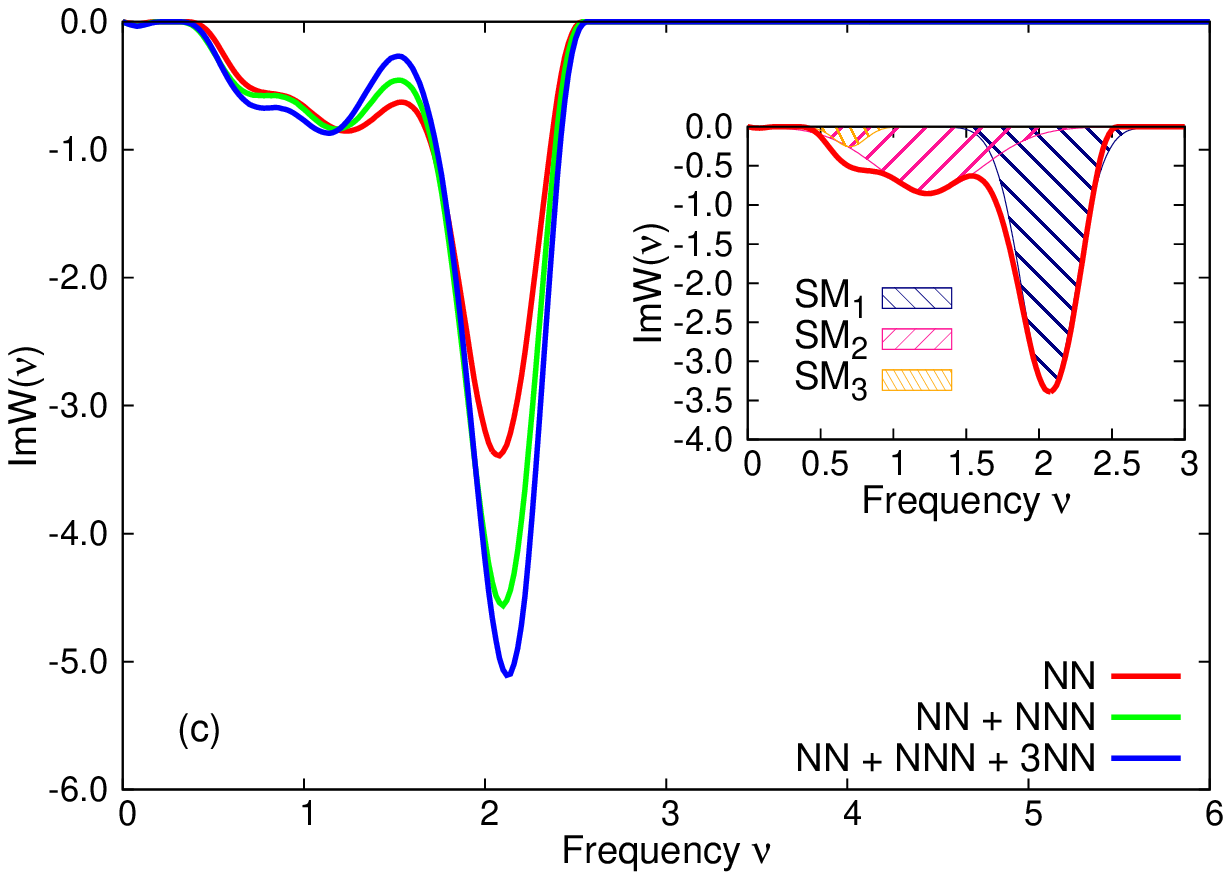}
\includegraphics[width=0.32\textwidth]{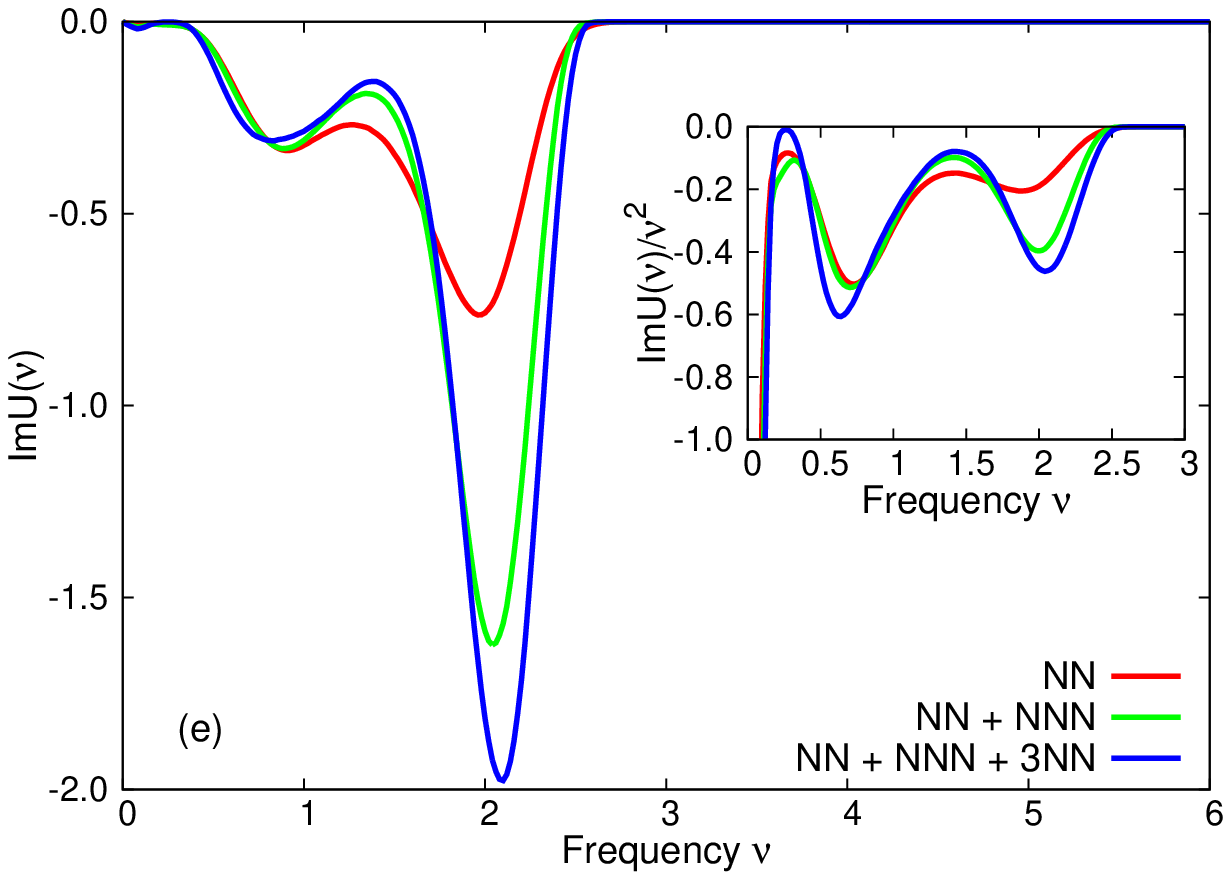}
\includegraphics[width=0.32\textwidth]{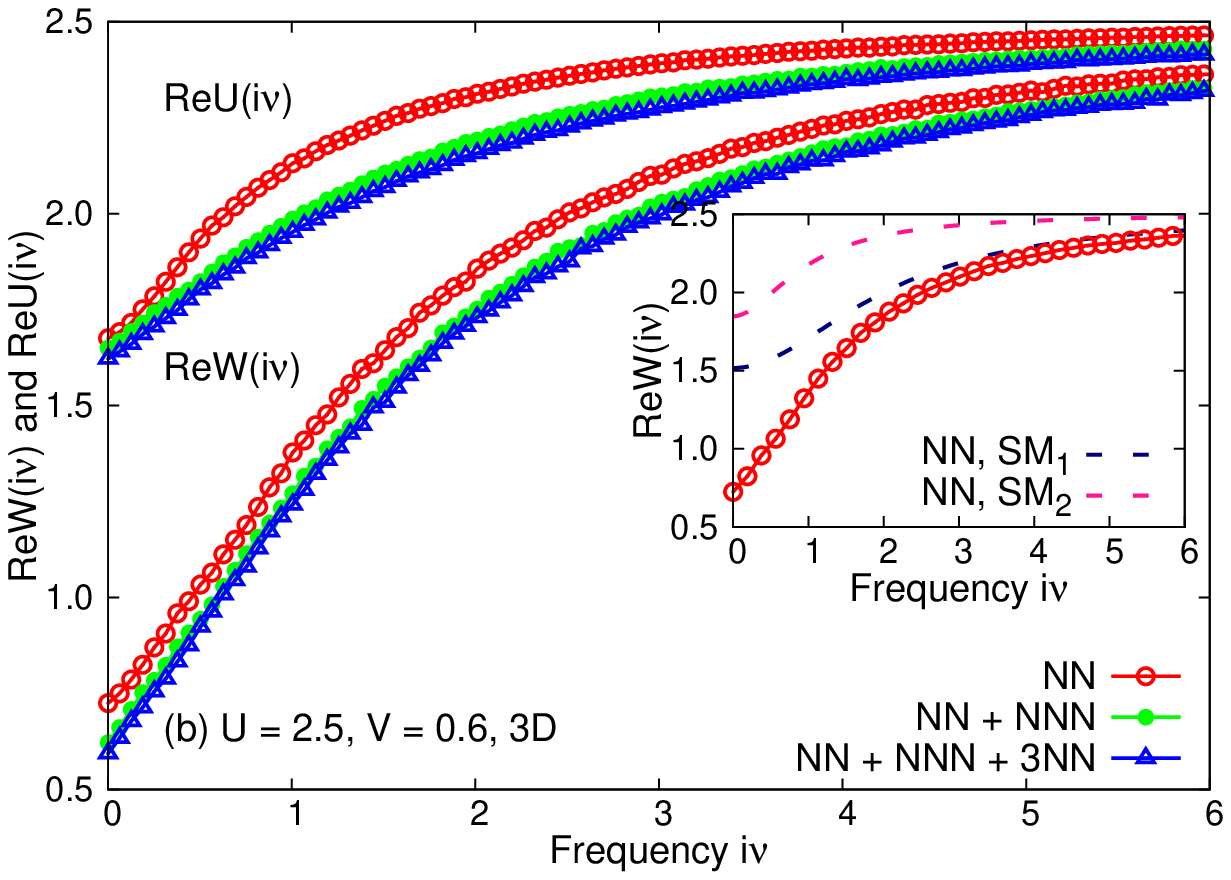}
\includegraphics[width=0.32\textwidth]{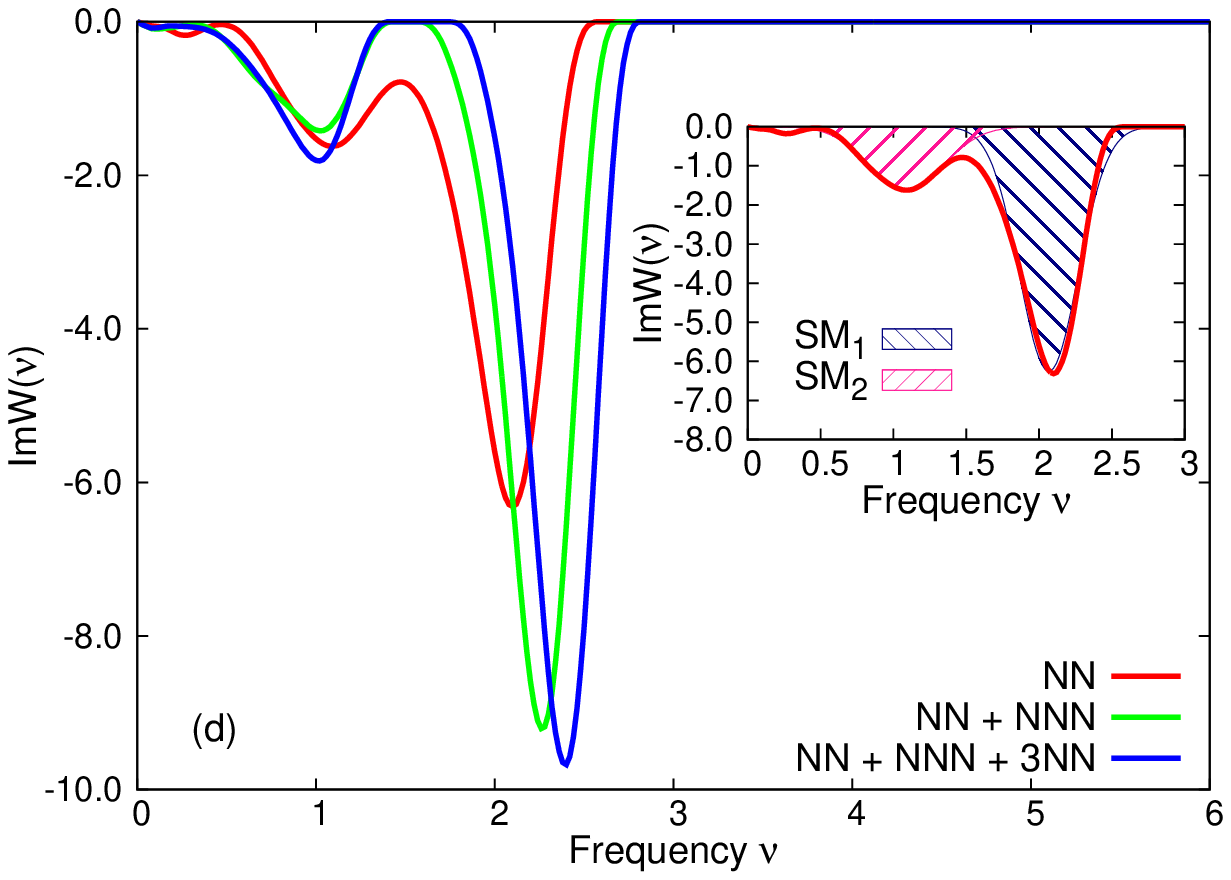}
\includegraphics[width=0.32\textwidth]{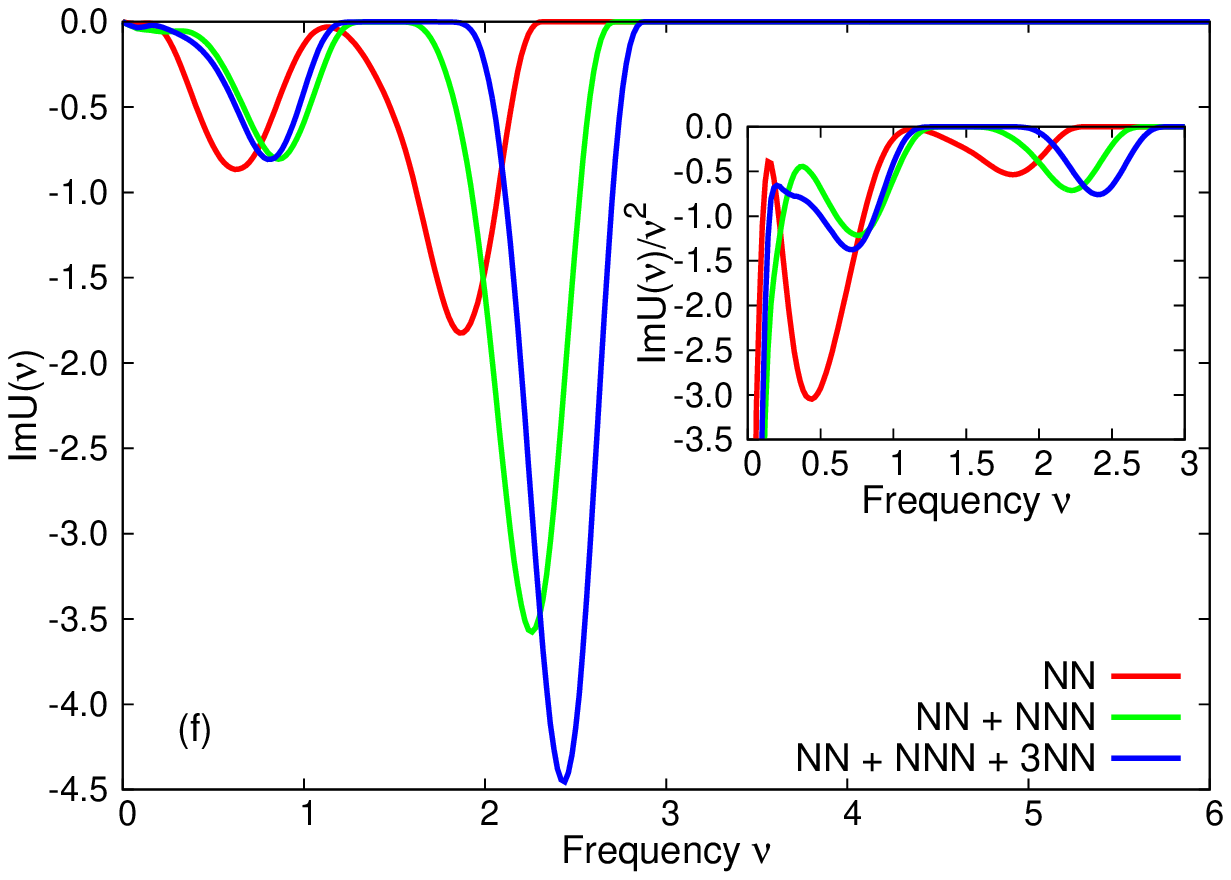}
\caption{(Color online) Real part of fully screened interactions Re$W(i\nu)$ and partially screened interaction Re$\mathcal{U}(i\nu)$, imaginary part of real frequency fully screened interaction Im$W(\nu)$ and partially screened interactions Im$\mathcal{U}(\nu)$ for the extended Hubbard model solved by EDMFT. (a), (c) and (e) Results for the square lattice, $U = 2.5$ and $V = 0.8$. (b), (d) and (e) Results for the simple cubic lattice, $U = 2.5$ and $V = 0.6$. In this figure, SM means screening mode. In the insets of (a) and (b) panels, the SM-resolved Re$W(i\nu)$, together with full Re$W(i\nu)$ are shown for the NN case. In (c) and (d) panels, the Im$W(\nu)$ for the NN case is approximated by Gaussian-type functions. The fitted results are shown in the insets. Each Gaussian peak denotes a SM. The insets in (e) and (f) panels show the Im$\mathcal{U}(\nu)/\nu^2$ functions. Here Im$W(\nu)$ and Im$\mathcal{U}(\nu)$ are extracted using a modified maximum entropy method. See Appendix \ref{app:mem} for more details. \label{fig:wloc23}}
\end{figure*}

\begin{figure}[tp]
\centering
\includegraphics[width=\columnwidth]{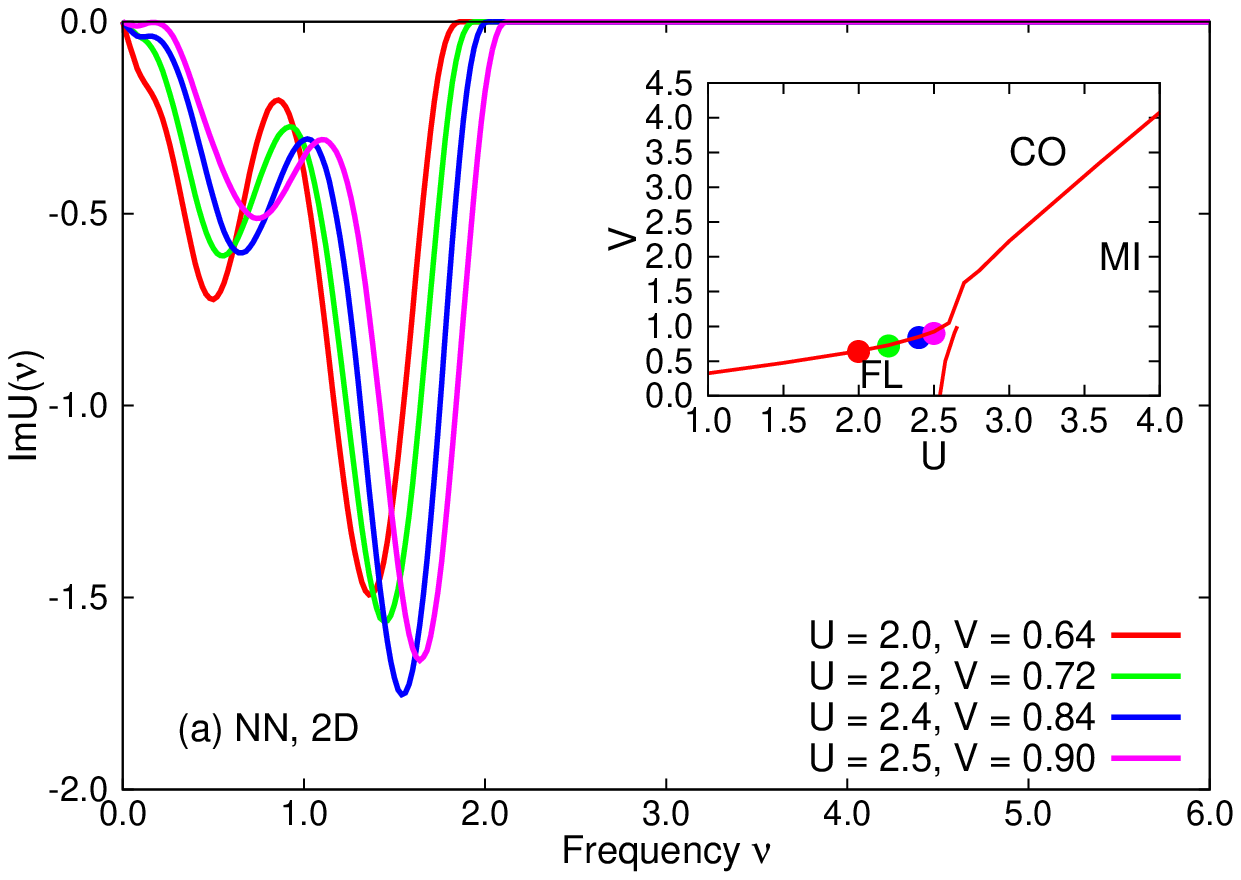}
\includegraphics[width=\columnwidth]{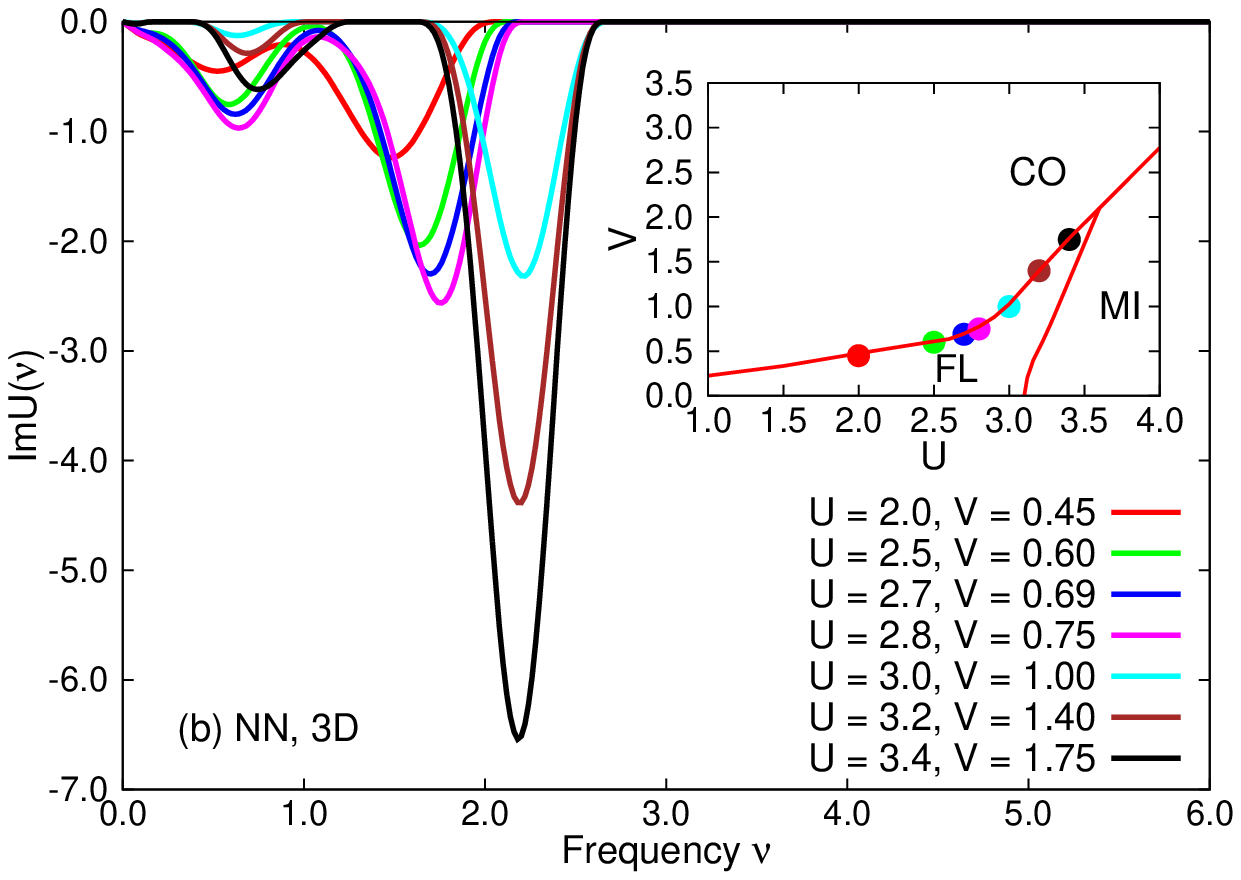}
\includegraphics[width=0.47\columnwidth]{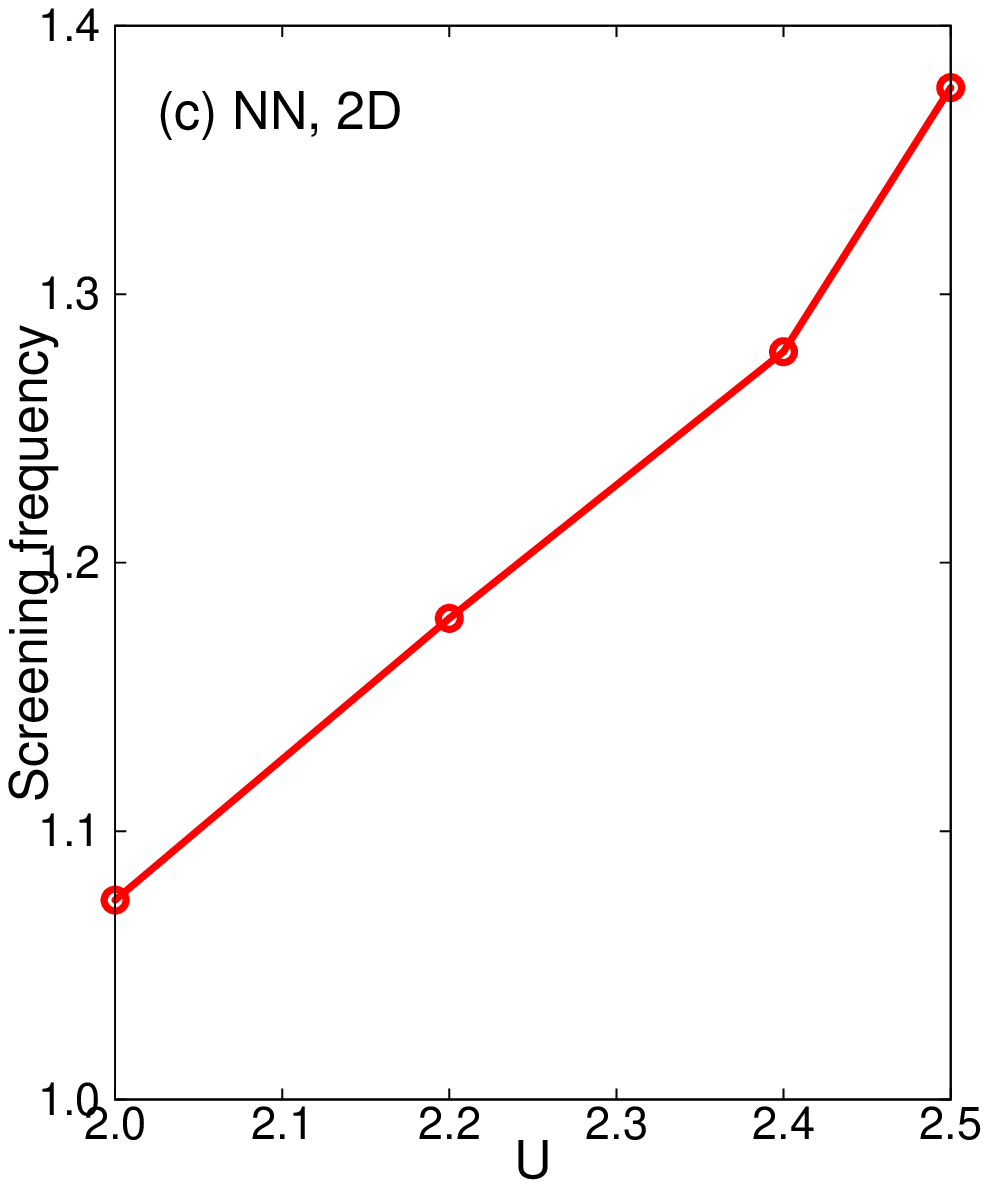}
\includegraphics[width=0.47\columnwidth]{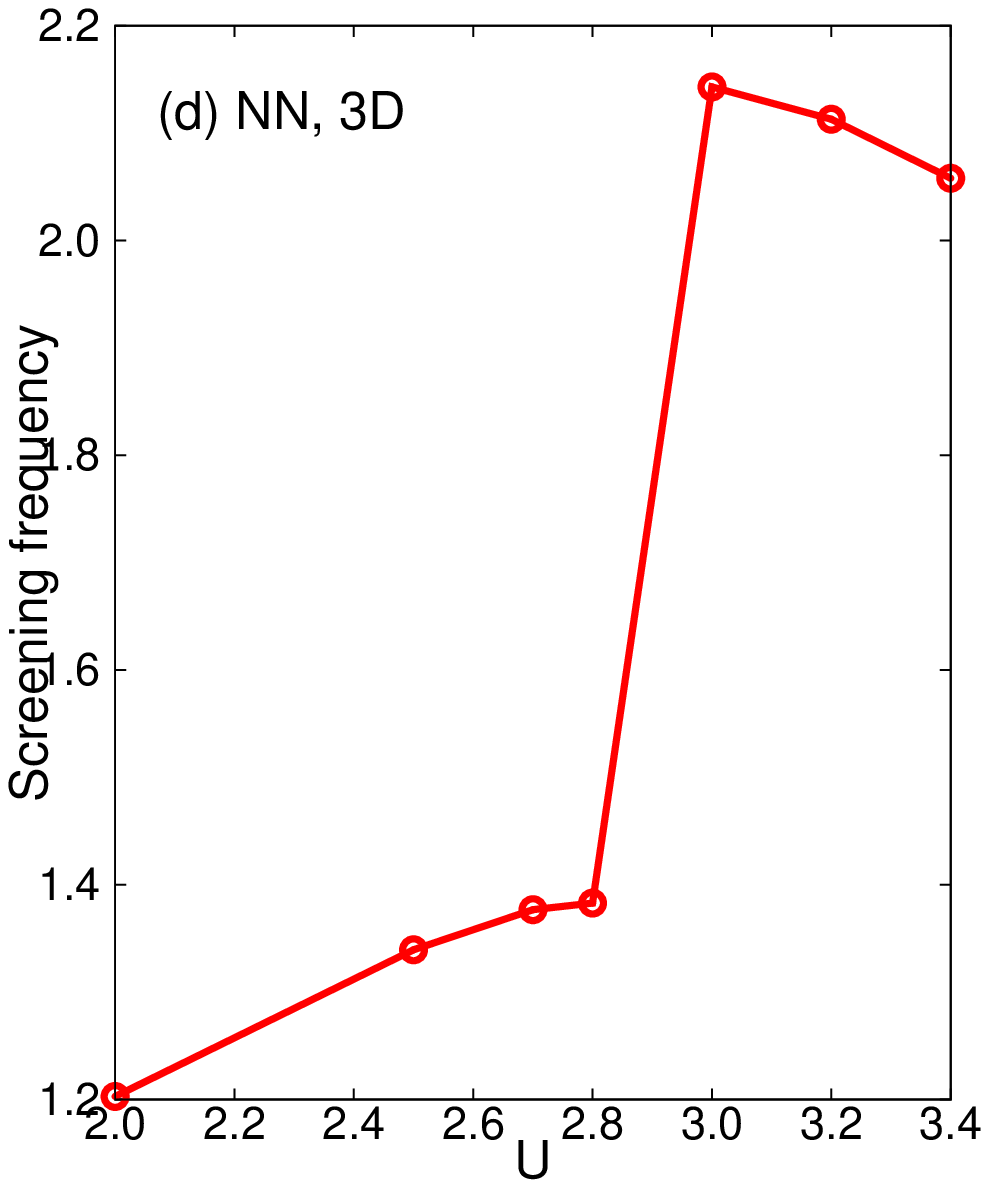}
\caption{(Color online) Imaginary part of real frequency partially screened interactions Im$\mathcal{U}(\nu)$ for the extended Hubbard model with the NN interactions solved by EDMFT. (a) Results for the square lattice. (b) Results for the simple cubic lattice. The $U$ and $V$ parameters are shown as color-filled circles in the insets. In (c) and (d), the corresponding effective screening frequencies $\nu_0$ are shown. \label{fig:uloc_scan23}}
\end{figure}

\begin{table*}[tp]
\begin{tabular}{ccccccccccc}
\hline
\hline
\multicolumn{11}{c}{Metallic state} \\
\hline
& \multicolumn{5}{c}{Square lattice} & \multicolumn{5}{c}{Simple cubic lattice} \\
\hline
mode            & $V$  & $U$  & Re$\mathcal{U}(\nu = 0)$  & Re$W(\nu = 0)$  & $\nu_0$     & $V$  & $U$  & Re$\mathcal{U}(\nu = 0)$  & Re$W(\nu = 0)$  & $\nu_0$ \\
\hline
NN              & 0.80 & 2.50 & 2.14 (2.36)               & 1.51 (1.96)     & 1.61 (1.11) & 0.60 & 2.50 & 1.68 (2.21)               & 0.73 (1.23)     & 1.44 (1.06) \\
NN + NNN        & 0.80 & 2.50 & 2.03 (2.31)               & 1.34 (1.91)     & 1.77 (1.12) & 0.60 & 2.50 & 1.65 (2.06)               & 0.62 (1.15)     & 1.96 (1.12) \\
NN + NNN + 3NN  & 0.80 & 2.50 & 1.98 (2.28)               & 1.27 (1.86)     & 1.84 (1.10) & 0.60 & 2.50 & 1.62 (2.03)               & 0.59 (1.14)     & 2.14 (1.16) \\
\hline
\multicolumn{11}{c}{Mott insulating state} \\
\hline
& \multicolumn{5}{c}{Square lattice} & \multicolumn{5}{c}{Simple cubic lattice} \\
\hline
mode            & $V$  & $U$  & Re$\mathcal{U}(\nu = 0)$  & Re$W(\nu = 0)$  & $\nu_0$     & $V$  & $U$  & Re$\mathcal{U}(\nu = 0)$  & Re$W(\nu = 0)$  & $\nu_0$ \\
\hline
NN              & 1.50 & 3.00 & 2.75 (2.82)               & 2.54 (2.61)     & 2.48 (1.75) & 1.50 & 3.60 & 3.24 (3.33)               & 2.98 (3.08)     & 2.88 (2.16) \\
NN + NNN        & 1.50 & 3.00 & 2.63 (2.75)               & 2.40 (2.55)     & 2.52 (1.66) & 1.50 & 3.60 & 2.81 (3.04)               & 2.50 (2.79)     & 2.84 (2.00) \\
NN + NNN + 3NN  & 1.50 & 3.00 & 2.56 (2.71)               & 2.34 (2.51)     & 2.54 (1.66) & 1.50 & 3.60 & 2.56 (2.98)               & 2.27 (2.74)     & 2.87 (2.00) \\
\hline
\hline
\end{tabular}
\caption{Summary of Re$\mathcal{U}(\nu = 0)$, Re$W(\nu = 0)$ and effective screening frequency $\nu_0$ for $U$ and $V$ parameters in the metallic and Mott insulating regime. The $\nu_0$ is defined by Eq.~(\ref{eq:scr_frq}). The results in parentheses are from fully self-consistent $GW$ + EDMFT calculations (see Sec.~\ref{sub:gw_edmft} for further details), while the others are from self-consistent EDMFT calculations. \label{tab:square_cubic}}
\end{table*}

In the top panels of Fig.~\ref{fig:wloc23}, we plot the real parts of $W(i\nu)$ and $\mathcal{U}(i\nu)$, and the imaginary parts of $W(\nu)$ and $\mathcal{U}(\nu)$ for the square lattice with selected $U$ and $V$ parameters. The counterparts for the simple cubic lattice are shown in the bottom panels of Fig.~\ref{fig:wloc23}. We concentrate here on the FL region for both the 2D and 3D lattices. When $\nu \to \infty$, both the fully screened interactions Re$W(i\nu)$ and partially screened interactions Re$\mathcal{U}(i\nu)$ [see Fig.~\ref{fig:wloc23}(a) and (b)] asymptotically approach the bare interaction $U$. As the frequency $\nu$ is lowered, Re$W(i\nu)$ and Re$\mathcal{U}(i\nu)$ decrease monotonously. Longer range intersite interactions produce a stronger screening effect, and lead to lower values of the static interactions Re$W(i\nu=0)$ and Re$\mathcal{U}(i\nu=0)$. 

Let us take a closer look at the Im$W(\nu)$ and Im$\mathcal{U}(\nu)$ spectra, which we have obtained from a modified maximum entropy procedure\cite{mem:1996} (see Appendix~\ref{app:mem}). To analyze the spectra, we fit Im$W(\nu)$ with multiple Gaussians. Each peak can be regarded as a screening mode (abbreviated as SM), and the position of the peak corresponds to the screening frequency. Figures~\ref{fig:wloc23}(c) and (d) show that the Im$W(\nu)$ spectra feature two prominent SMs, whose screening frequencies differ by about a factor of two. The insets of Fig.~\ref{fig:wloc23}(a) and (b) show the contributions of these modes to the frequency dependence of Re$W(i\nu)$. In the Im$\mathcal{U}(\nu)$ spectra, one can also distinguish two humps, and the locations and weights of these screening modes are similar to the Im$W(\nu)$ counterparts. In both cases, the weight of the high-energy screening mode depends on the range of the intersite interaction. In the 3D case, the high-energy mode also seems to shift in energy, as longer range interactions are included. 

The physical interpretation of the two screening modes is somewhat subtle. As we will see in the following section, the spectral function in the metallic phase essentially exhibits a three-peak structure consisting of two Hubbard bands and a renormalized quasiparticle band. One can therefore distinguish screening processes stemming from transitions between the Hubbard bands, between the quasiparticle peak and one of the Hubbard bands, and within the quasiparticle band.\cite{PhysRevB.87.125149} It is natural to associate the high-energy screening mode with inter-Hubbard band transitions and the low-energy mode with transitions from the quasiparticle peak to either Hubbard band. Consistent with this interpretation is the fact that the energy difference between the two modes is roughly a factor of two. Even the energy values associated with the two modes are in good agreement with the energy separation between the two Hubbard bands and between the quasiparticle and the Hubbard bands, respectively (see Fig.~\ref{fig:aw23} below). One may however wonder why the bosonic spectra do not exhibit a low-energy mode related to transitions within the renormalized quasiparticle band. There is in fact no necessity for this to happen: even in the metallic phase, where $\text{Im}\chi_\text{imp}(\omega)$ has a Drude-like contribution $\pi\alpha\delta(\omega)$ and hence, by the Kramers-Kronig relation, $\text{Re}\chi_\text{imp}= \alpha/\omega$, the polarization $\Pi_\text{imp}$ does not have a pole at $\omega=0$. Indeed, taking $\mathcal{U}=U$ for simplicity, we have $\Pi_\text{imp}=-\chi_\text{imp}/(1-\mathcal{U}\chi_\text{imp})=-\alpha/(\omega-\alpha U)$. As a result, the screened interaction does not have a pole at $\omega=0$ either: $W_\text{loc}=\sum_q v_q/(1-v_q\Pi_\text{imp})=\sum_q v_q(\omega-\alpha  U)/(\omega-\alpha (U-v_q))$.

It is worth noting that the structures in the Im$\mathcal{U}(\nu)/\nu^2$ function, which are shown in the insets of Fig.~\ref{fig:wloc23}(e) and (f), determine the most relevant screening modes and the associated energies of satellites in the local spectral function $A(\omega)$.\cite{Werner2012} Therefore, despite the smaller weight, the low-energy mode is equally or even more important than the high-energy mode. In order to quantify the evolution of the screening modes by a single number, we define the effective screening frequency $\nu_0$ as follows:\cite{PhysRevLett.109.126408}
\begin{equation}
\label{eq:scr_frq}
\nu_0 = \int^{\infty}_{0} \text{d}\nu \nu \text{Im} \mathcal{U}(\nu) \Big\slash \int^{\infty}_{0} \text{d}\nu \text{Im} \mathcal{U}(\nu).
\end{equation}
In Tab.~\ref{tab:square_cubic}, the static retarded interaction Re$\mathcal{U}(\nu=0)$, fully screened interaction Re$W(\nu=0)$, and the effective screening frequency $\nu_0$ are listed for some representative regions in the phase diagrams (see Fig.~\ref{fig:phase}). Re$\mathcal{U}(\nu=0)$ and Re$W(\nu=0)$ are two key quantities that can be used to quantify the screening effect. They decrease for longer range intersite interactions, irrespective of the strength of the bare interaction $U$, the strength of the intersite interaction $V$, and the lattice dimension. This is to be expected, since a longer ranged interaction increases the number of sites which participate in the screening process. In addition, Re$W(\nu = 0)$ is always smaller than Re$\mathcal{U}(\nu=0)$, since the former incorporates the screening effects not only from the nonlocal processes, but also from the local processes. As is seen in Tab.~\ref{tab:square_cubic}, the effective screening frequency increases with increasing range of the intersite interaction in the metallic phase, while it is almost independent of the range of the interaction in the Mott insulating phase. The larger the bare interaction, the larger the effective screening frequency, which is consistent with previous EDMFT calculations.\cite{PhysRevB.87.125149}

It is instructive to look at the evolution of the SM along the metallic side of the $V_c(U)$ phase boundary, especially in the $U$-region where this phase boundary exhibits a slope change. The results for the two and three dimensional lattices with the nearest neighbor interactions are shown in Fig.~\ref{fig:uloc_scan23}. In the case of the simple cubic lattice [Fig.~\ref{fig:uloc_scan23}(d)], the slope change is smooth and occurs quite a bit before $U$ reaches the $V=0$ Mott transition value $U_c$. The slope change therefore occurs within the metallic phase, and is not directly associated with the Mott transition. Nevertheless, there is a sudden increase in the effective screening frequency at $U \approx 2.9$, originating from a simultaneous shift in the energy of both screening modes. In the square lattice case [Fig.~\ref{fig:uloc_scan23}(c)], where the slope change occurs simultaneously with the Mott transition, the effective screening frequency does not exhibit such a jump within the metallic phase. These results, and the comparison with the phase diagram of the Holstein-Hubbard model\cite{PhysRevLett.104.146401} show that the slope change, which cannot be understood within a simple mean-field picture, is related to correlation induced changes in the effective screening frequency.

\subsubsection{Effective static interaction}
\begin{figure*}[tp]
\centering
\includegraphics[width=\columnwidth]{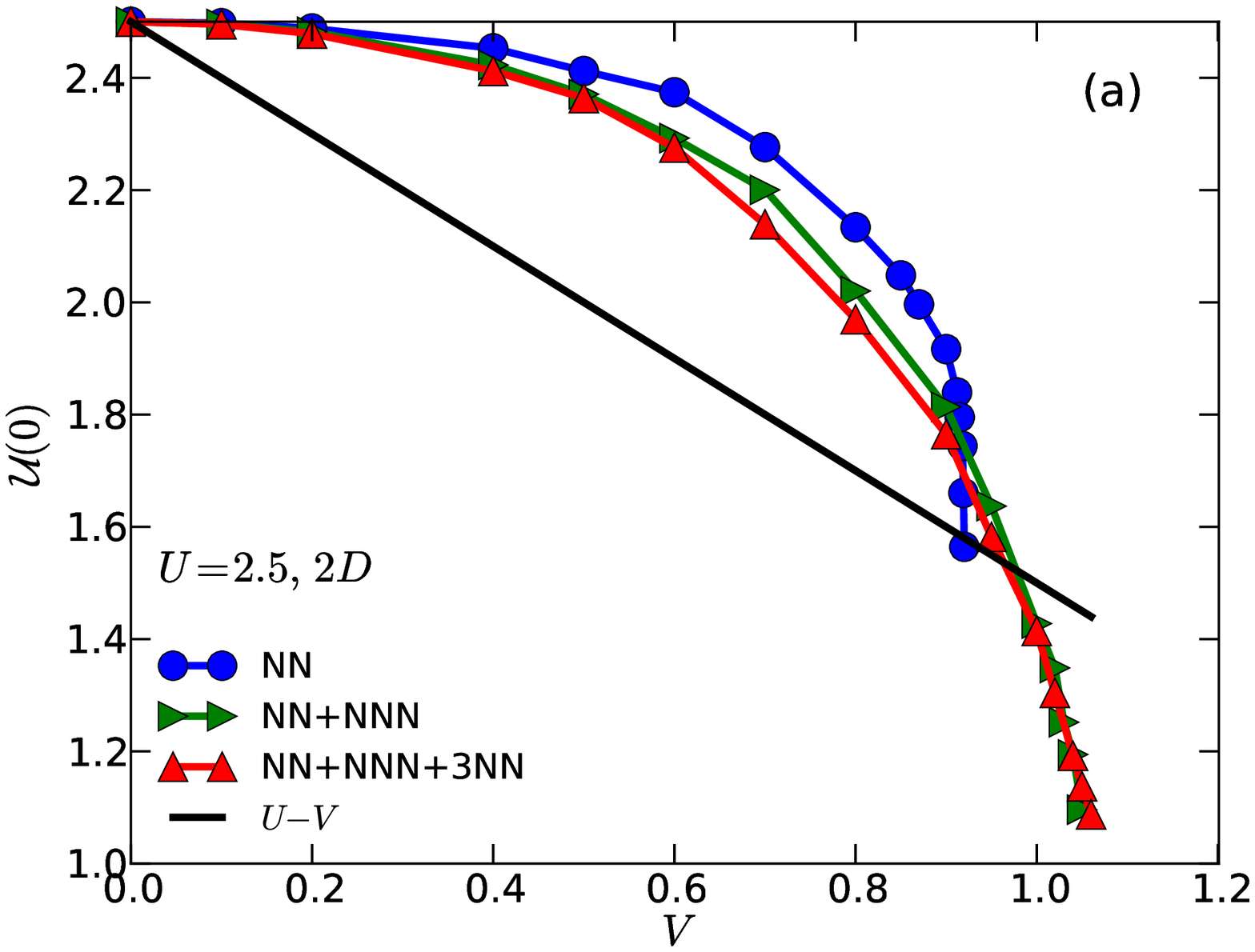}
\includegraphics[width=\columnwidth]{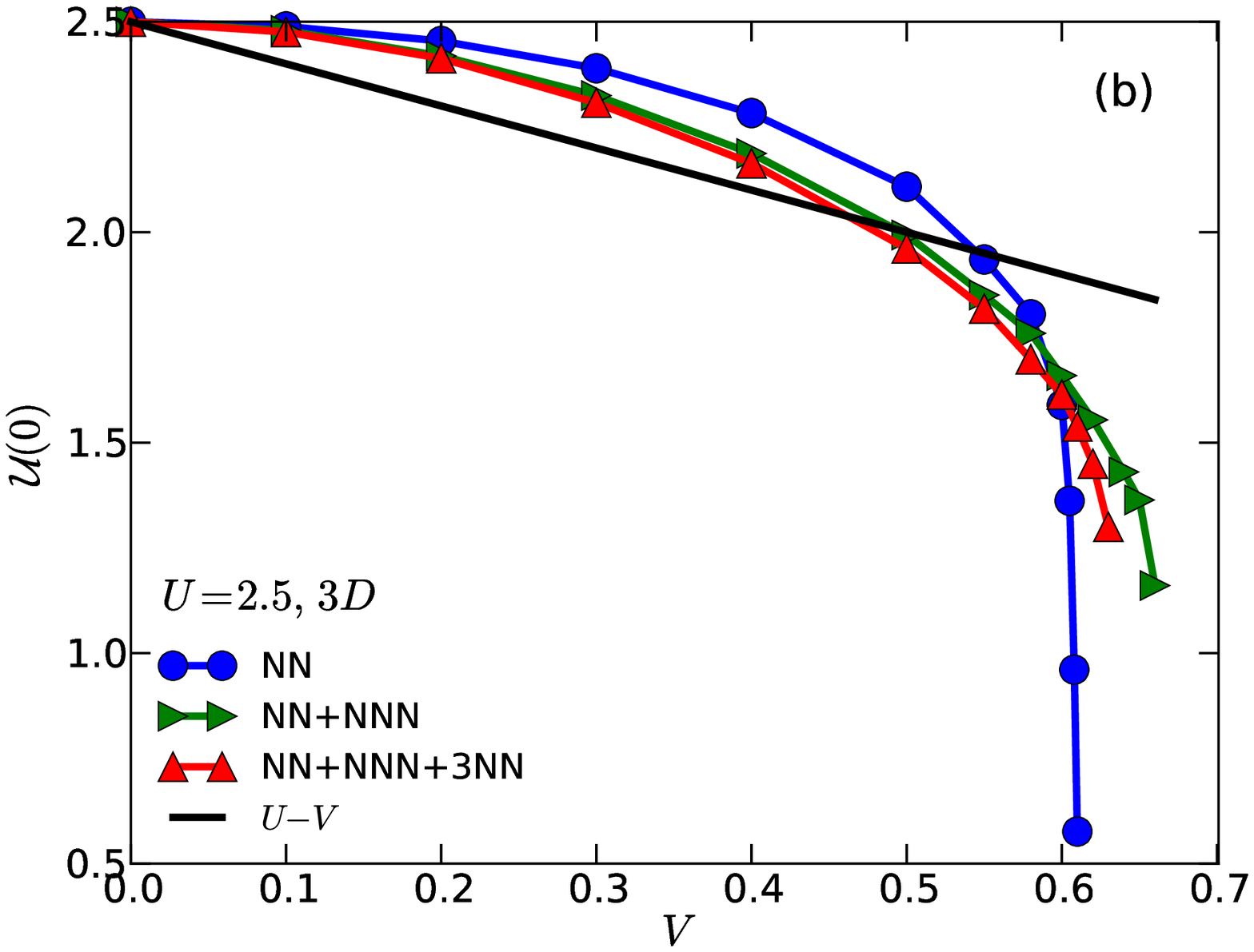}
\caption{(Color online) Comparison of the effective static interaction $\mathcal{U}(0)$ and the simple estimate $U - V$ [see Eq.~(\ref{Ueff2})]. (a) Results for the 2D model with $U=2.5$. (b) Results for the 3D model with $U=2.5$. \label{fig:u_min_v}} 
\end{figure*}

EDMFT provides an elegant means of constructing a model with purely local -- though dynamical -- interactions incorporating the effects of the nonlocal interactions in an effective manner. Furthermore, Ref.~\onlinecite{PhysRevLett.109.126408} demonstrated that -- at least in the anti-adiabatic limit -- a model with dynamical interactions can to a first approximation be thought of as a model with static interactions corresponding to the zero-frequency limit of the dynamical ones and a renormalized one-body Hamiltonian. These facts motivate a comparison of the zero-frequency limit of the effective dynamical interaction with attempts in the literature of constructing low-energy Hamiltonians with effective local static interactions, incorporating some of the screening effects stemming from longer range interactions. In Ref.~\onlinecite{PhysRevLett.111.036601}, it was shown that the {\it best} Hubbard model with purely local interactions mimicking the physics of a model with long-range interactions is one with modified local interactions. ``Best" is here defined in the sense of the Peierls-Feynman-Bogoliubov variational principle, leading to a free energy closest to the one of the original system. The result is an effective interaction where the bare interaction $U$ is modified by a weighted average of the nonlocal interaction matrix elements $V_{ij}$: 
\begin{eqnarray}
U_\text{eff} = U + \frac{1}{2} 
\sum_{i \ne j, \sigma, \sigma^{\prime}}
V_{ij}  \frac{\partial_{U_\text{eff}} \langle 
n_{i \sigma} n_{j \sigma^{\prime}} \rangle}
{\sum_l
\partial_{U_\text{eff}} \langle 
n_{l \uparrow} n_{l \downarrow} \rangle}.
\label{Ueff}
\end{eqnarray}
Here, the sums are over lattice sites and spins, and $\langle n_{i \sigma} n_{j \sigma^{\prime}} \rangle$ denotes the density-density correlator between sites $i$ and $j$. Assuming that a variation of $U$ leads to a displacement of charge only to the nearest neighbor sites, charge conservation leads to a further simplification. Eq.~(\ref{Ueff}) then reduces to 
\begin{eqnarray}
U_\text{eff} = U - V_{01},
\label{Ueff2}
\end{eqnarray}
that is, screening by nonlocal interactions results in a simple reduction of the onsite interaction by the nearest neighbor one. Numerical calculations for graphene, silicene and benzene in Ref.~\onlinecite{PhysRevLett.111.036601} indeed found values for the effective interactions close to the simple estimate given by Eq. (\ref{Ueff2}). Inspection of the calculations of Ref.~\onlinecite{PhysRevB.87.125149} for an extended Hubbard model in two dimensions with NN interactions reveals another interesting aspect: in these calculations screening was found to be strongly dependent on the regime, with barely any screening in the Mott phase (as expected) but a strong reduction of the effective local interaction in the correlated metal. Interestingly, however, the simple estimate of Eq.~(\ref{Ueff2}) was found to provide a lower bound with $U_\text{eff}$ coming closer to $U-V_{01}$ or $U$ depending on the proximity to the metallic or Mott phase, respectively.

Here, we address the question of the generic character of this observation. In Fig.~\ref{fig:u_min_v}, we plot the static part of the effective local interaction obtained from EDMFT as a function of $V$. As expected, this quantity is strongly reduced when approaching the phase boundary to the CO phase where strong charge fluctuations dominate. In the two-dimensional case with onsite and NN interactions, the effective interaction remains bounded by Eq. (\ref{Ueff2}), while for longer-ranged interactions, $\mathcal{U}(0)$ drops below this bound as one approaches the phase boundary. In three dimensions we find a drastic drop of the effective interaction even for the NN case, invalidating any simple estimate. Some of the differences between the 2D and 3D results are presumably due to the fact that the 2D system is closer to the Mott transition. 

\subsubsection{Local spectral properties}
\begin{figure*}[tp]
\centering
\includegraphics[width=0.32\textwidth]{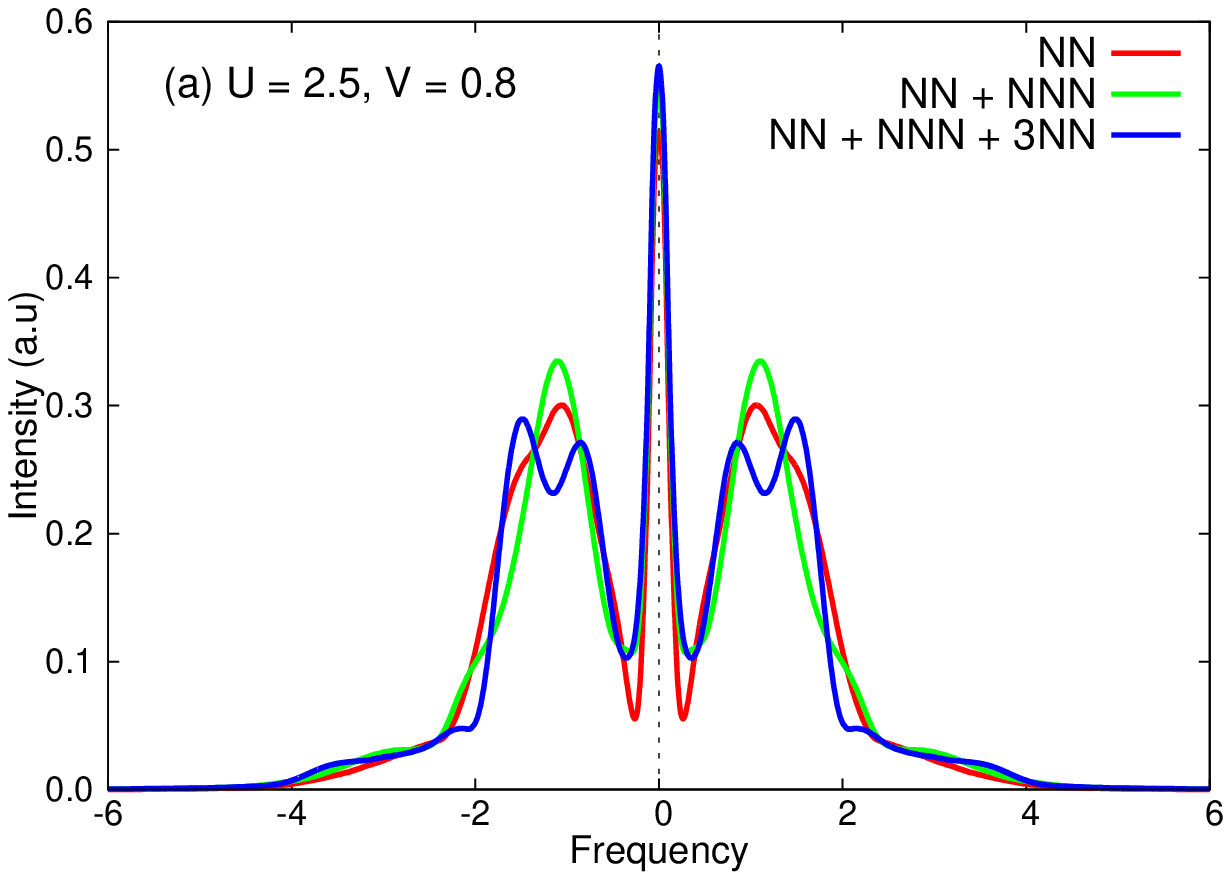}    
\includegraphics[width=0.32\textwidth]{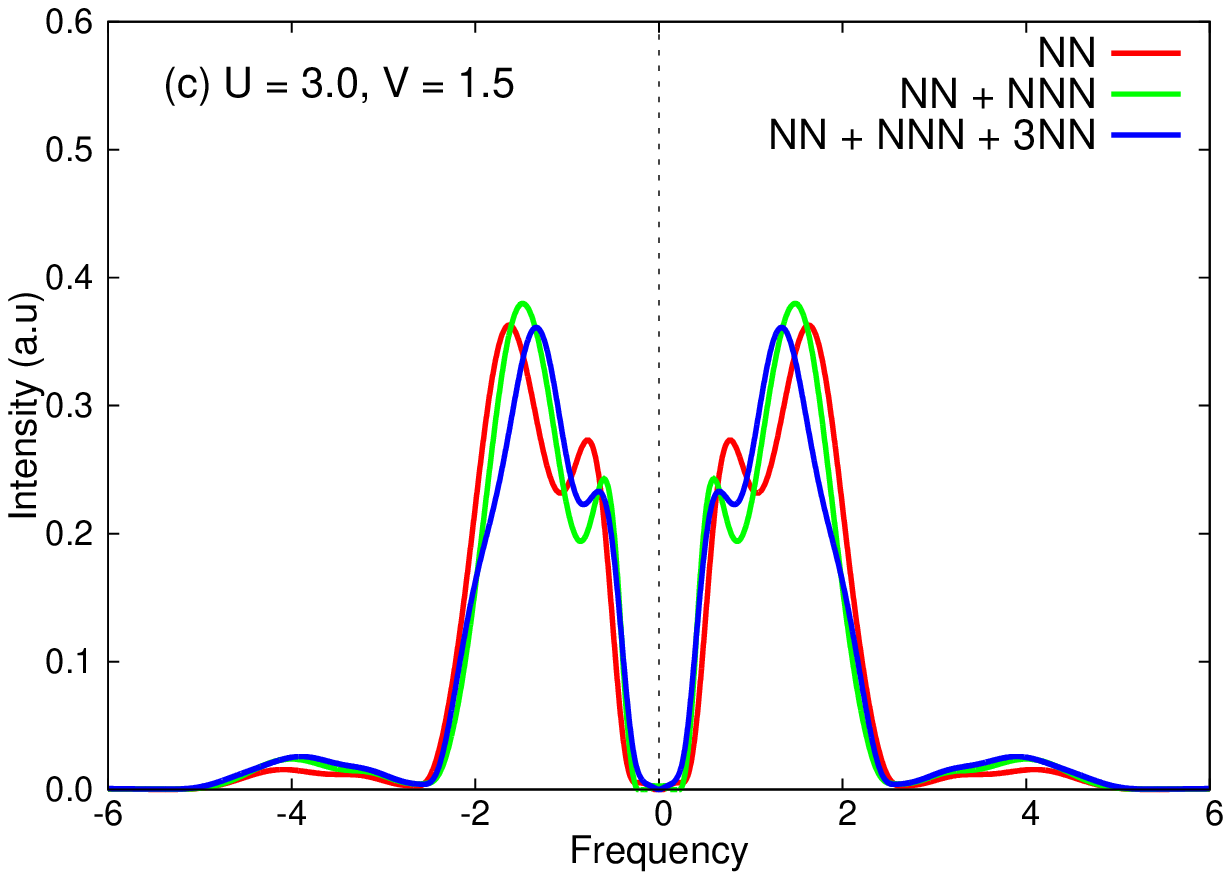}    
\includegraphics[width=0.32\textwidth]{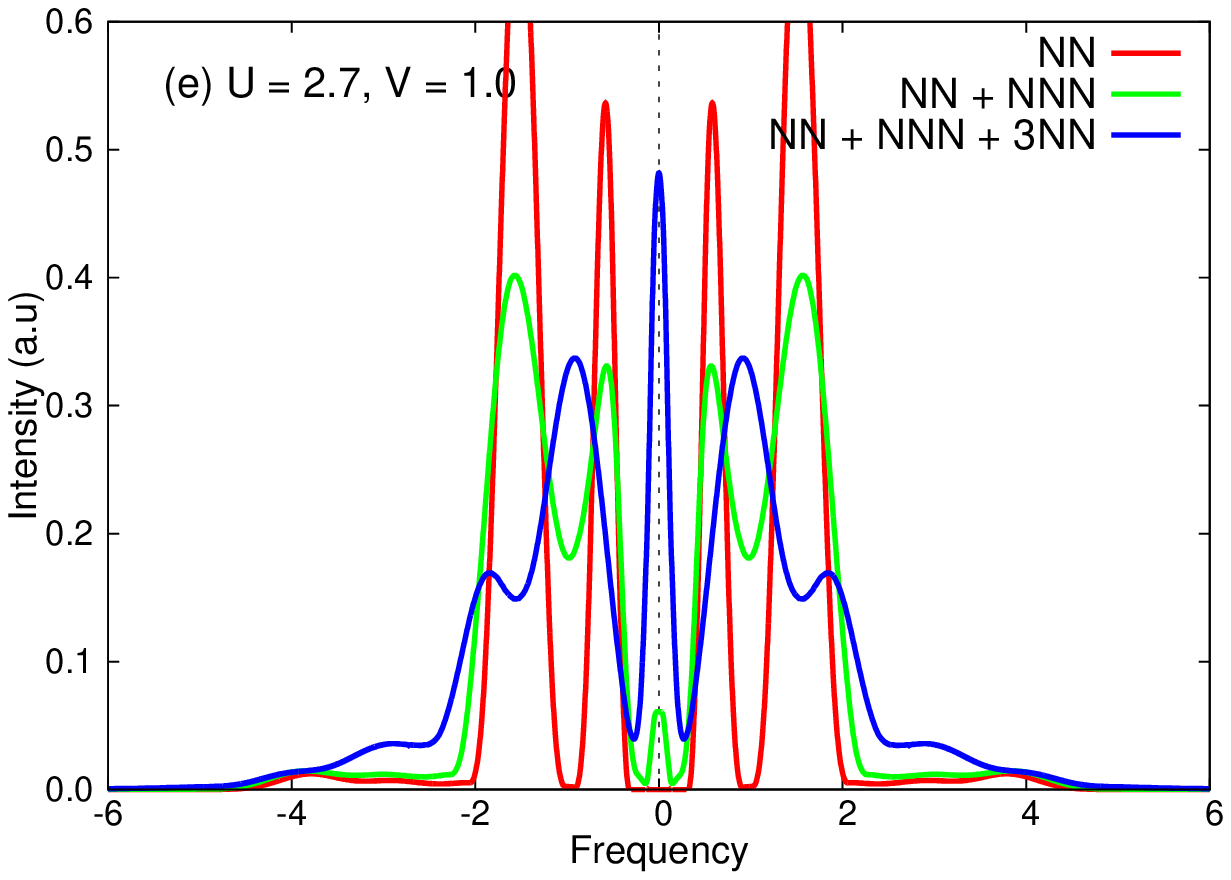}
\includegraphics[width=0.32\textwidth]{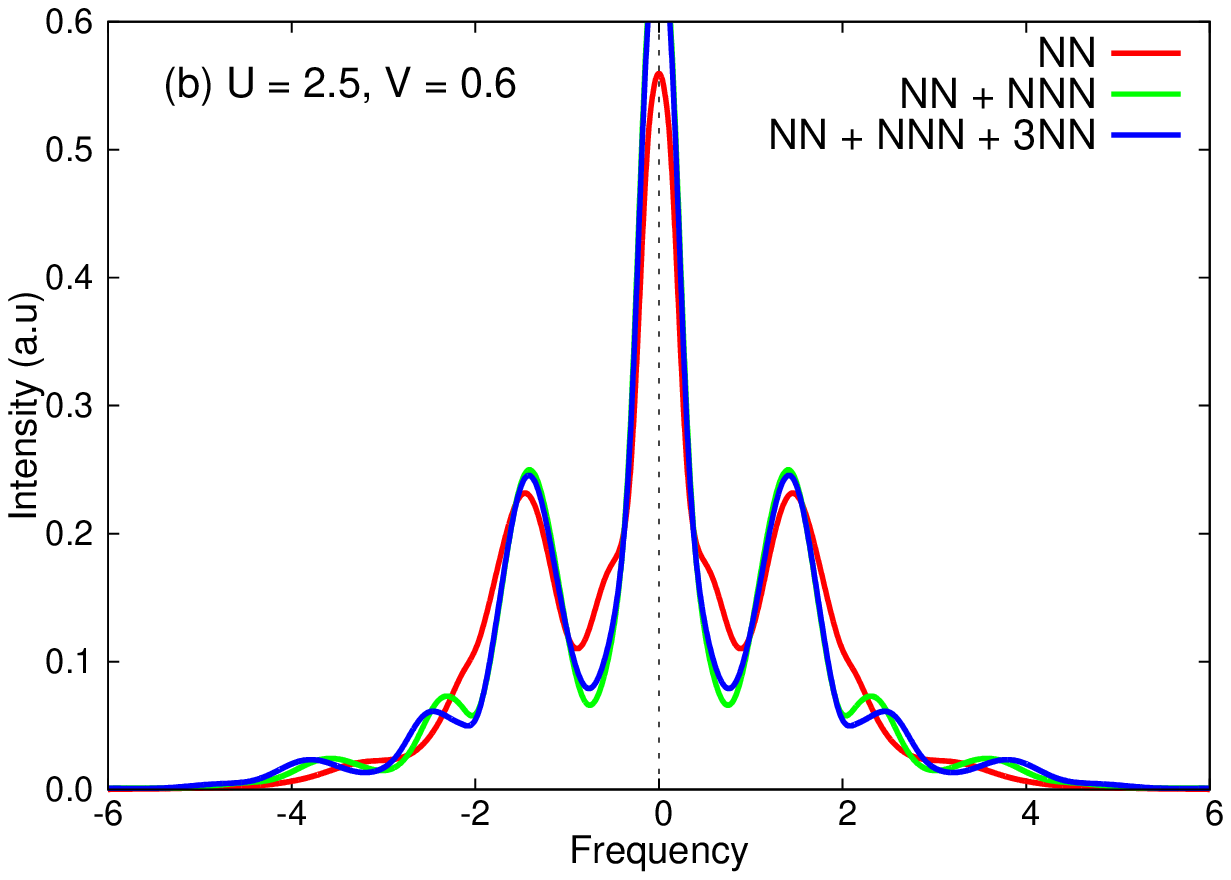}    
\includegraphics[width=0.32\textwidth]{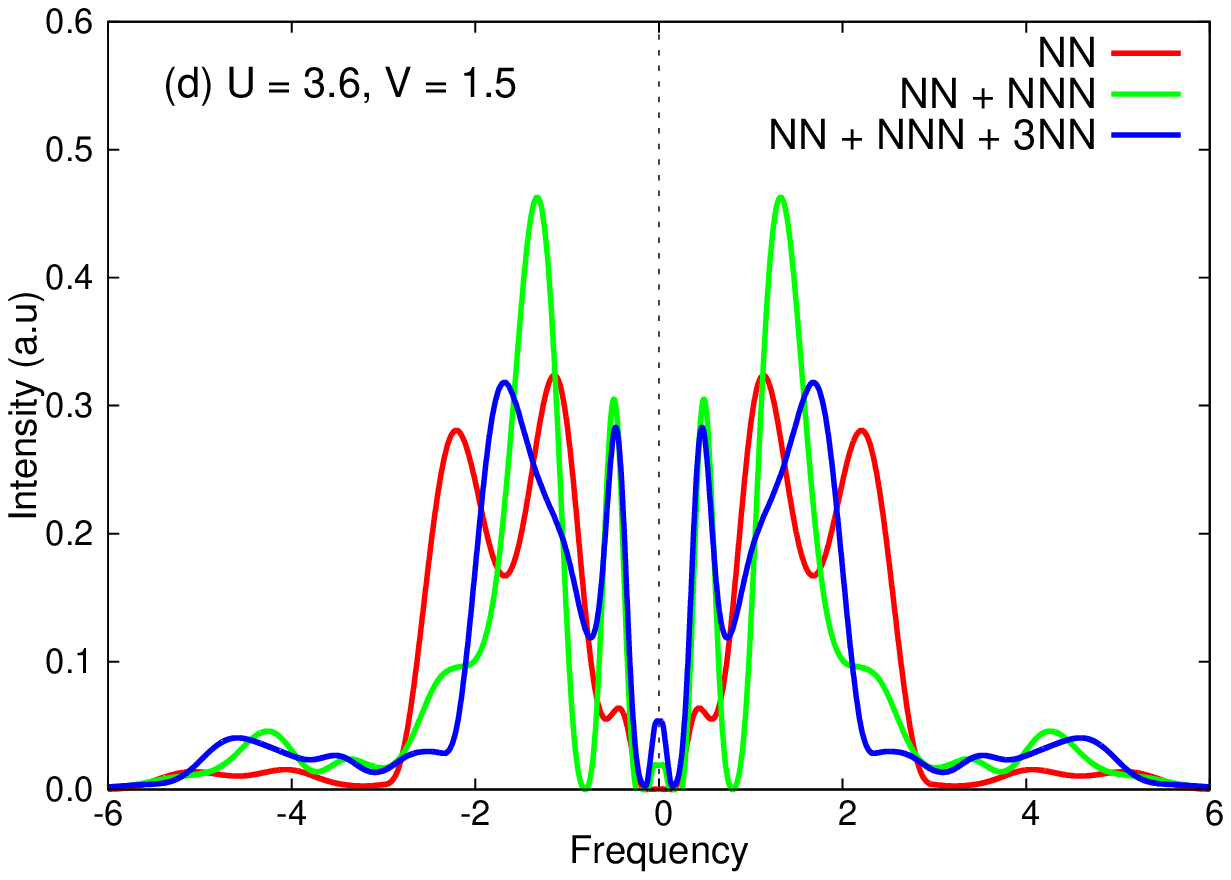}    
\includegraphics[width=0.32\textwidth]{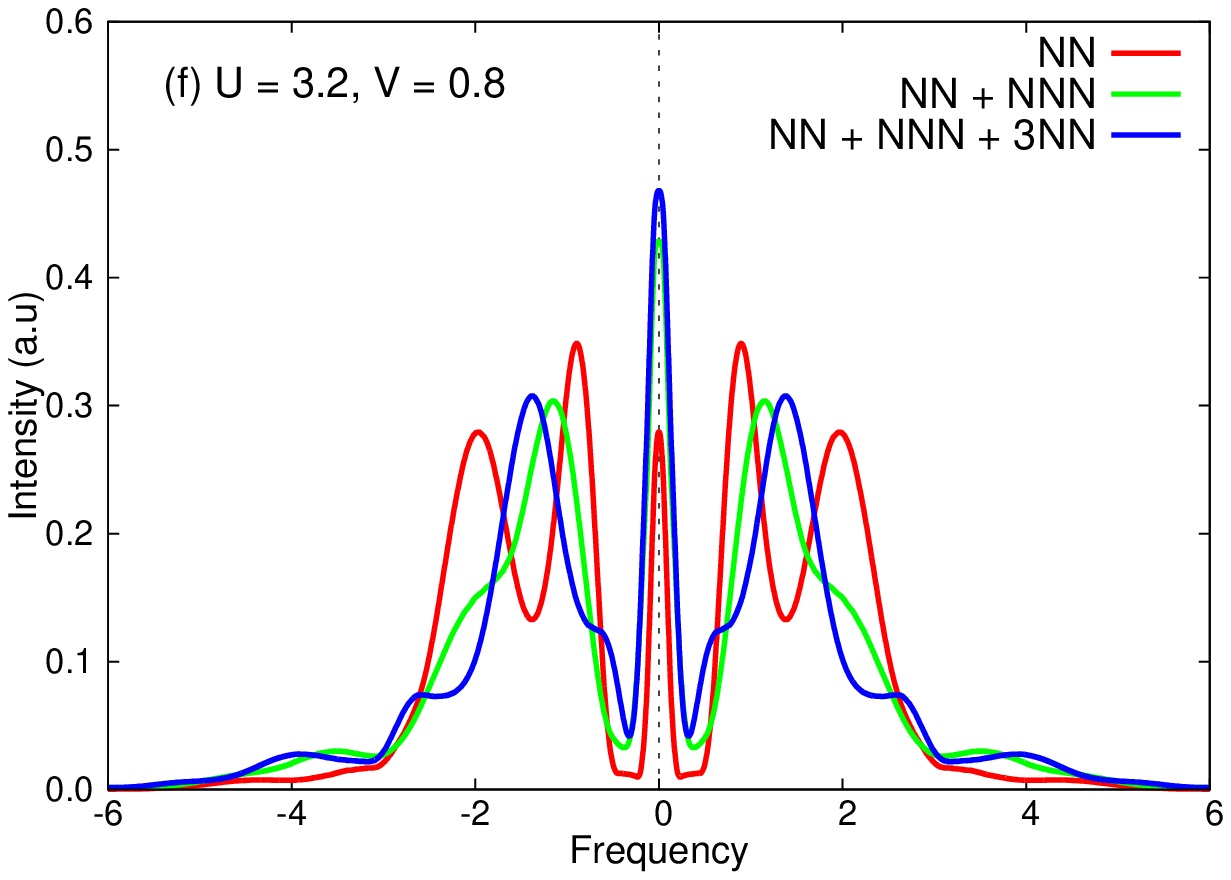}
\caption{(Color online) Spectral functions at selected points for the single-band half-filled extended Hubbard model solved by EDMFT. (a), (c), and (e) Results for the square lattice. (b), (d), and (e) Results for the simple cubic lattice. The parameters are as follows: (a) Metallic region, $U = 2.5$ and $V = 0.8$; (b) Metallic region, $U = 2.5$ and $V = 0.6$; (c) Mott insulating region, $U = 3.0$ and $V = 1.5$; (d) Mott insulating region, $U = 3.6$ and $V = 1.5$; (e) ``Triangle" zone, $U = 2.7$ and $V = 1.0$; (f) ``Triangle" zone, $U = 3.2$ and $V = 0.8$. The impurity spectral functions are obtained using the analytical continuation method proposed in Ref.~\onlinecite{PhysRevB.85.035115}. \label{fig:aw23}}
\end{figure*}

We focus on three characteristic regions in the phase diagrams: the FL metallic phase, the MI phase, and the metallic region between the CO and MI phases (or ``triangle zone" in between the $V_c(U)$ and $U_c(V)$ lines). We computed the local spectral functions in these zones via analytical continuation of the impurity Green's function $G(\tau)$. For the calculations, we use the method described in Sec.~\ref{sub:maxent}, with the bosonic factor $B(\tau)$ obtained from the maximum entropy result for Im$\mathcal{U}(\nu)$.\cite{PhysRevB.85.035115,mem:1996} In the calculations of $B(\tau)$, we introduced a cutoff at small frequencies to prevent an unphysical divergence of $\text{Im}\mathcal{U}(\nu)/\nu^2$ [see insets in Fig.~\ref{fig:wloc23}(e) and (f)]. The spectral functions $A(\omega)$ for the square lattice are displayed in the top panels of Fig.~\ref{fig:aw23}, while those for the simple cubic lattice are shown in the bottom panels.

We found that the screening effects resulting from long range intersite interactions affect the impurity spectral functions in several ways. In the FL regime, the onsite interaction is weak. The major effect of longer range intersite interactions is to transfer spectral weight from the Hubbard bands to the quasiparticle peak, and to small satellites, which are shifted from the Hubbard bands by roughly the effective screening frequency $\nu_0$. In the triangle zone, where the onsite interaction is moderate, the longer range intersite interactions can trigger an insulator-metal phase transition. Let's look at Fig.~\ref{fig:aw23}(e), which illustrates the evolution of the spectral functions across such a metal-insulator transition. For the NN case, the system is an insulator with sharp Hubbard bands and sizable gap. However, for the NN + NNN case, spectral weight appears at the Fermi level, which indicates a strongly renormalized metallic state. While the Hubbard bands are smeared out, their position is almost unchanged. When the 3NN intersite interaction is added, the system turns into a good metal with a large quasiparticle peak and the Hubbard bands are shifted to higher energy. In the MI phase in which the onsite interaction is strong, the spectral functions are less affected by longer range intersite interactions. It seems that the longer range intersite interactions do not significantly shrink the gaps. The main effect is to redistribute the weight within the Hubbard bands. At the beginning, the upper and lower Hubbard bands are broad and smooth. When longer range intersite interactions are included, the Hubbard bands turn sharper and thinner, and spectral weight is transfered to the edges of the gap and high-frequency features [see Fig.~\ref{fig:aw23}(d)].

As mentioned before, the structures in Im$\mathcal{U}(\nu)/\nu^2$ produce satellites in the local spectral functions $A(\omega)$. For example, the screening modes displayed in Fig.~\ref{fig:wloc23}(e) and (f) explain the broad tails in the energy range $|\omega|\gtrsim 2$ in Fig.~\ref{fig:aw23}(a) and (b). 

\subsubsection{Away from half-filling}
\begin{figure*}[tp]
\centering
\includegraphics[width=\columnwidth]{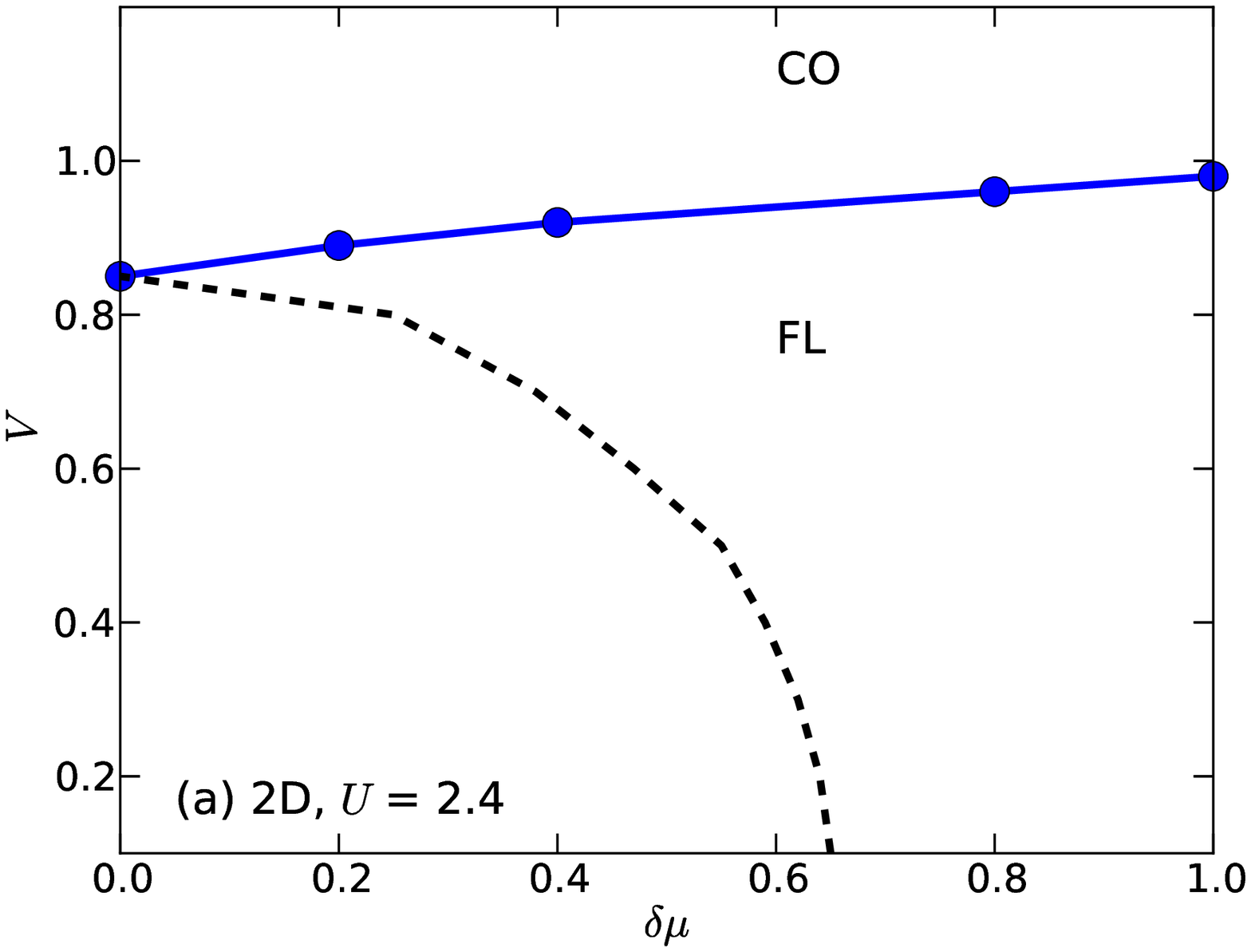}
\includegraphics[width=\columnwidth]{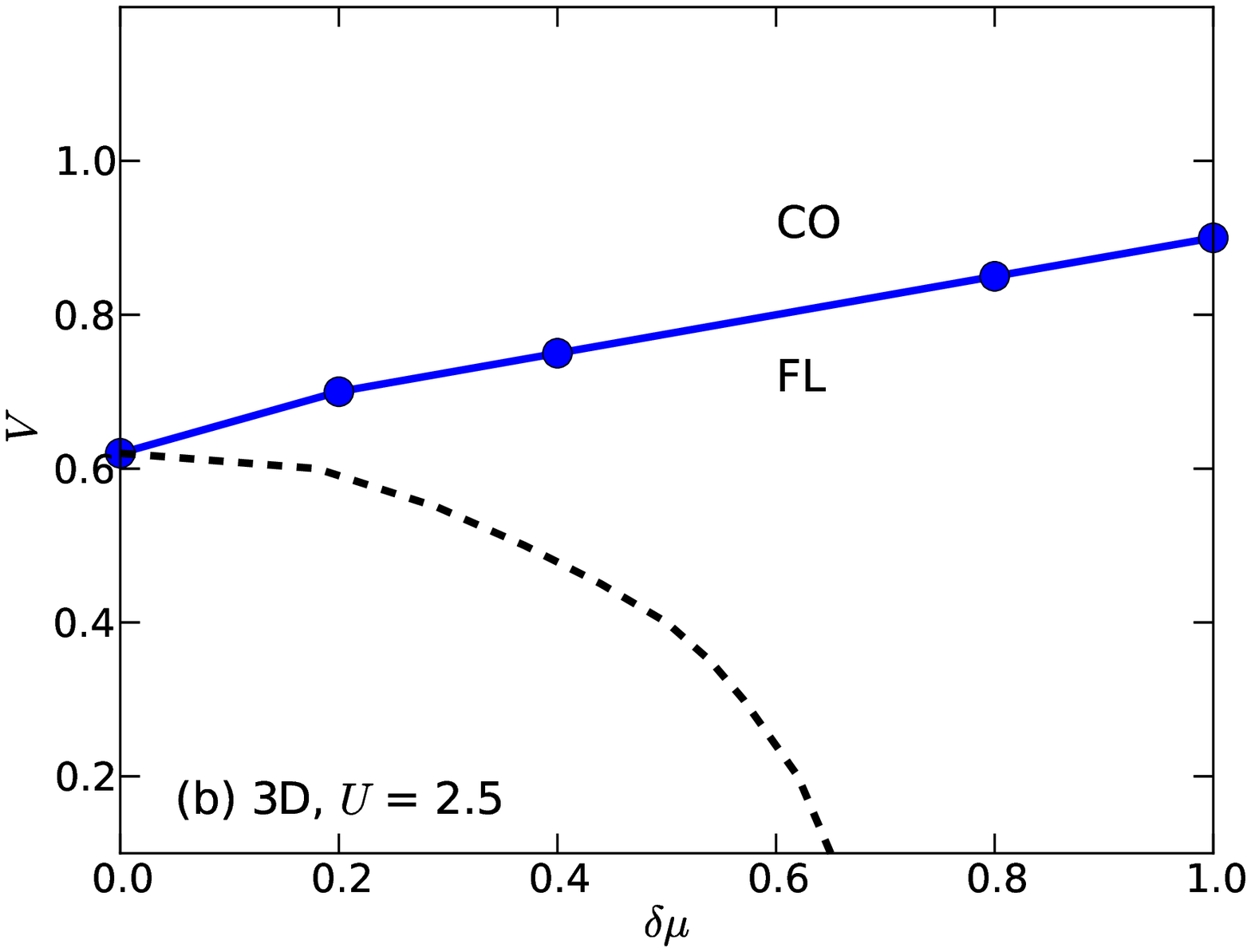}
\includegraphics[width=\columnwidth]{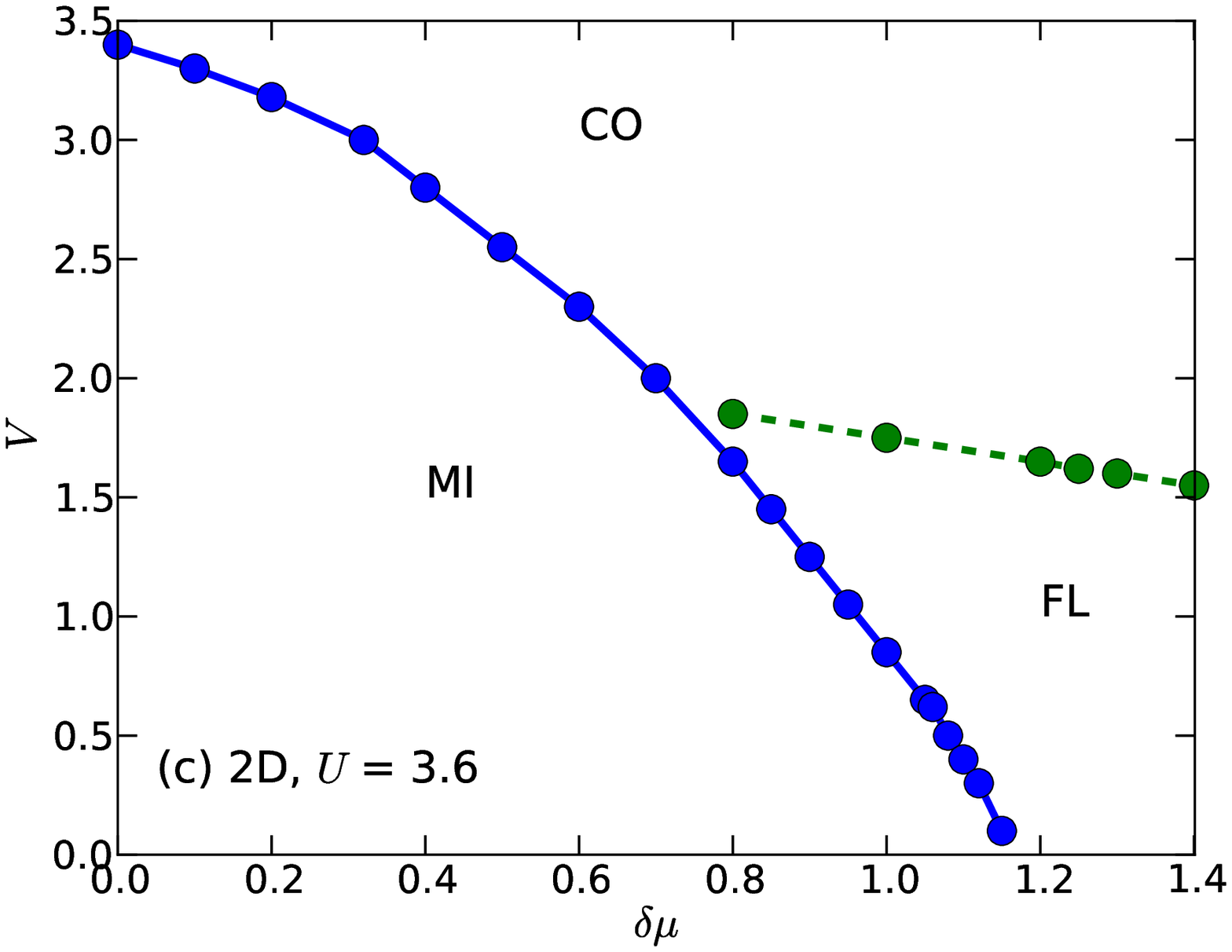}
\includegraphics[width=\columnwidth]{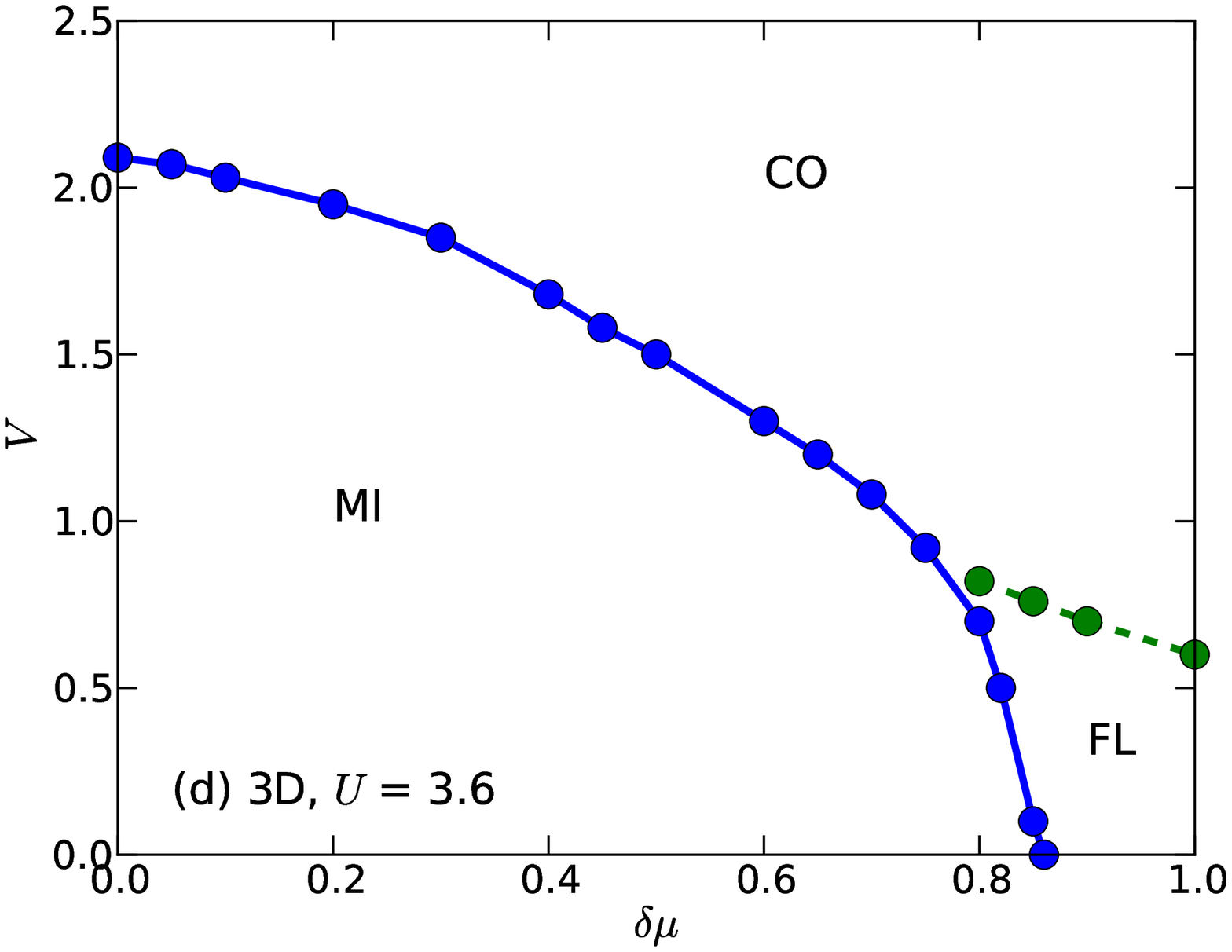}
\caption{(Color online) $V$-$\mu$ phase diagrams for the single-band extended Hubbard model with NN interactions, determined by EDMFT calculations. Here $\delta \mu = \mu - U/2$. Panels (a) and (c) show results for the 2D square lattice which at half-filling is in the FL or MI regime. Panels (b) and (d) show similar results for the 3D simple cubic lattice. The black dashed lines in (a) and (b) show the location of $W(0)=0$, i.e. on the right side of this boundary, the static screened interaction is negative. \label{fig:doping_diagram}}
\end{figure*}

\begin{figure*}[tp]
\centering
\includegraphics[width=0.32\textwidth]{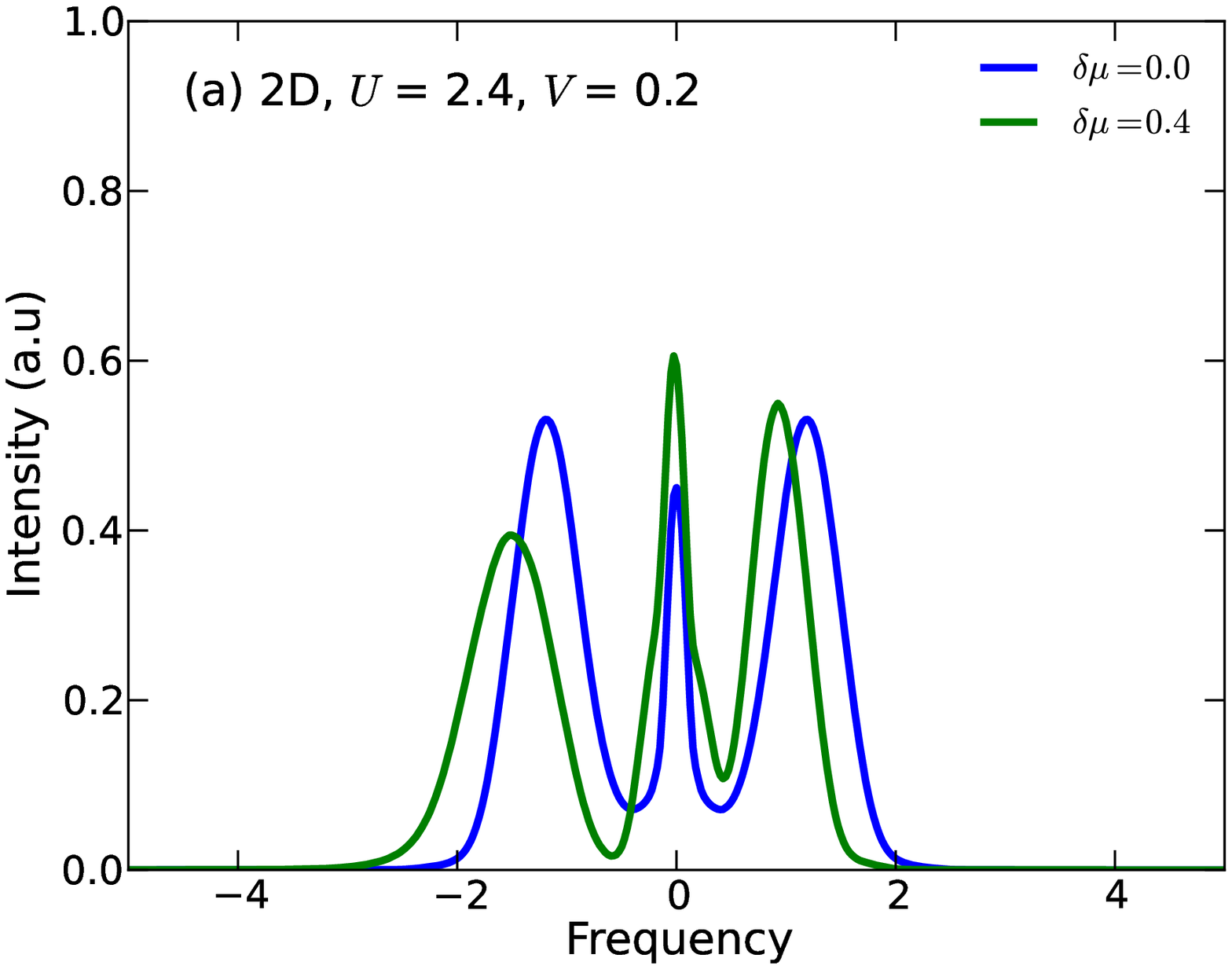}
\includegraphics[width=0.32\textwidth]{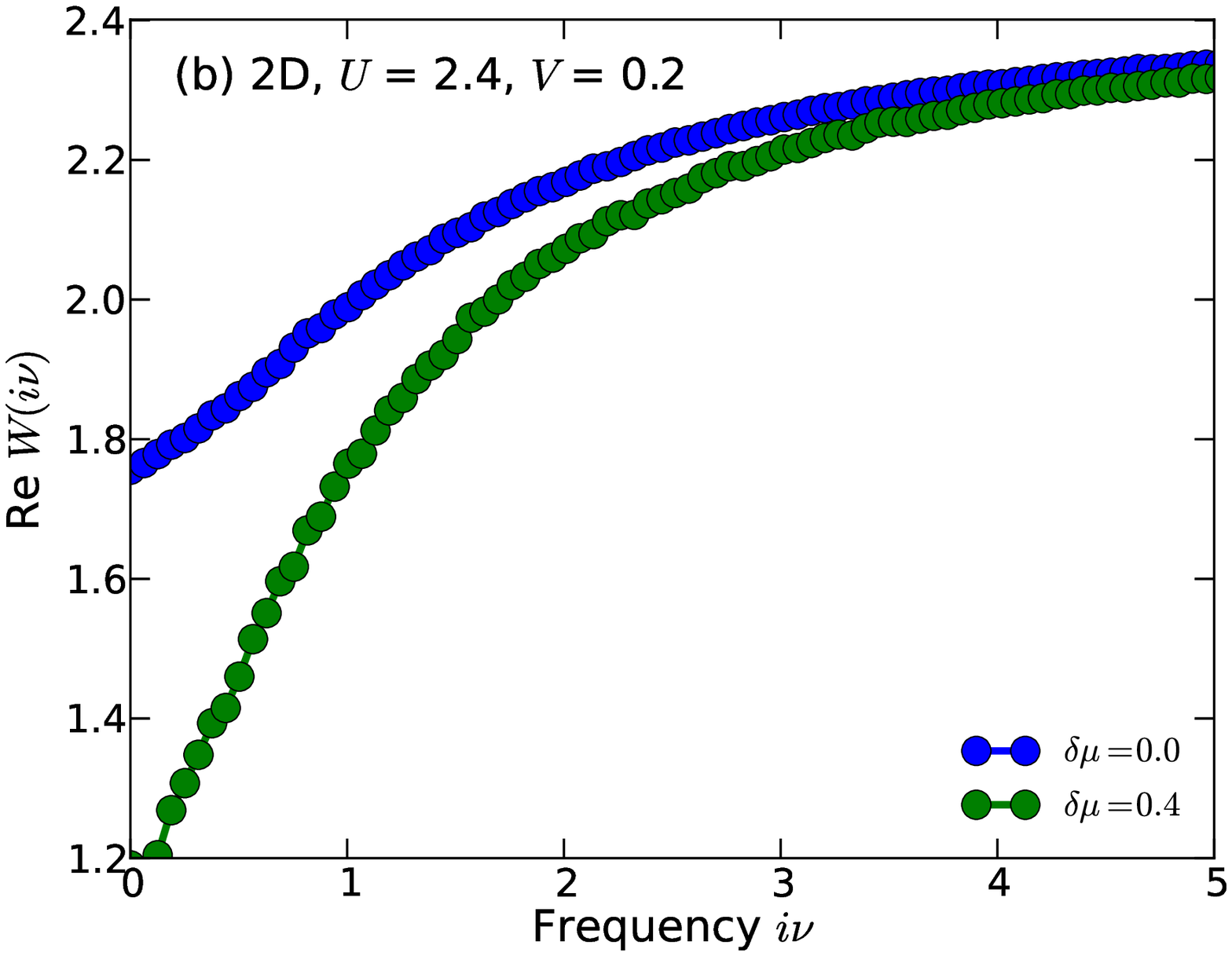}
\includegraphics[width=0.32\textwidth]{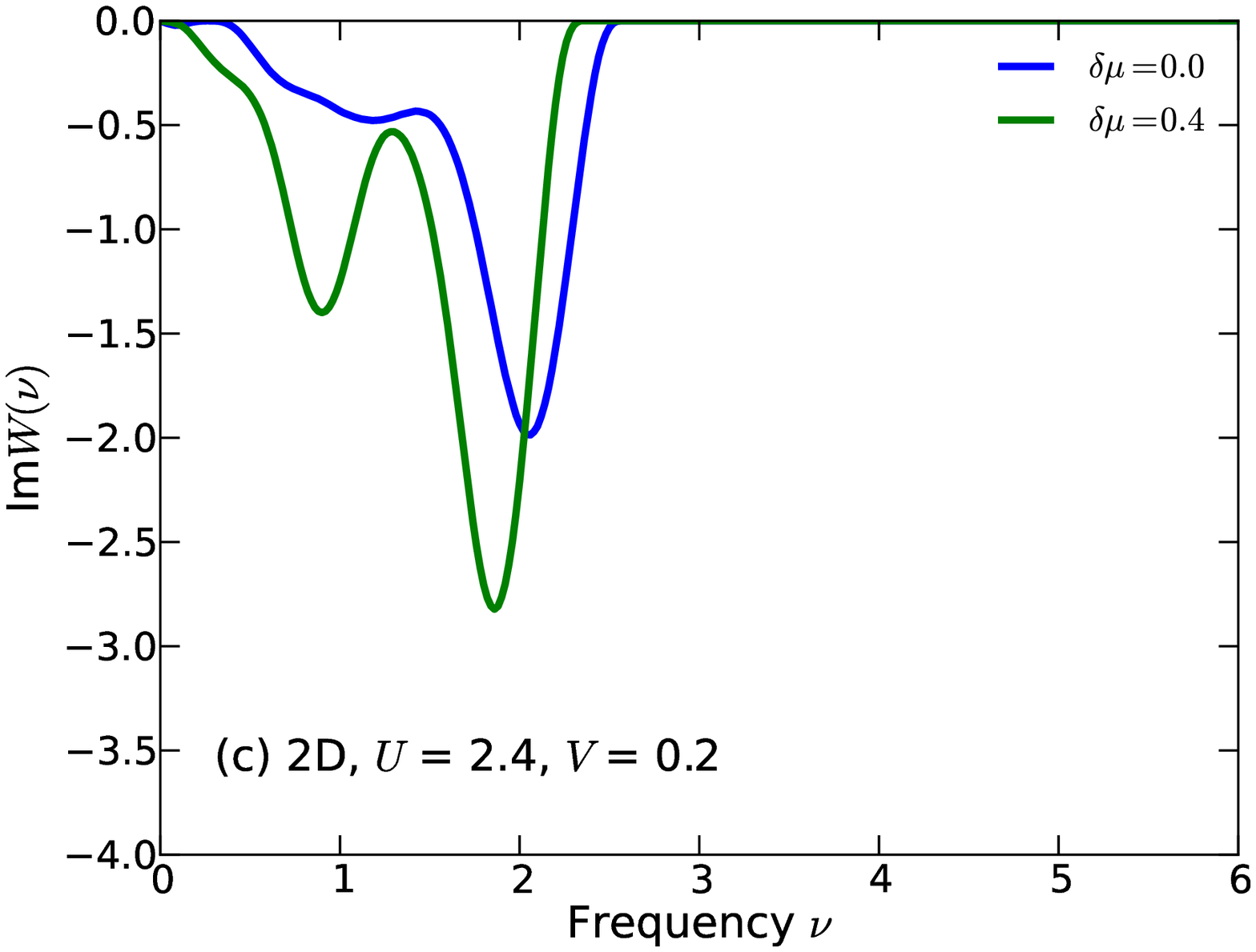}
\includegraphics[width=0.32\textwidth]{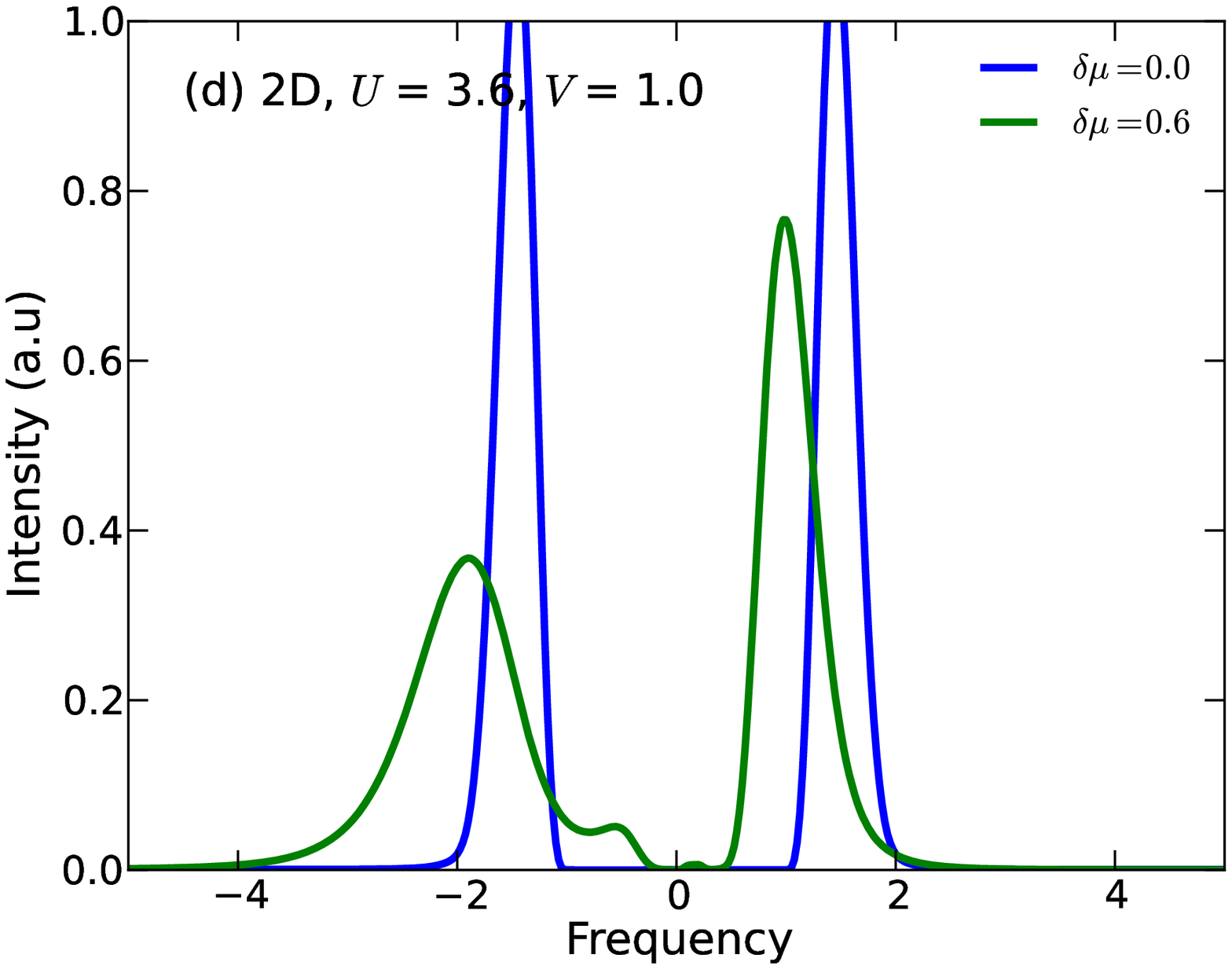}
\includegraphics[width=0.32\textwidth]{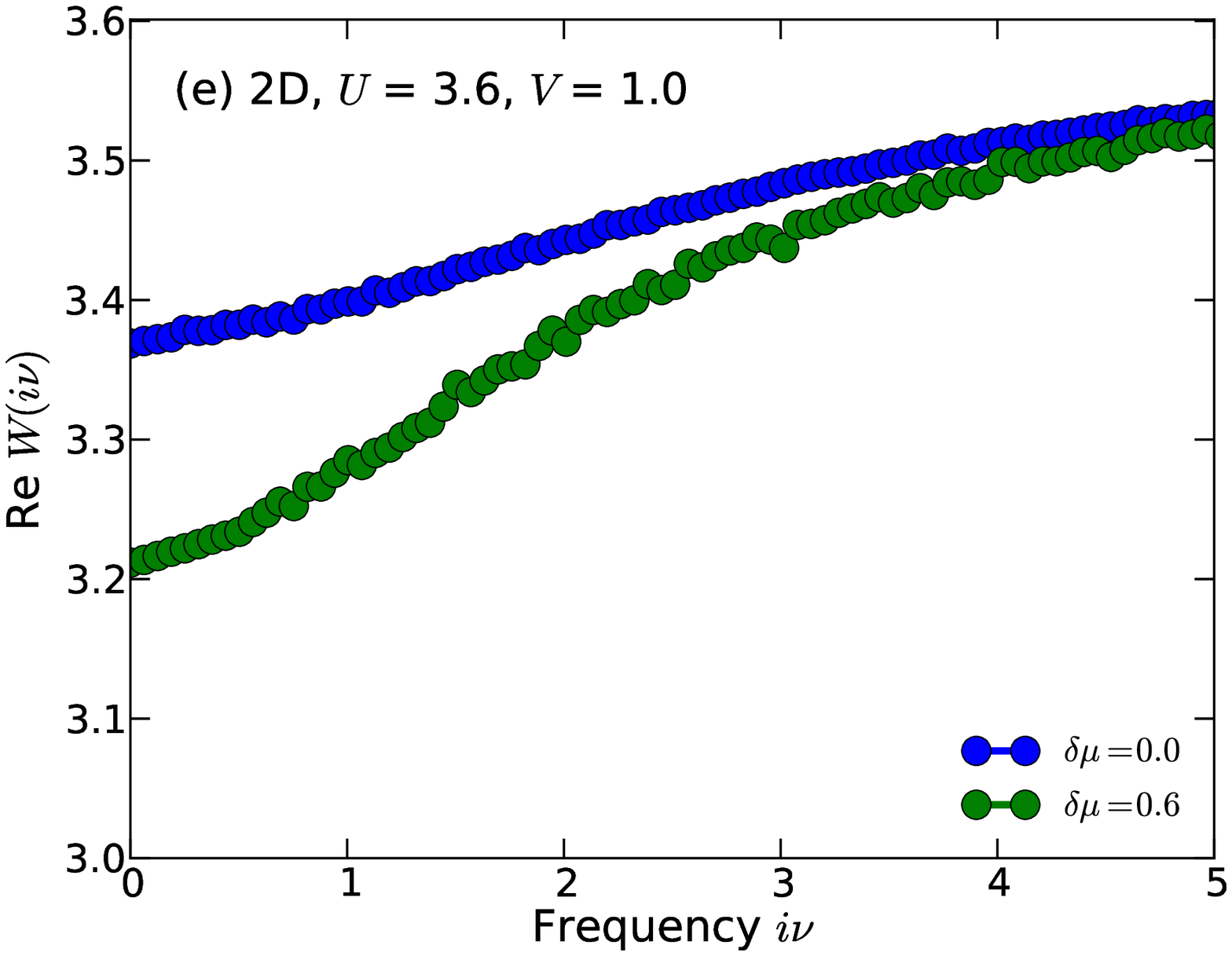}
\includegraphics[width=0.32\textwidth]{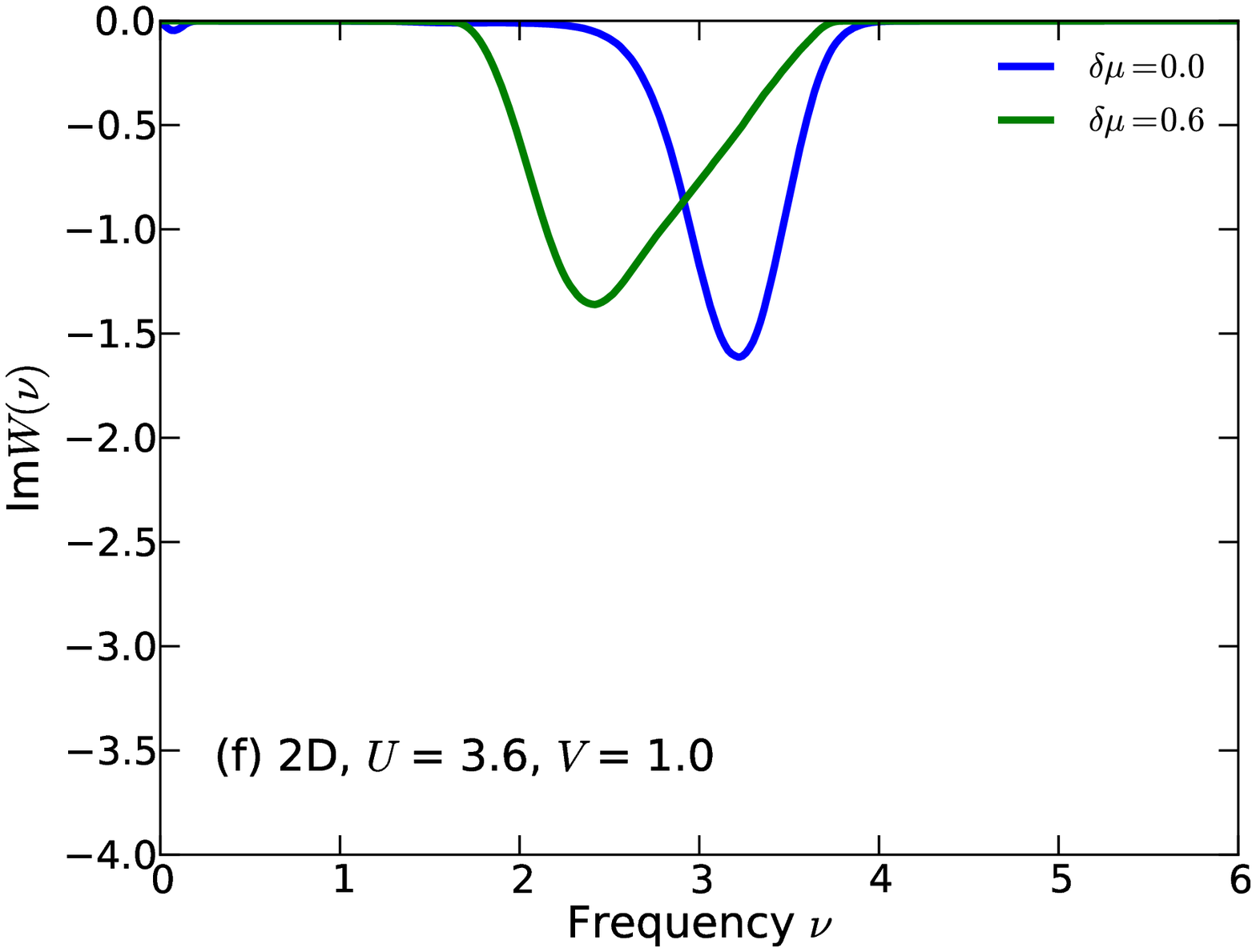}
\includegraphics[width=0.32\textwidth]{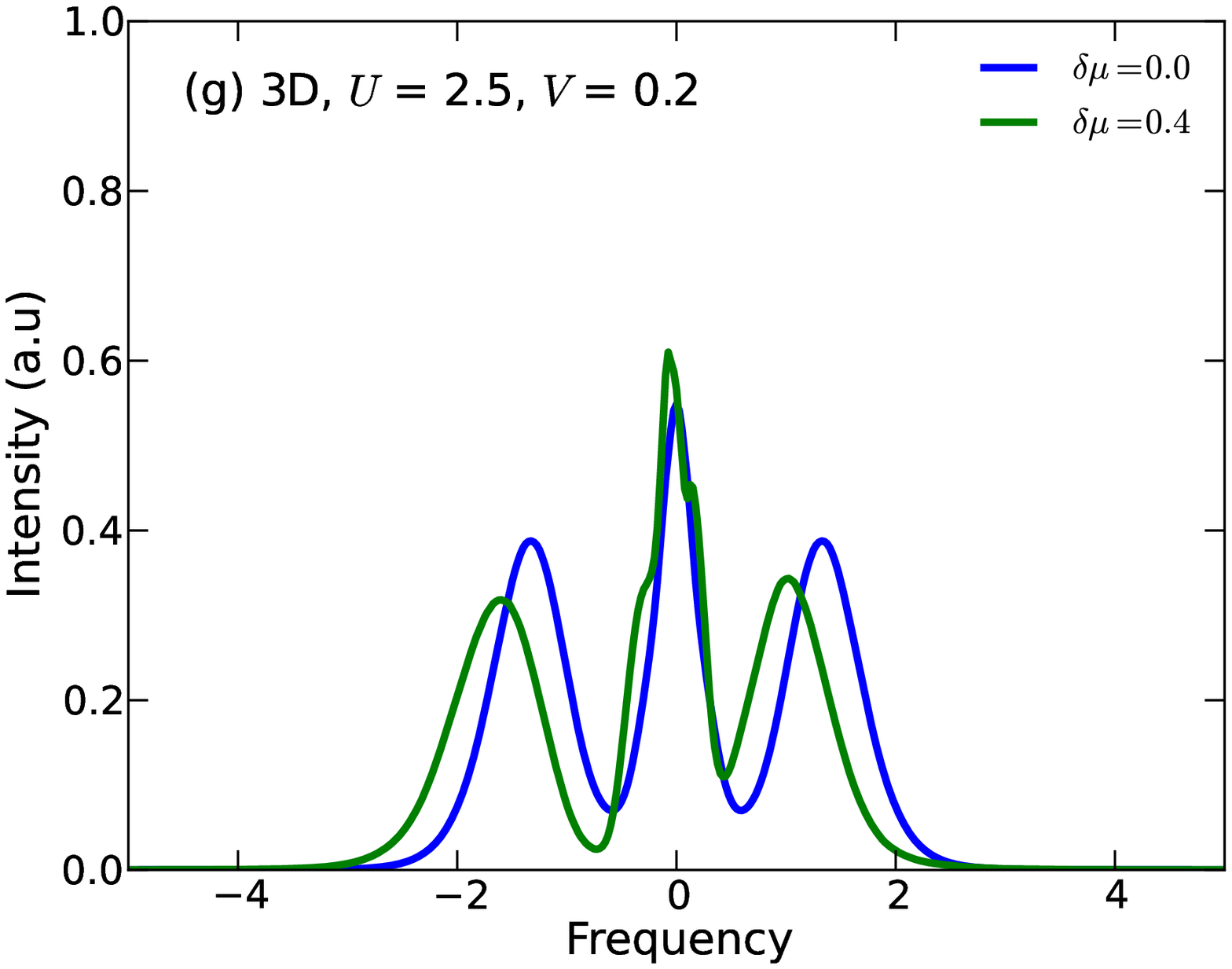}
\includegraphics[width=0.32\textwidth]{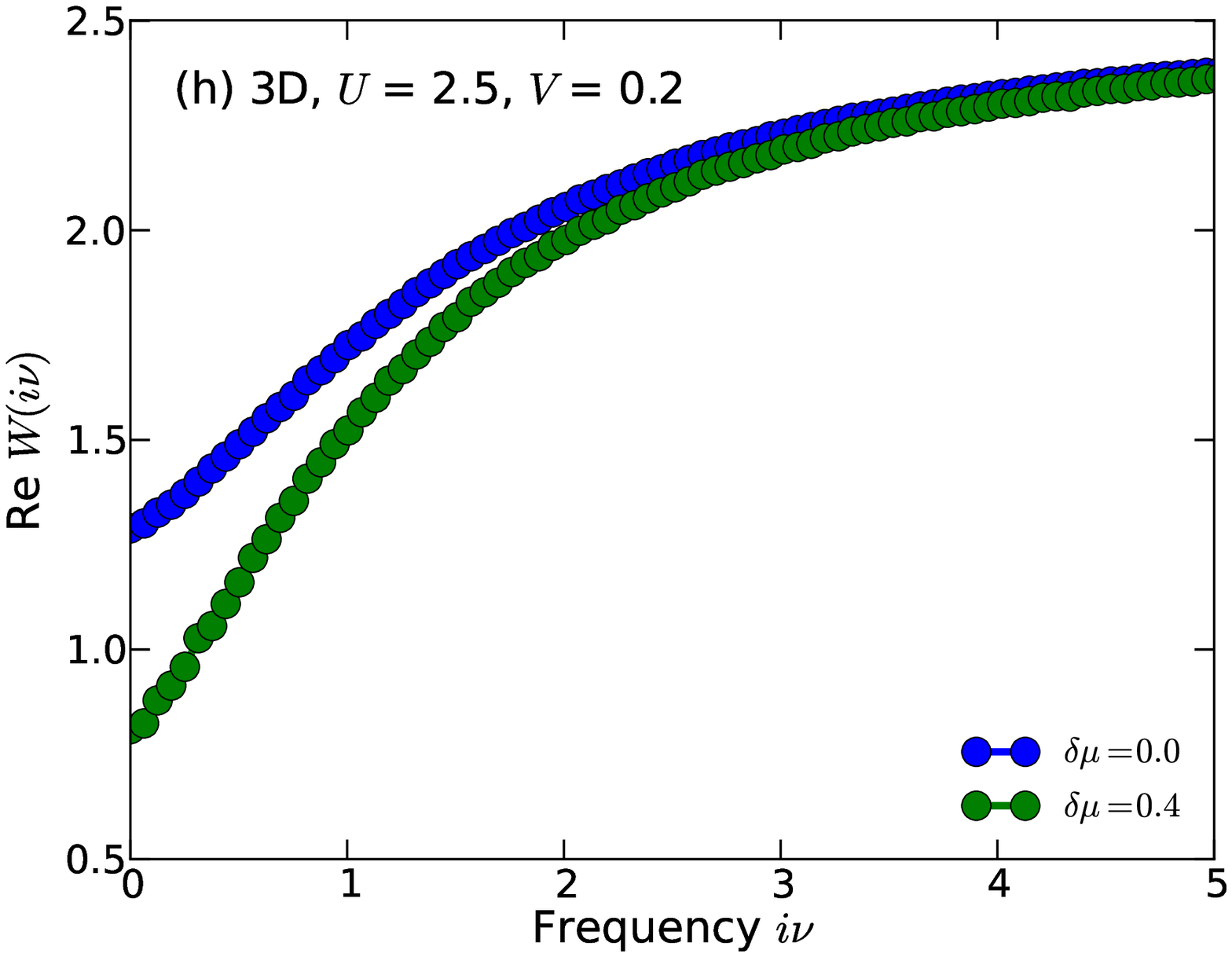}
\includegraphics[width=0.32\textwidth]{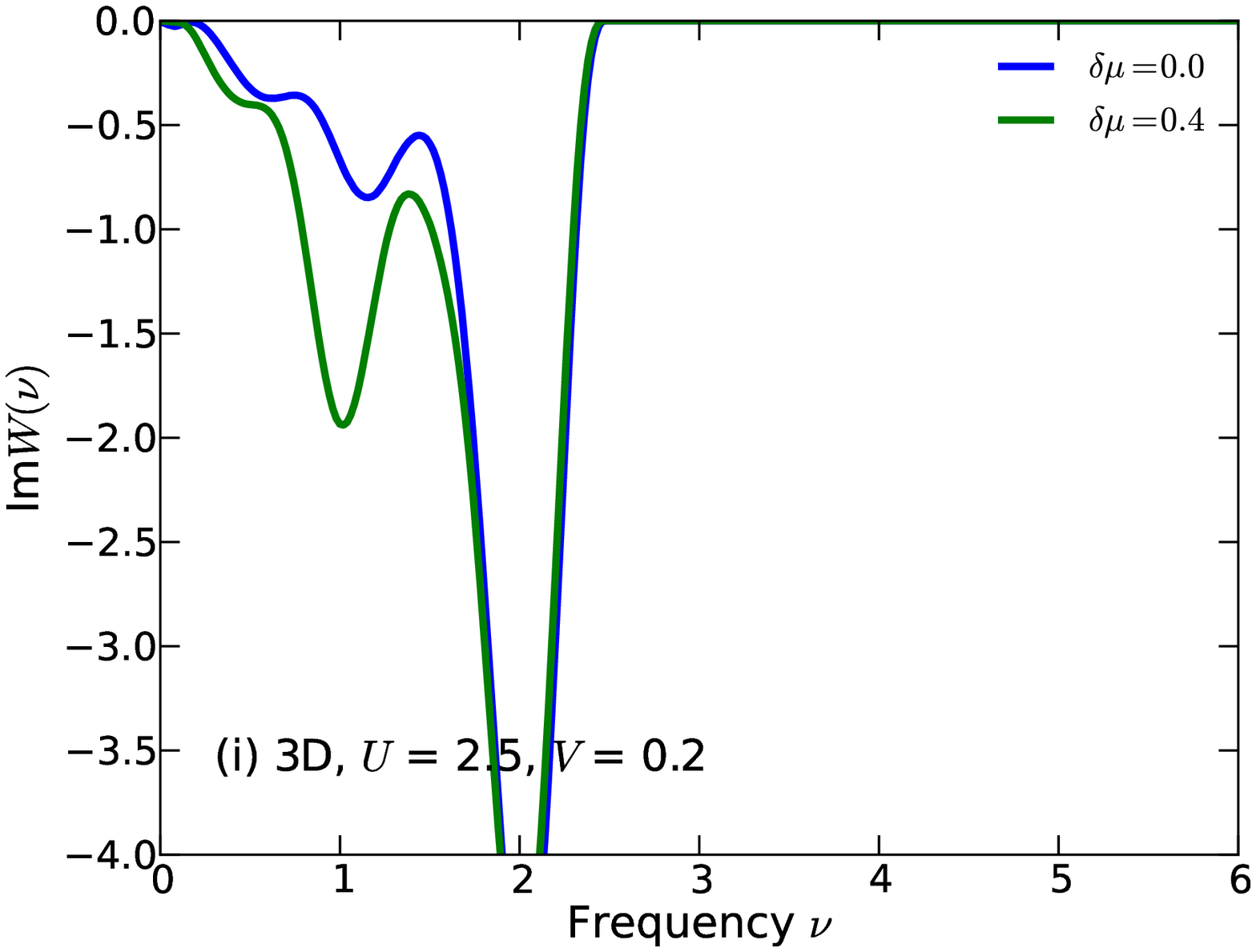}
\includegraphics[width=0.32\textwidth]{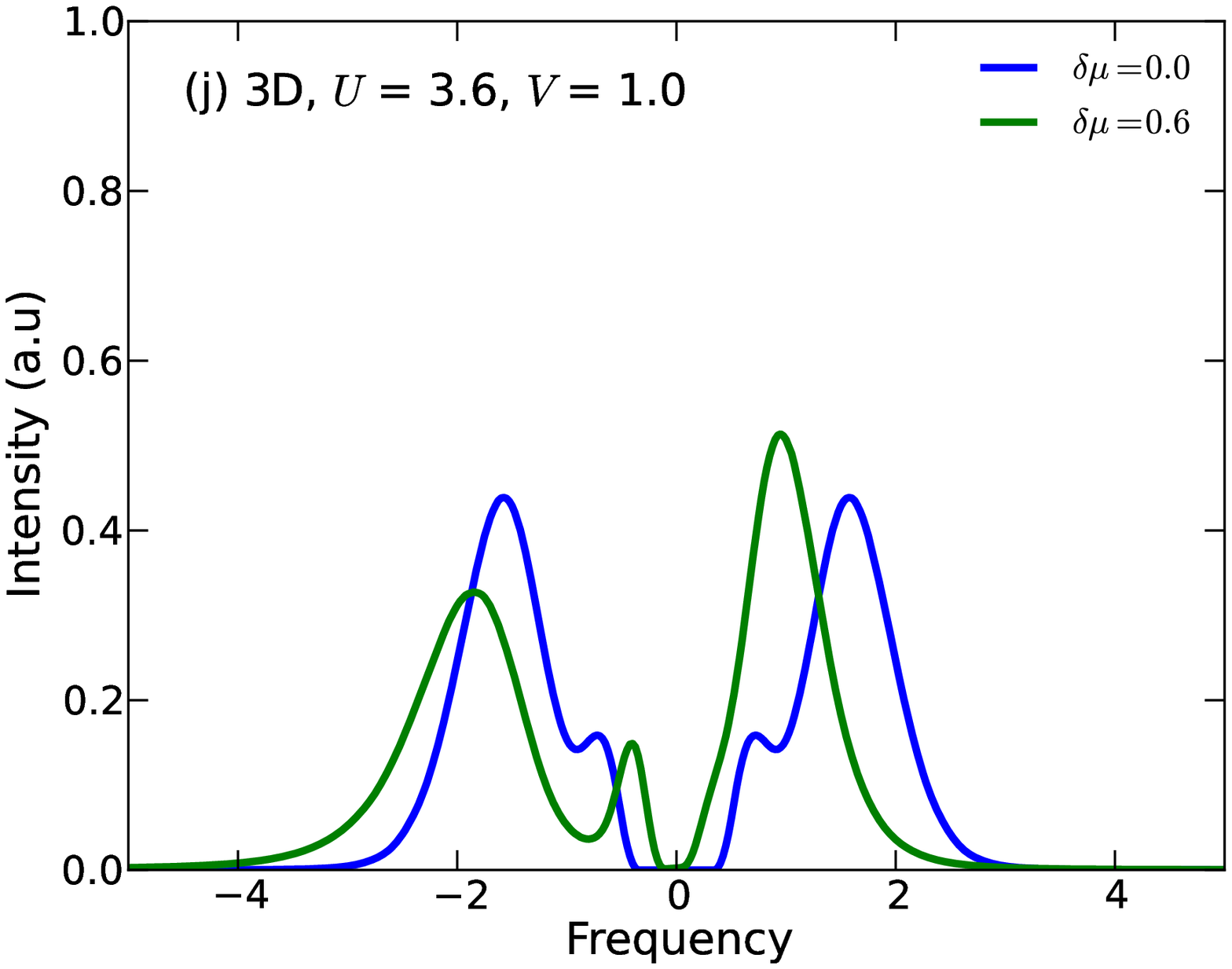}
\includegraphics[width=0.32\textwidth]{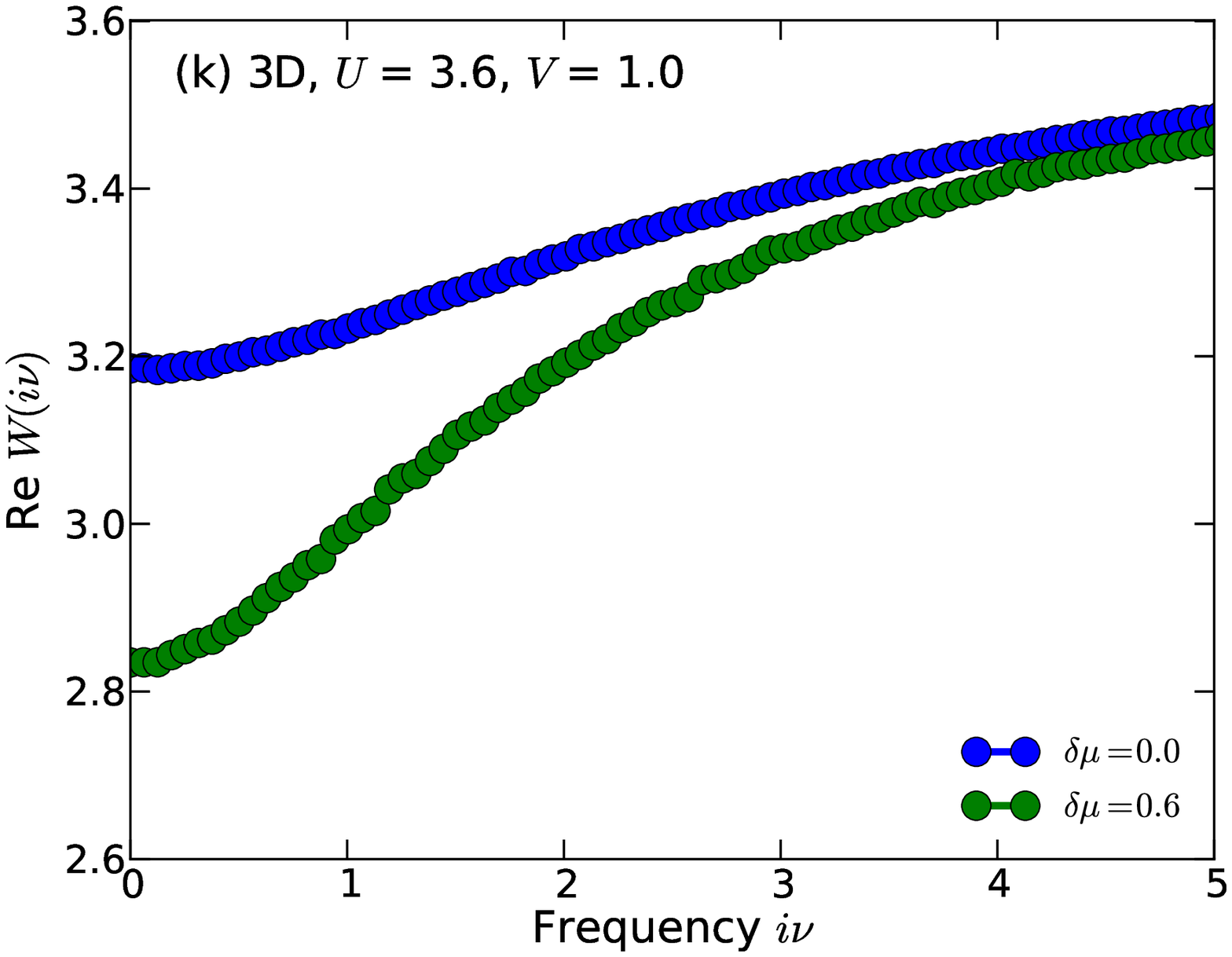}
\includegraphics[width=0.32\textwidth]{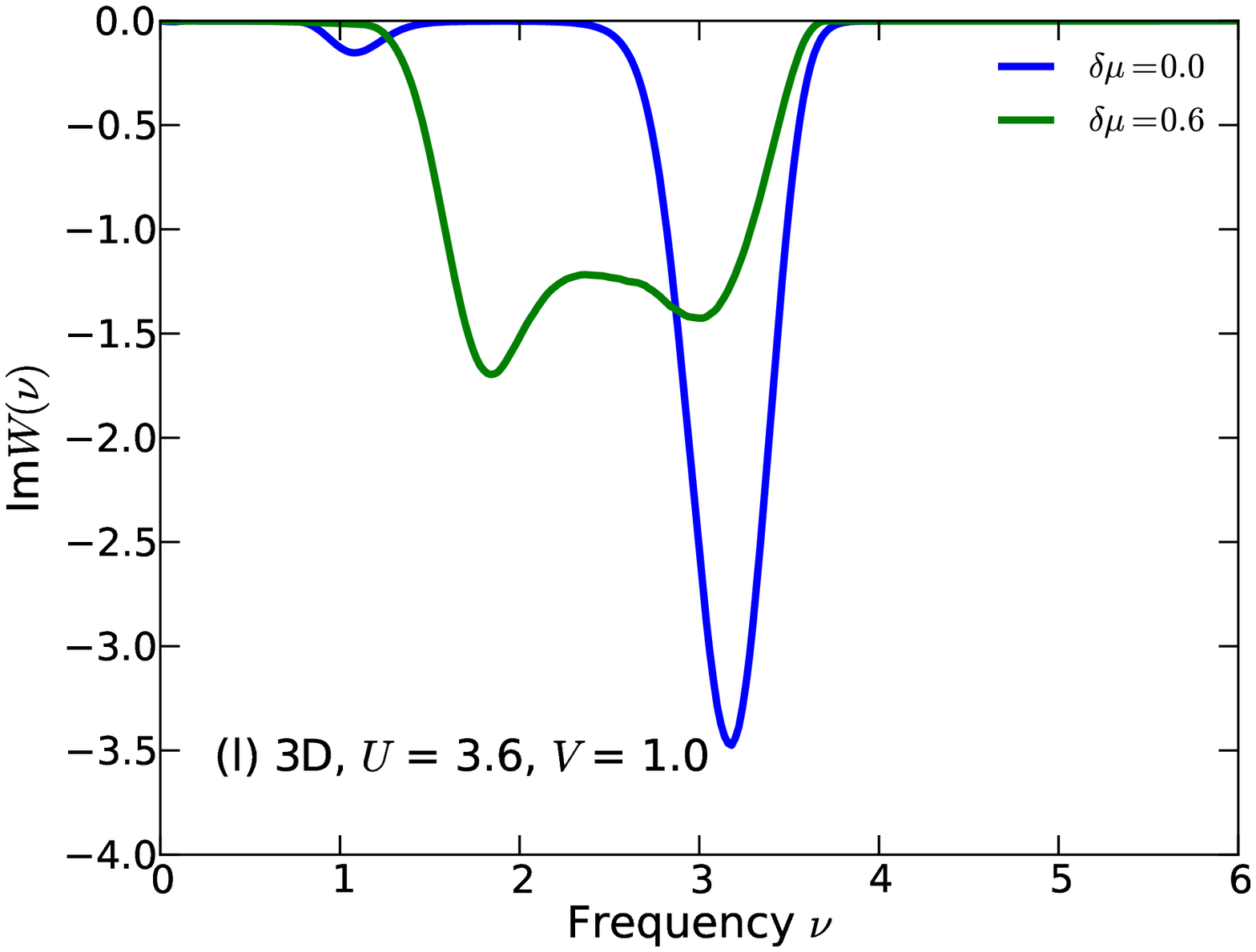}
\caption{(Color online) Spectral functions for the Hubbard model with NN interactions away from half-filling. (a)-(f) Results for the square lattice. (g)-(i) Results for the cubic lattice. In the left column, the impurity spectral functions $A(\omega)$ are shown. In the middle and right columns, we show the screened interaction $W(i\nu)$ and corresponding Im$ W(\nu)$. The parameters are as follows: (a)-(c) $U = 2.4$, $V = 0.2$, 2D lattice. (d)-(f) $U = 3.6$, $V = 1.0$, 2D lattice. (g)-(i) $U = 2.5$, $V = 0.2$, 3D lattice. (j)-(l) $U = 3.6$, $V = 1.0$, 3D lattice. \label{fig:doping_spectra}}
\end{figure*}

Having identified the dominant screening modes in the half-filled system and their interpretation in terms of the spectral function, it is interesting to look also at the evolution of these quantities away from half-filling. In this section, we present some results for the 2D and 3D lattices with onsite and NN intersite interactions. First, we show the phase diagrams for fixed $U$ in the space of $V$ and $\delta\mu=\mu-U/2$. In the 2D (3D) case we choose $U=2.4$ and $U=3.6$ ($U=2.5$ and $U=3.6$). For the smaller onsite interaction, the system at half-filling ($\delta\mu=0$) and small enough $V$ is metallic, while for the larger $U$ it is Mott insulating. As the filling of the metallic system is increased, the phase boundary to the CO phase shifts to larger $V$, i.e., in the small-$U$ regime, the CO instability is a nesting-type phenomenon. We also plot, as dashed lines, the location where the screened interaction $W(0)$ changes sign. We note that this $W(0)=0$ line is very different from the FL-CO phase boundary. In the heavily doped region, one can still obtain a stable metallic solution even though $W(0) < 0$. 

The situation is quite different for the larger $U$, where the half-filled solution is either MI or CO. Here, the MI solution is destabilized by doping. In the 3D case, one observes a transition into a doped metal phase for $V\lesssim 1.0$, while in the 2D system, a similar transition occurs for $V \lesssim 2.0$. We note that the phase diagrams of the 2D/3D system are qualitatively very similar to those of the Holstein-Hubbard model.\cite{PhysRevLett.99.146404} 

Both the electron spectral function and the screened interaction depend sensitively on $\delta\mu$. Some representative results are shown in Fig.~\ref{fig:doping_spectra}. For $\delta\mu>0$, the electron spectral function (left panels) becomes asymmetric. In the metallic phase, the quasiparticle peak grows and shifts closer to the upper Hubbard band, while in the insulating phase, the gap shrinks due to a broadening of the lower Hubbard band. These changes in the electron spectral function qualitatively explain the changes in the bosonic spectra (right panels). In the metallic case, the main effect of increasing $\delta\mu$ is a growing low-energy feature in $\text{Im}W(\nu)$. This can be explained by the larger number of states in the quasiparticle band. In the Mott insulating case, where the bosonic spectra for the half-filled system show a single peak at an energy given by the gap, the shrinking of the gap with increasing $\delta\mu$ leads to a broadening and shift of this peak to lower energies. In the 3D case, where the gap size for $\delta\mu=0.6$ is small and electron spectral function has a peak at the lower gap edge, we also find a low-energy mode in $\text{Im}W(\nu)$ which is associated with transitions between this peak and the upper Hubbard band. Since the low-energy mode in $\text{Im}W(\nu)$ produces the largest screening effect, it is not surprising that increasing $\delta\mu$ has a large effect on the screened interaction (middle panels). As we saw in Fig.~\ref{fig:doping_diagram} (dashed line), in the metallic phase, doping quickly leads to an overscreening of the local interaction. 

\subsection{$GW$ + EDMFT results\label{sub:gw_edmft}}
In this subsection, we present the $GW$ + EDMFT results. Since the computational cost of fully self-consistent $GW$ + EDMFT calculations is much higher than in the case of EDMFT calculations, we do not map out the whole $U$-$V$ phase diagram. Instead, we performed $GW$ + EDMFT calculations for selected $U$ and $V$ parameters. As a starting point for the self-consistent $GW$ + EDMFT calculation, we used the converged EDMFT results. 

\subsubsection{Nonlocal and local self-energy and polarization}
\begin{figure*}[tp]
\centering
\includegraphics[width=0.155\textwidth,bb=120 66 330 280]{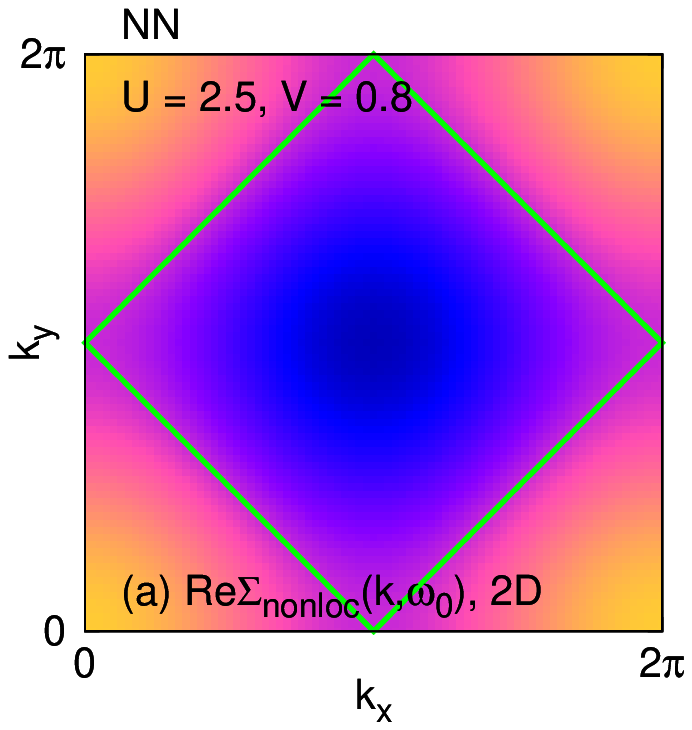}
\includegraphics[width=0.155\textwidth,bb=120 66 330 280]{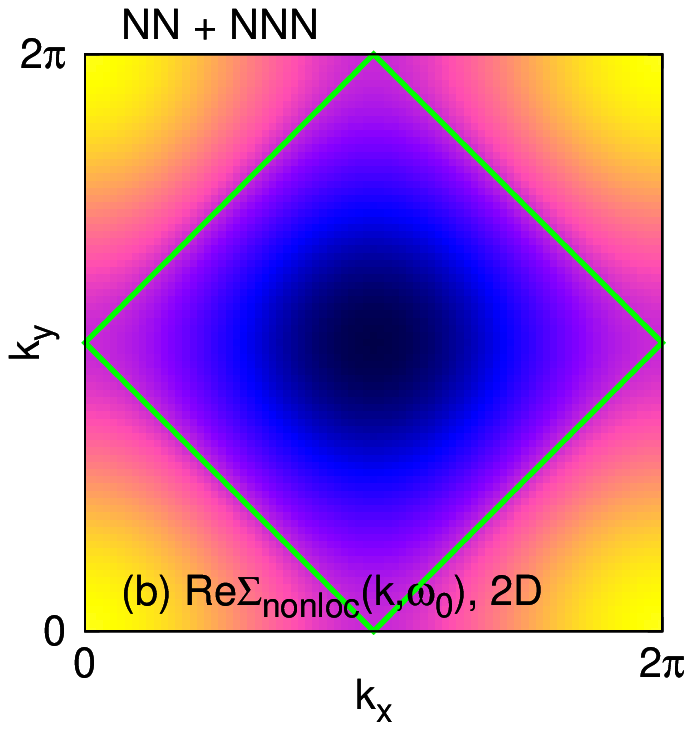}
\includegraphics[width=0.175\textwidth,bb=120 66 360 280]{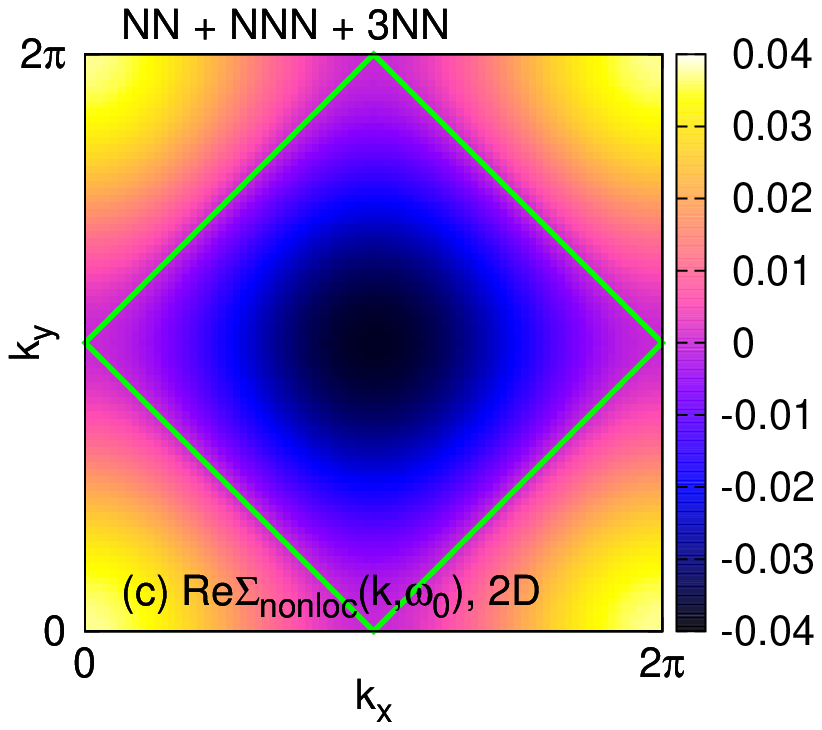}
\includegraphics[width=0.155\textwidth,bb=120 66 330 280]{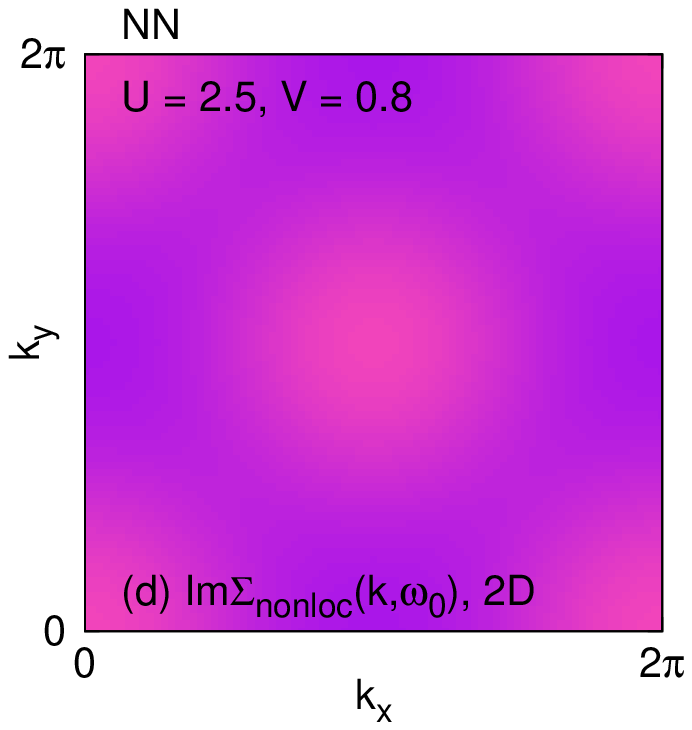}
\includegraphics[width=0.155\textwidth,bb=120 66 330 280]{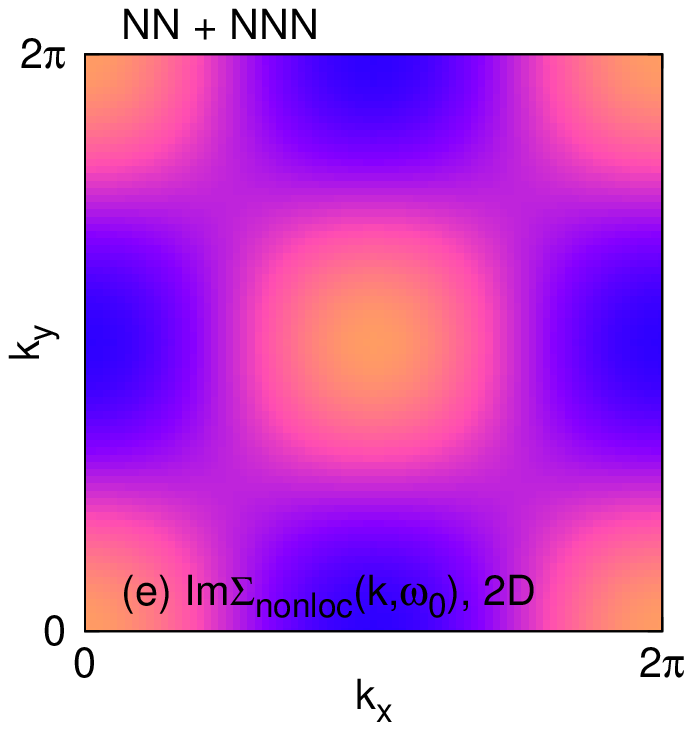}
\includegraphics[width=0.175\textwidth,bb=120 66 360 280]{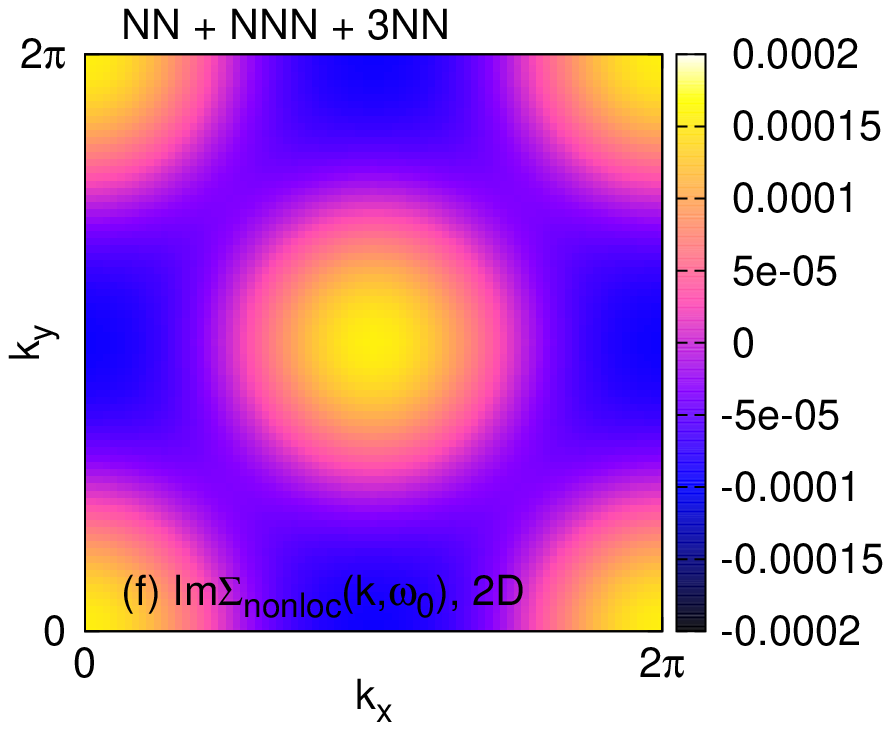}
\includegraphics[width=0.155\textwidth,bb=120 66 330 280]{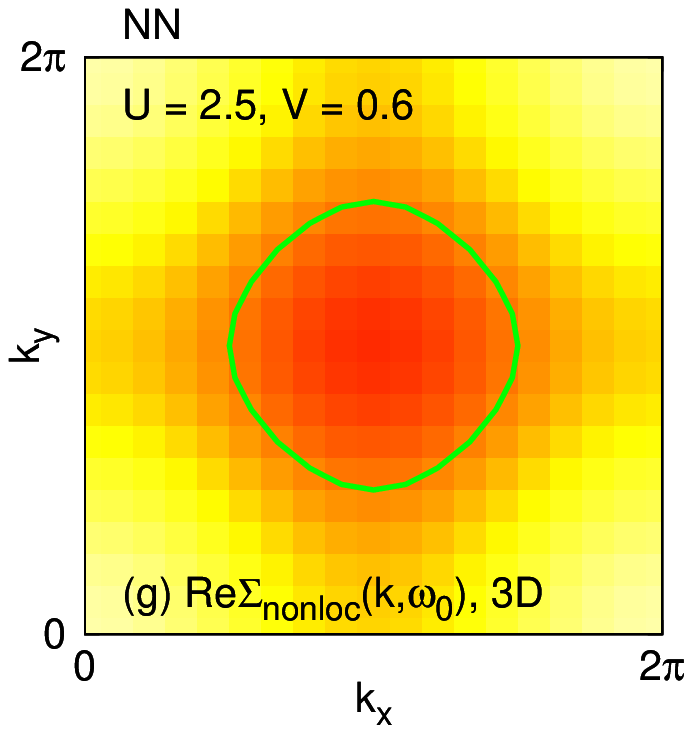}
\includegraphics[width=0.155\textwidth,bb=120 66 330 280]{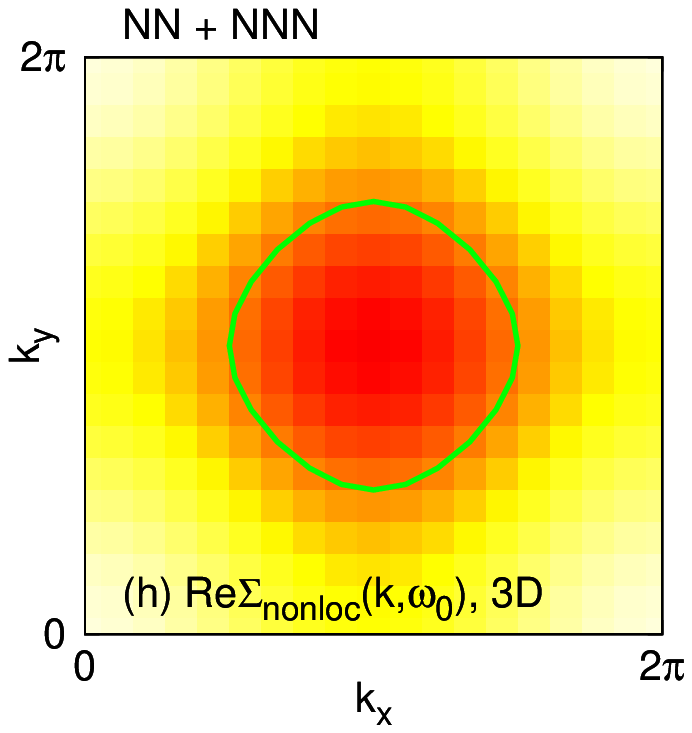}
\includegraphics[width=0.175\textwidth,bb=120 66 360 280]{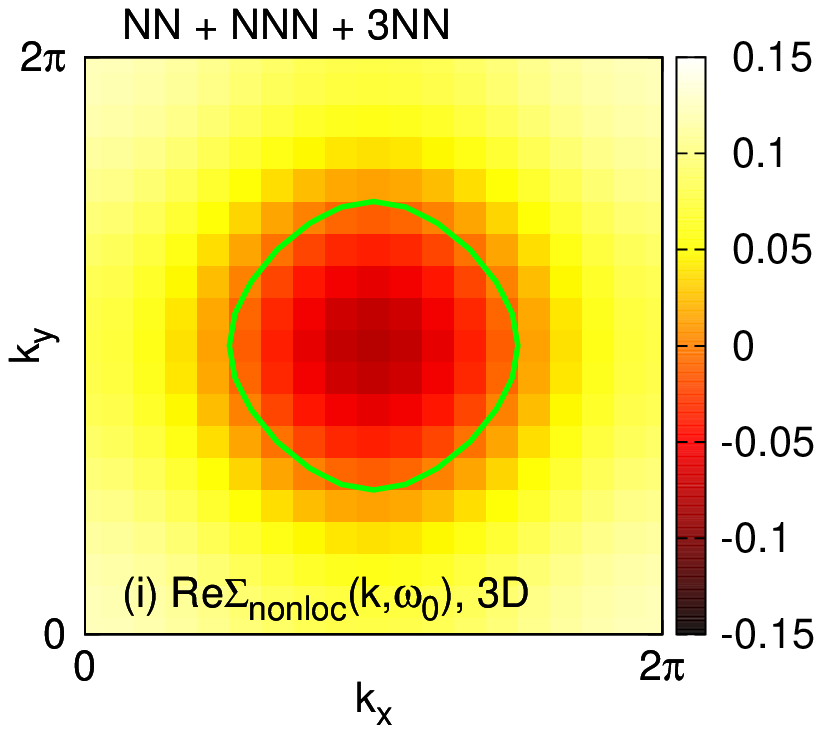}
\includegraphics[width=0.155\textwidth,bb=120 66 330 280]{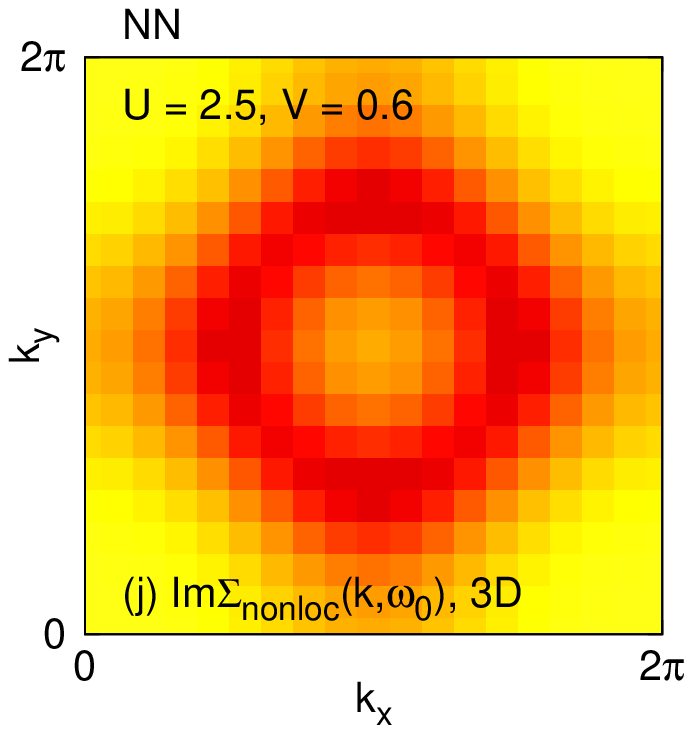}
\includegraphics[width=0.155\textwidth,bb=120 66 330 280]{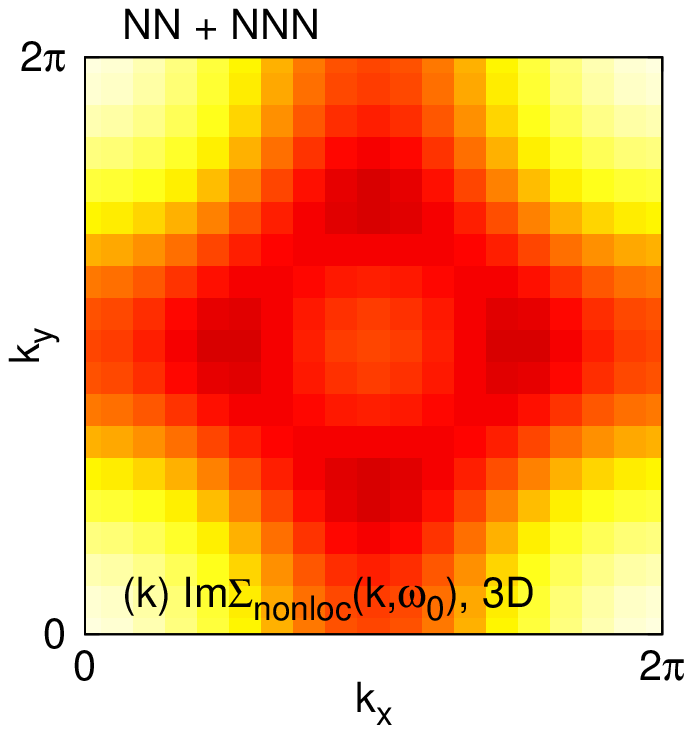}
\includegraphics[width=0.175\textwidth,bb=120 66 360 280]{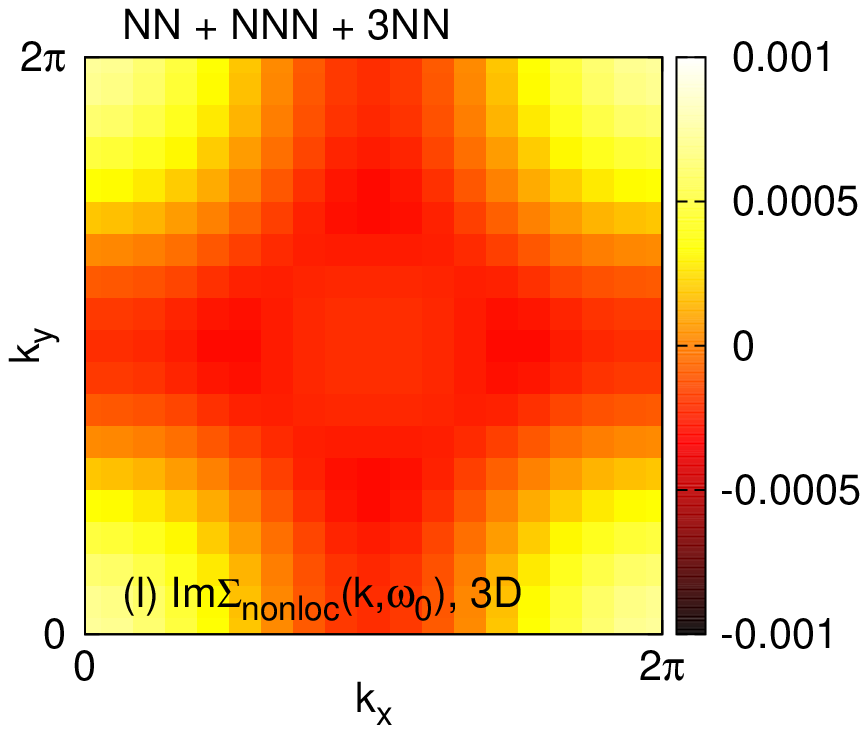}
\caption{(Color online) $\Sigma_{\text{nonloc}}(k, i\omega_0)$ for the extended Hubbard model from $GW$ + EDMFT. (a)-(f) Results for the square lattice, $U$ = 2.5, $V$ = 0.80. (g)-(l) Results for the simple cubic lattice, $U$ = 2.5, $V$ = 0.60. We only show the $k_z = 0$ plane. (a)-(c) and (g)-(i) Re$\Sigma_{\text{nonloc}}(k, i\omega_0)$. (d)-(f) and (j)-(l) Im$\Sigma_{\text{nonloc}}(k, i\omega_0)$. The green curves in (a)-(c) and (g)-(i) panels denote the EDMFT Fermi surface. \label{fig:gw_s}}
\end{figure*}

\begin{figure*}[tp]
\centering
\includegraphics[width=\columnwidth]{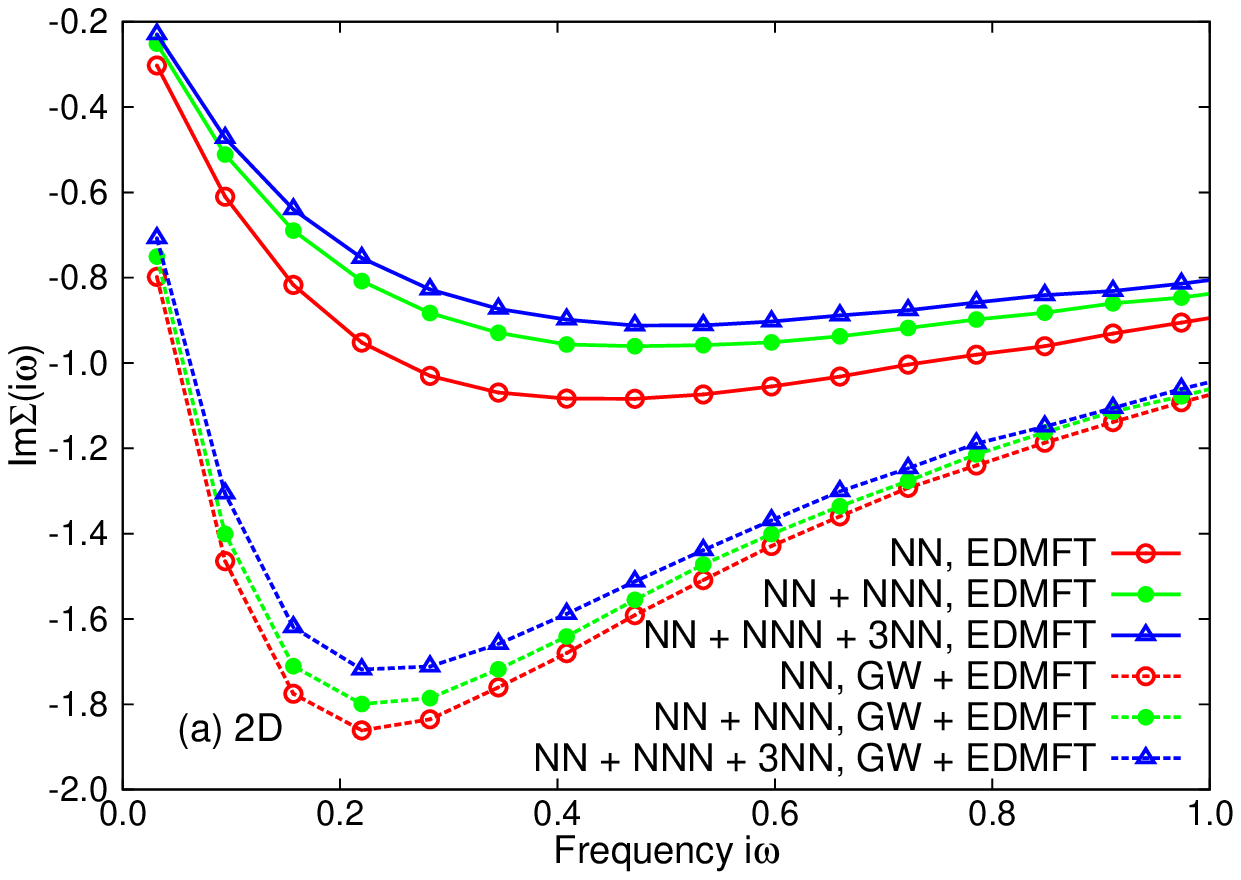}
\includegraphics[width=\columnwidth]{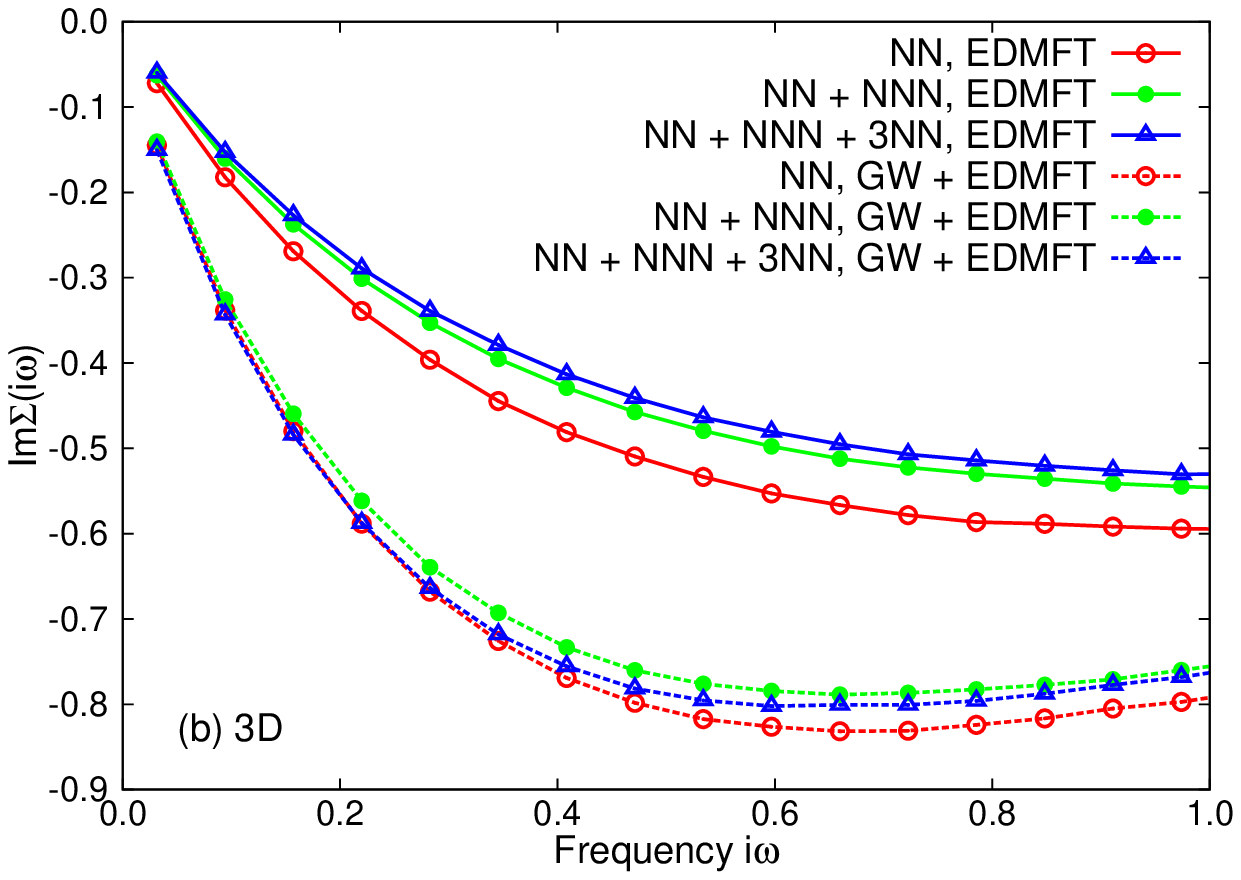}
\caption{(Color online) Imaginary part of the local self-energy function Im$\Sigma(i\omega)$ for the extended Hubbard model solved with EDMFT and $GW$ + EDMFT. (a) Results for the square lattice, $U = 2.5$ and $V = 0.8$. (b) Results for the simple cubic lattice, $U = 2.5$ and $V = 0.6$. \label{fig:gw_sloc23}}
\end{figure*}

The $GW$ + EDMFT method incorporates nonlocal correlations by adding the nonlocal components of the $GW$ self-energy and polarization functions to the EDMFT result.\cite{PhysRevB.66.085120,PhysRevLett.109.226401,PhysRevLett.90.086402,PhysRevB.87.125149} Hence, the $GW$ + EDMFT self-energy and polarization functions are not only frequency-dependent but also momentum-dependent. 

In Fig.~\ref{fig:gw_s} the nonlocal parts of the self-energy for the lowest Matsubara frequency $\omega_0$ are shown. These data have been obtained using Eq.~(\ref{eq:nonlocal_s}). For the square lattice, we plot $\Sigma_{\text{nonloc}}(k, i\omega_0)$ for $k_x$ and $k_y$ $\in [0,2\pi]$. In the case of the simple cubic lattice, we show a cut of $\Sigma_{\text{nonloc}}(k, i\omega_ 0)$ in the $k_z = 0$ plane. Consistent with previous $GW$ + EDMFT calculations for the square lattice with NN interations,\cite{PhysRevB.87.125149} we find that the $GW$ contribution to the imaginary part of the nonlocal self-energy is negligible with respect to the local self-energy. The real part of the nonlocal self-energy is relatively large away from the EDMFT Fermi surface, but does not alter this Fermi surface. Longer range interactions do increase the $k$-dependence, but they do not significantly affect the conclusion that the $k$-dependence of the self-energy both for the 2D and 3D lattice models is not very strong in the $GW$ + EDMFT scheme. Even in the vicinity of the Mott transition (for instance, $U$ = 2.5 and $V$ = 0.8 for the square lattice is very close to the Mott transition, see Fig.~\ref{fig:phase}), the momentum differentiation is weak. This result is in contrast to the strong momentum dependence observed in the self-energy functions obtained from dynamical cluster approximation (DCA)\cite{PhysRevB.80.045120,PhysRevB.80.245102} and cellular dynamical mean-field theory (CDMFT)\cite{PhysRevB.73.205106,RevModPhys.77.1027} calculations for the two-dimensional Hubbard model as one approaches the Mott transition. This discrepancy suggests that additional nonlocal diagrams, such as ladder diagrams, should be included to provide a better description of the momentum dependence of the self-energy functions (and other $k$-dependent quantities). 

As for the nonlocal polarization function for the first bosonic Matsubara frequency $\Pi_{\text{nonloc}}(k, i\nu = 0)$ (not shown in this figure), we observe a stronger momentum dependence, especially when one approaches the charge-ordering transition.\cite{PhysRevB.87.125149} However, it seems that longer range intersite interactions do not enhance this $k$-dependence prominently, which is contrary to the trend found for the nonlocal self-energy.

Finally, we plot  in Fig.~\ref{fig:gw_sloc23} some typical local self-energies in the FL phase. $|\text{Im}\Sigma(i\omega_0)|$ is considerably enhanced in the $GW$ + EDMFT calculations, compared to the EDMFT result. These observations show that local correlations become stronger if the $k$-dependent $GW$ contributions are added to the self-energy and polarization functions in the self-consistency loop. More evidence for this change will be presented in the following section. In Fig.~\ref{fig:gw_sloc23}, we also compare the local self-energies for intersite interactions of different range. The effect of the longer ranged interactions is to reduce the self-energy. In the calculations with long range interactions and nonlocal self-energies we thus have a competition between the additional screening from long range interactions, which leads to weaker correlation effects, and the momentum dependence, which enhances local correlations. The latter effect seems to be dominant. 

\subsubsection{Screened and retarded interactions}
\begin{figure*}[tp]
\centering
\includegraphics[width=0.32\textwidth]{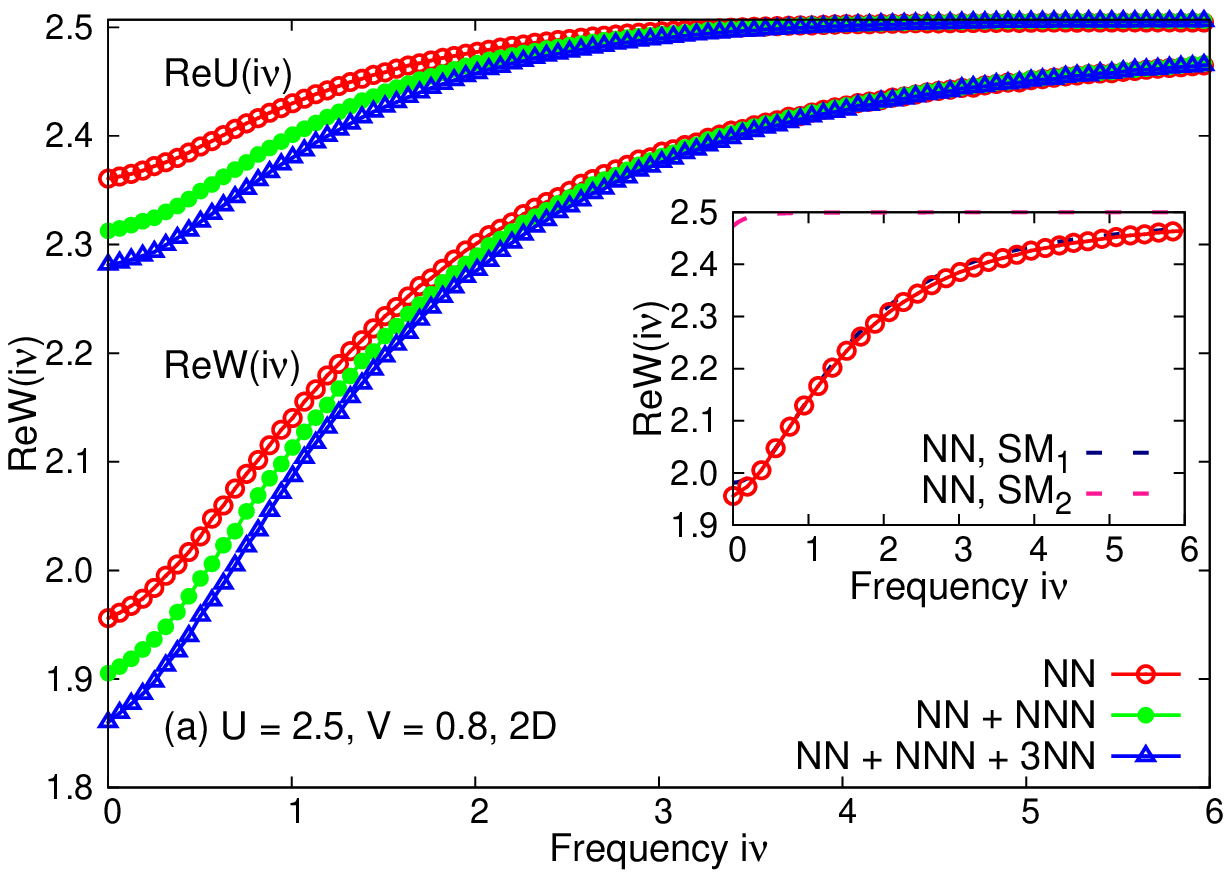}
\includegraphics[width=0.32\textwidth]{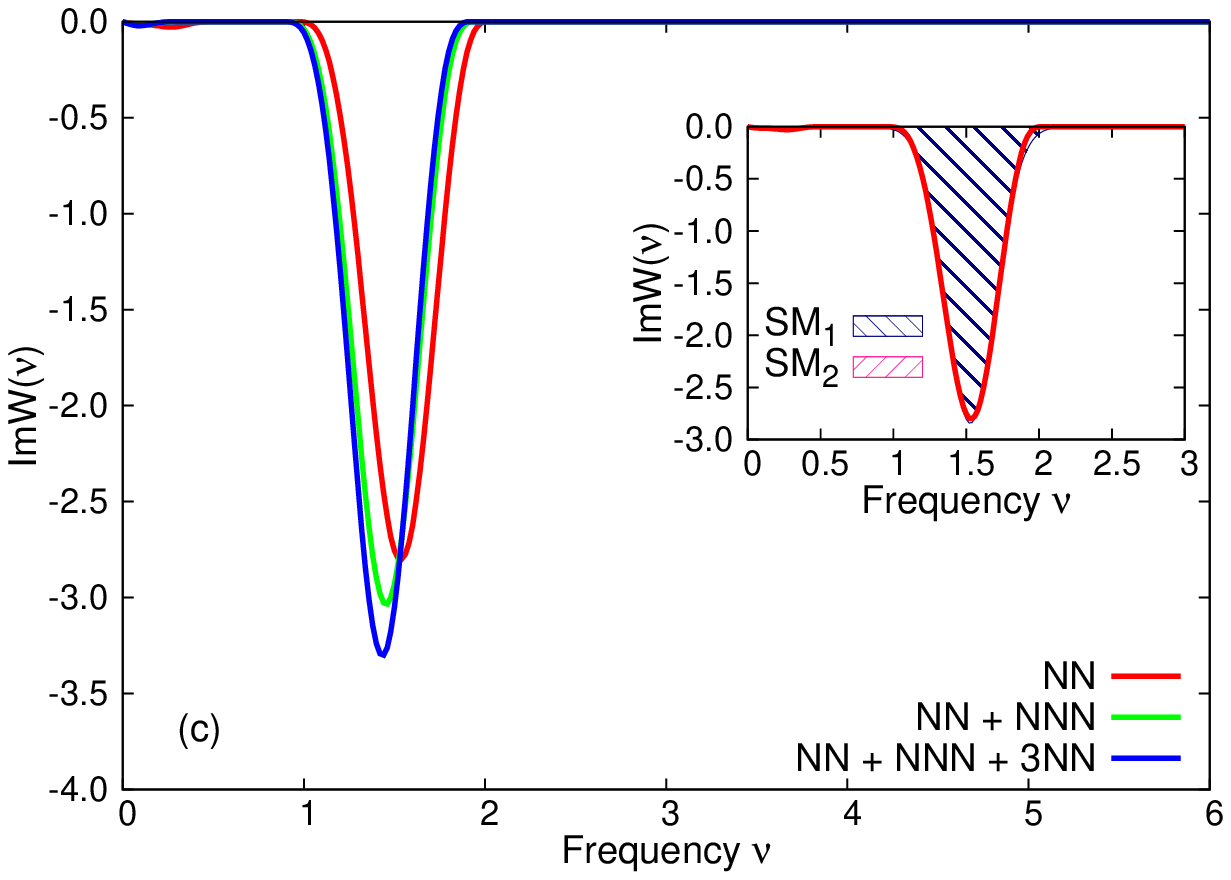}
\includegraphics[width=0.32\textwidth]{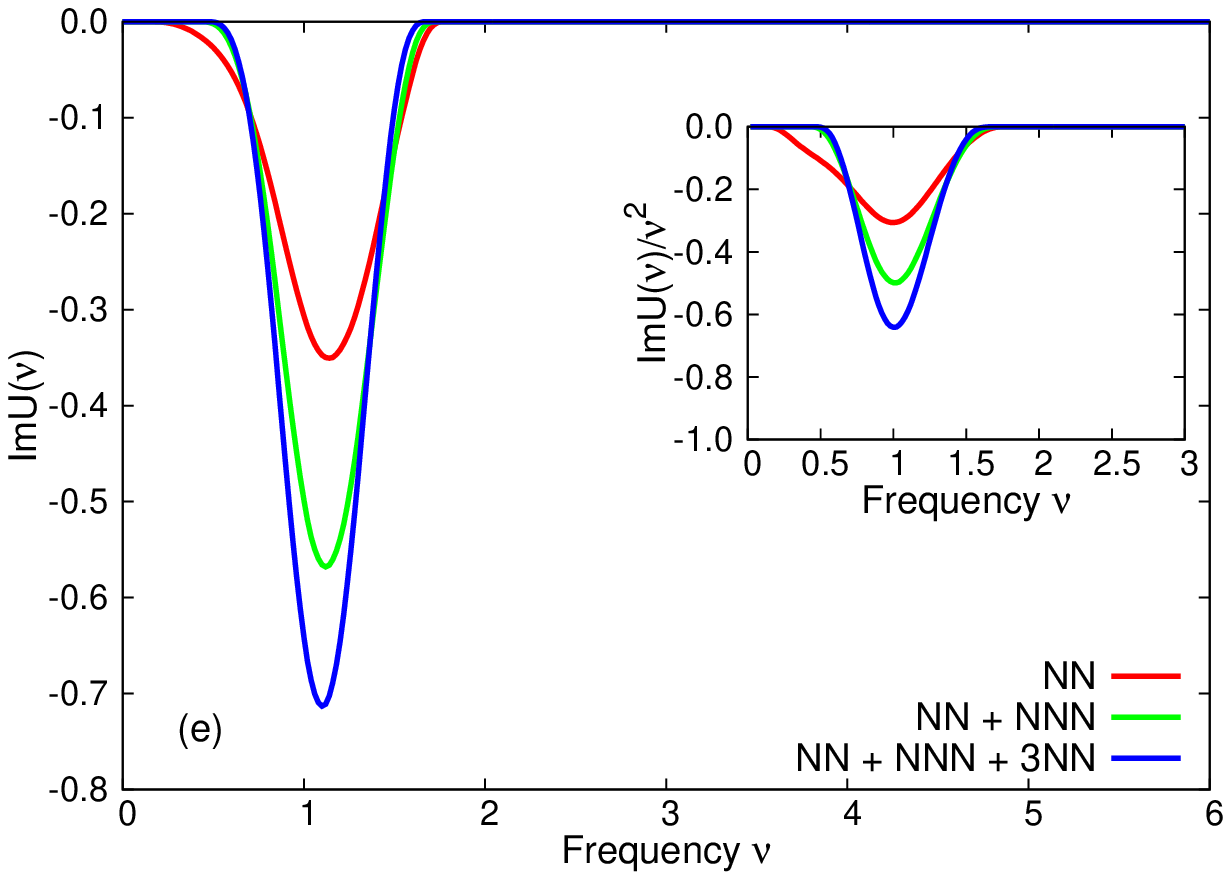}
\includegraphics[width=0.32\textwidth]{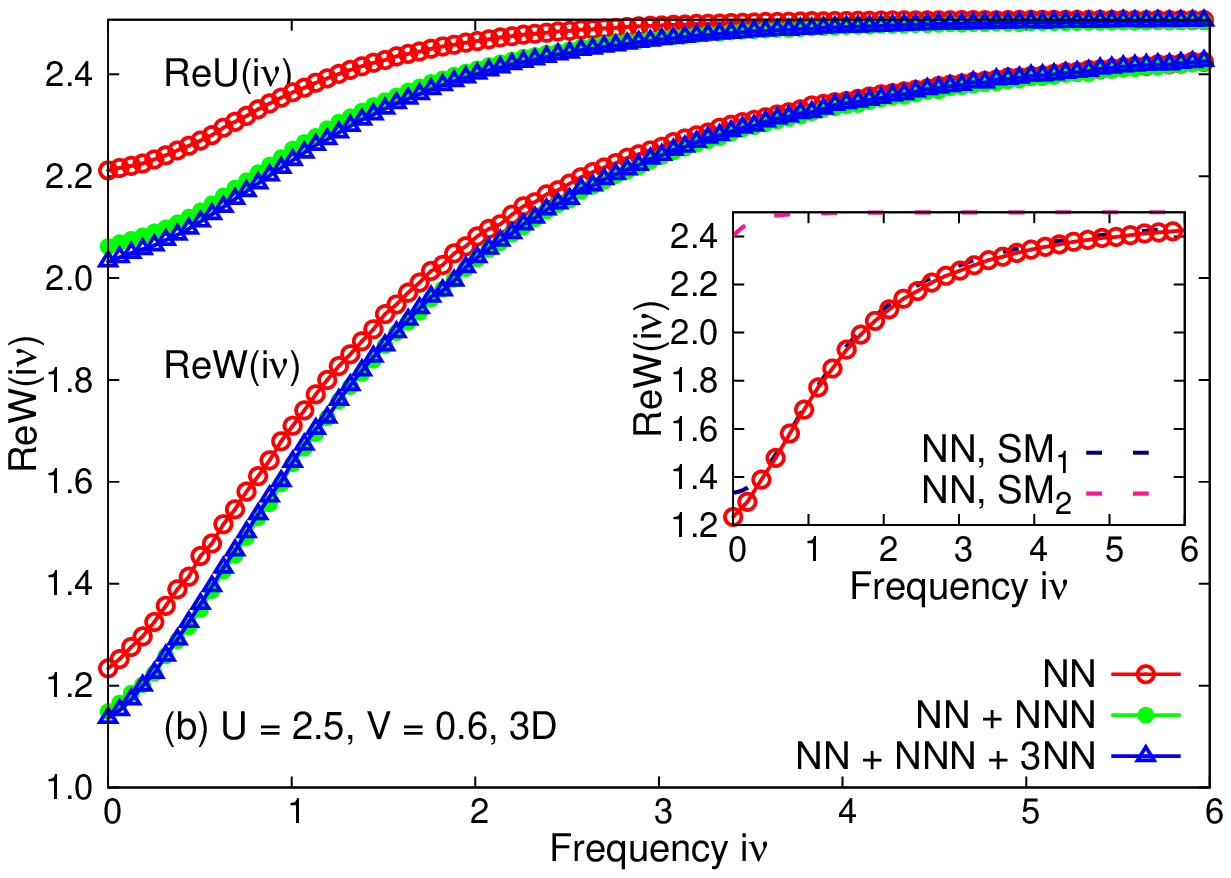}
\includegraphics[width=0.32\textwidth]{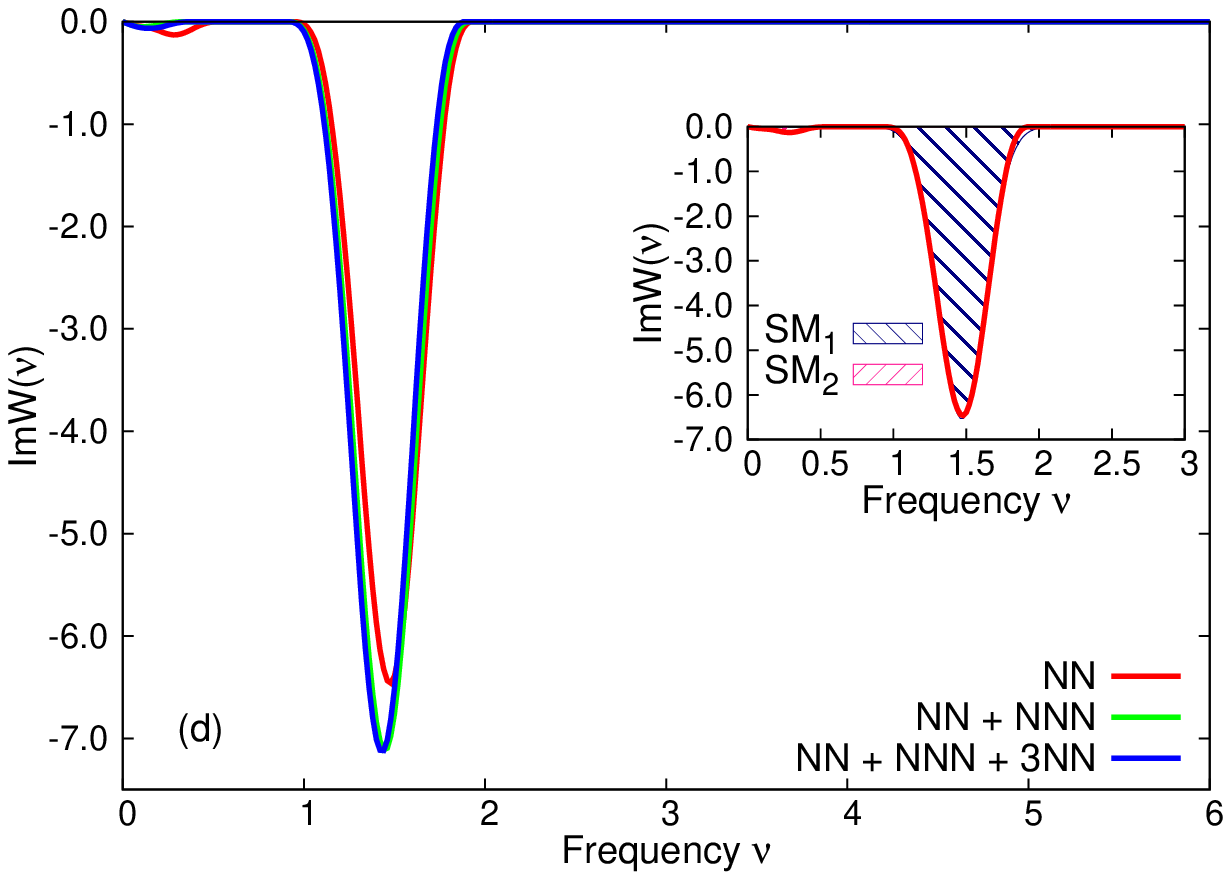}
\includegraphics[width=0.32\textwidth]{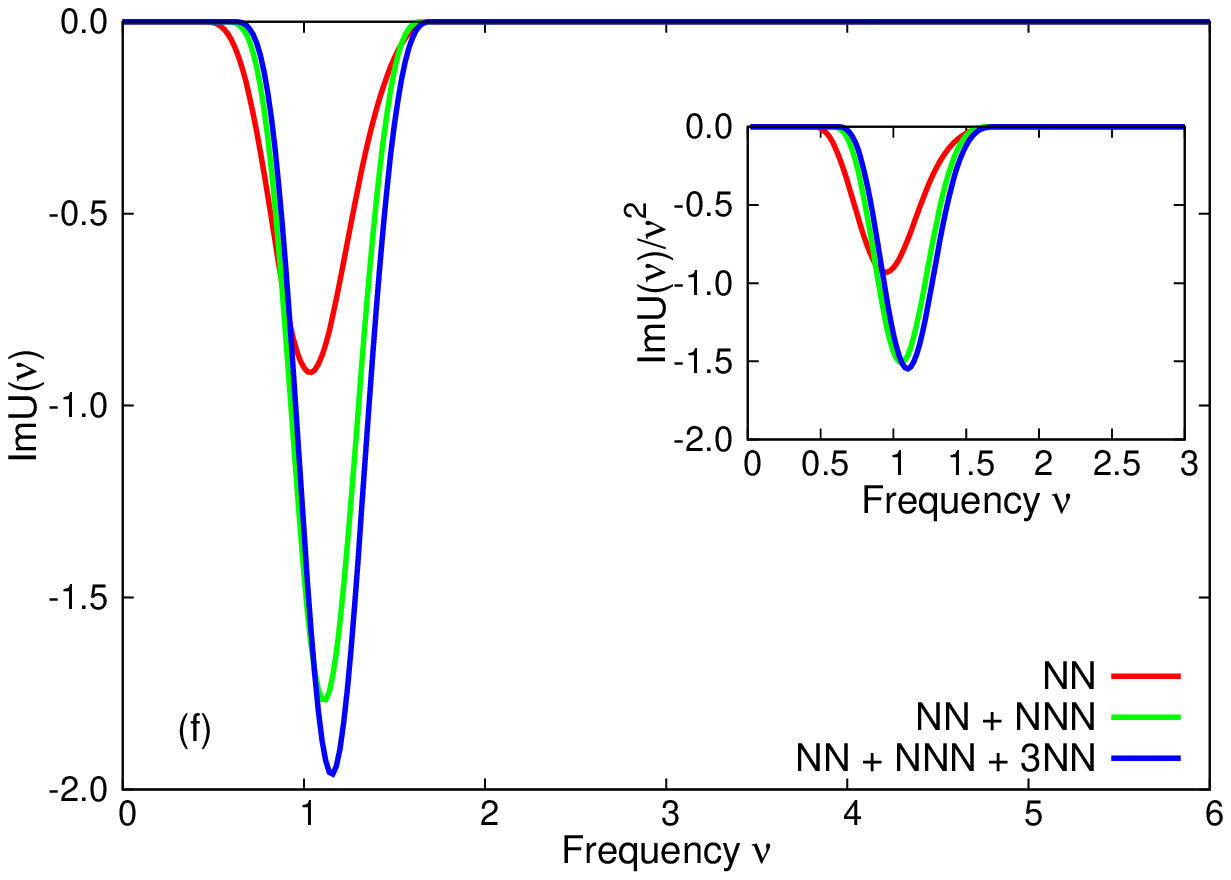}
\caption{(Color online) Real part of the fully screened interactions Re$W(i\nu)$ and partially screened interaction Re$\mathcal{U}(i\nu)$, imaginary part of the real frequency fully screened interaction Im$W(\nu)$ and partially screened interactions Im$\mathcal{U}(\nu)$ for the extended Hubbard model solved by $GW$ + EDMFT. (a), (c) and (e) Results for the square lattice, $U = 2.5$ and $V = 0.8$. (b), (d) and (e) Results for the simple cubic lattice, $U = 2.5$ and $V = 0.6$. In this figure, SM means screening mode. In the insets of panels (a) and (b), the SM-resolved Re$W(i\nu)$, together with full Re$W(i\nu)$ are shown for the NN case. In the (c) and (d) panels, the Im$W(\nu)$ for the NN case is approximated by Gaussian-type functions. The fitted results are shown in the insets. Each Gaussian peak corresponds to a SM. The insets in panels (e) and (f) show the Im$\mathcal{U}(\nu)/\nu^2$ functions. Here Im$W(\nu)$ and Im$\mathcal{U}(\nu)$ are extracted using a modified maximum entropy method. See Appendix \ref{app:mem} for more details. \label{fig:gw_wloc23}}
\end{figure*}

As we have seen in the previous subsection, the $GW$ + EDMFT scheme not only adds nonlocal contributions to the self-energy $\Sigma(k,i\omega_n)$ and polarization $\Pi(k,i\nu_n)$, but it also affects the local quantities through the self-consistency loop.\cite{PhysRevB.87.125149} Figure~\ref{fig:gw_wloc23} shows the fully screened local interaction Re$W(i\nu)$ and partially screened interaction Re$\mathcal{U}(i\nu)$, together with the corresponding spectral functions Im$W(\nu)$ and Im$\mathcal{U}(\nu)$, for the square lattice and simple cubic lattice in the FL metallic state. The related EDMFT data have been plotted in Fig.~\ref{fig:wloc23} and analyzed in Sec.~\ref{sub:edmft}. Again, our results are consistent with previous $GW$ + EDMFT studies for the 2D and 3D extended Hubbard model if available.\cite{PhysRevB.87.125149,PhysRevB.66.085120} 

Compared to the EDMFT result, both Re$W(i\nu = 0)$ and Re$\mathcal{U}(i\nu = 0)$ are greatly enhanced [see Fig.~\ref{fig:gw_wloc23}(a) and (b)], while $|\text{Im}G(i\omega_0)|$ (not shown in these figures) is reduced. This indicates that the local interactions are stronger in $GW$ + EDMFT than in EDMFT, i.e., that the screening effect is weaker. This can be understood in the following way:\cite{PhysRevB.87.125149} In the EDMFT approach, all of the screening and correlation effects are absorbed into the local self-energy. However, in the framework of $GW$ + EDMFT, some of these effects are carried by the nonlocal self-energy. In other words, the screening between local and nonlocal quantities is redistributed in the $GW$ + EDMFT scheme, and the result of this is that the local interaction becomes less screened. Let us also mention that Nomura \emph{et al}.\cite{PhysRevB.86.085117} have shown that the nonlocal polarization induces an anti-screening effect, which competes with the screening effect caused by the long range intersite interactions. Anyhow, the interplay between the local and nonlocal self-energy and polarization in $GW$ + EDMFT leads, after self-consistency, to a weaker screening effect.

Another interesting observation is that the Im$W(\nu)$ and Im$\mathcal{U}(\nu)$ spectra extracted from the self-consistent $GW$ + EDMFT calculations [see Fig.~\ref{fig:gw_wloc23}(c)-(f)] exhibit a single-hump structure, whereas the corresponding EDMFT results yield a two-hump structure [see Fig.~\ref{fig:wloc23}(c)-(f)]. Once again, we have fitted Im$W(\nu)$ with multiple Gaussians to extract the positions and weights of the dominant SMs. It seems that the Im$W(\nu)$ spectra obtained from the $GW$ + EDMFT calculations feature only one medium-frequency SM ($\sim$ 1.5 eV), while the low-frequency SMs ($\sim$ 0.5 eV) are extremely weak and the high-frequency SMs ($2 \sim 3$ eV) previously identified in the EDMFT results have disappeared. As for the Im$\mathcal{U}(\nu)$ spectra, analogous characteristics are observed. Since the satellite structures of the local spectral function $A(\omega)$ are determined by the function Im$\mathcal{U}(\nu)/\nu^2$,\cite{Werner2012} we conclude that the high-frequency features of $A(\omega)$ will be different in the $GW$ + EDMFT calculations, and more specifically that the satellites will be at lower energy. Though we only present results for the FL metallic phase in this figure, those for the Mott phase and the strongly correlated metal phase between the MI and CO states exhibit the same trend (see also Tab.~\ref{tab:square_cubic}). 

Next, we consider the influence of longer range intersite interactions on the static screened and retarded interactions obtained with the $GW$ + EDMFT scheme. Table~\ref{tab:square_cubic} also shows data collected from $GW$ + EDMFT calculations. Once more, we see that Re$\mathcal{U}(\nu = 0)$ and Re$W(\nu = 0)$ are reduced, and $|\text{Im}G(i\omega_0)|$ (not shown in the Table) is enhanced if longer range intersite interactions are present. The effects of longer range interactions and nonlocal correlations compete with each other: the longer range intersite interaction tends to enhance the screening and make the system less correlated, while including the $GW$ nonlocal self-energies and polarizations has the opposite effect. The latter effect is dominant. From Fig.~\ref{fig:gw_wloc23}(e) and (f), we can see that the weight of the hump in the Im$\mathcal{U}(\nu)$ spectra increases if longer range intersite interactions are added which means a larger screening effect. However, interestingly, the effective screening frequency $\nu_0$ is only little affected by the range of the interaction within the $GW$ + EDMFT approach, which is also seen in Tab.~\ref{tab:square_cubic}. 

\subsubsection{Local spectral properties}
\begin{figure*}[tp]
\centering
\includegraphics[width=0.32\textwidth]{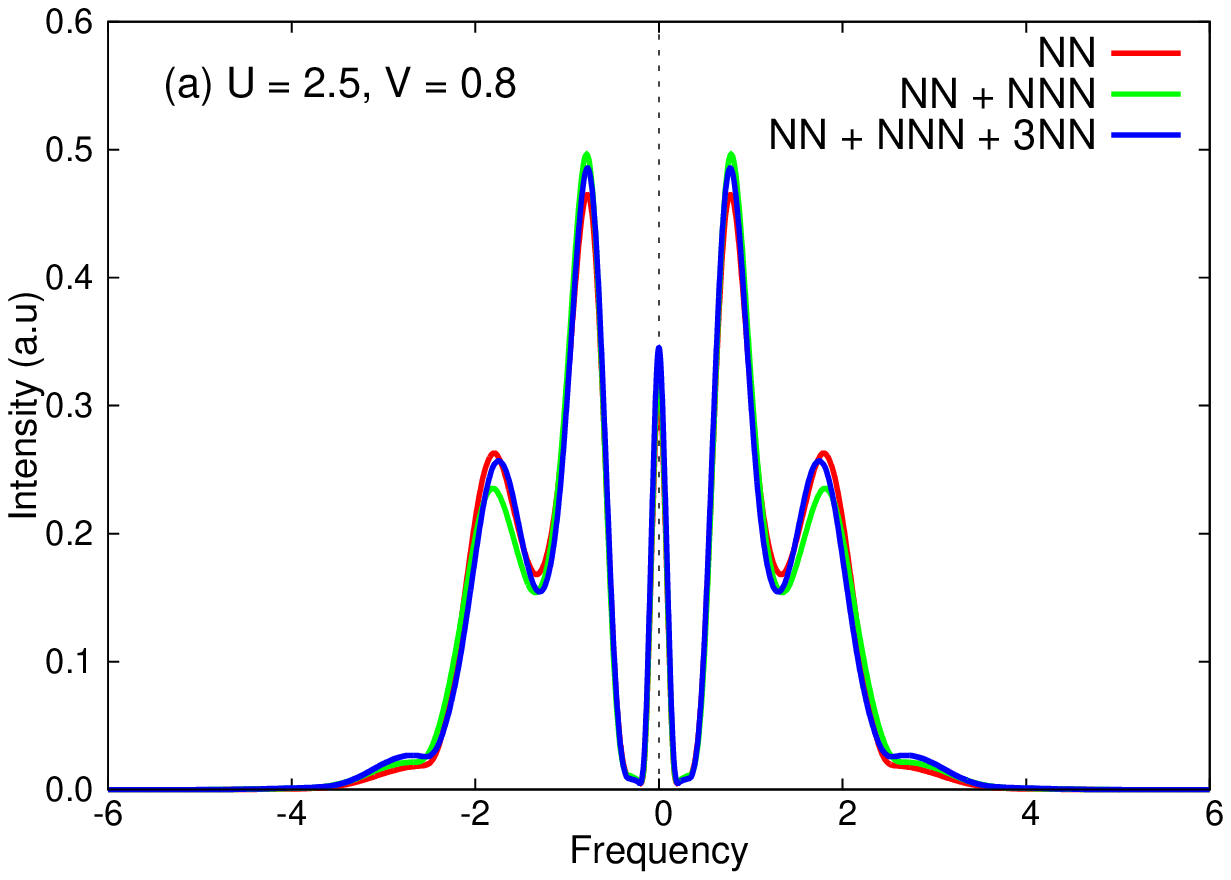}  
\includegraphics[width=0.32\textwidth]{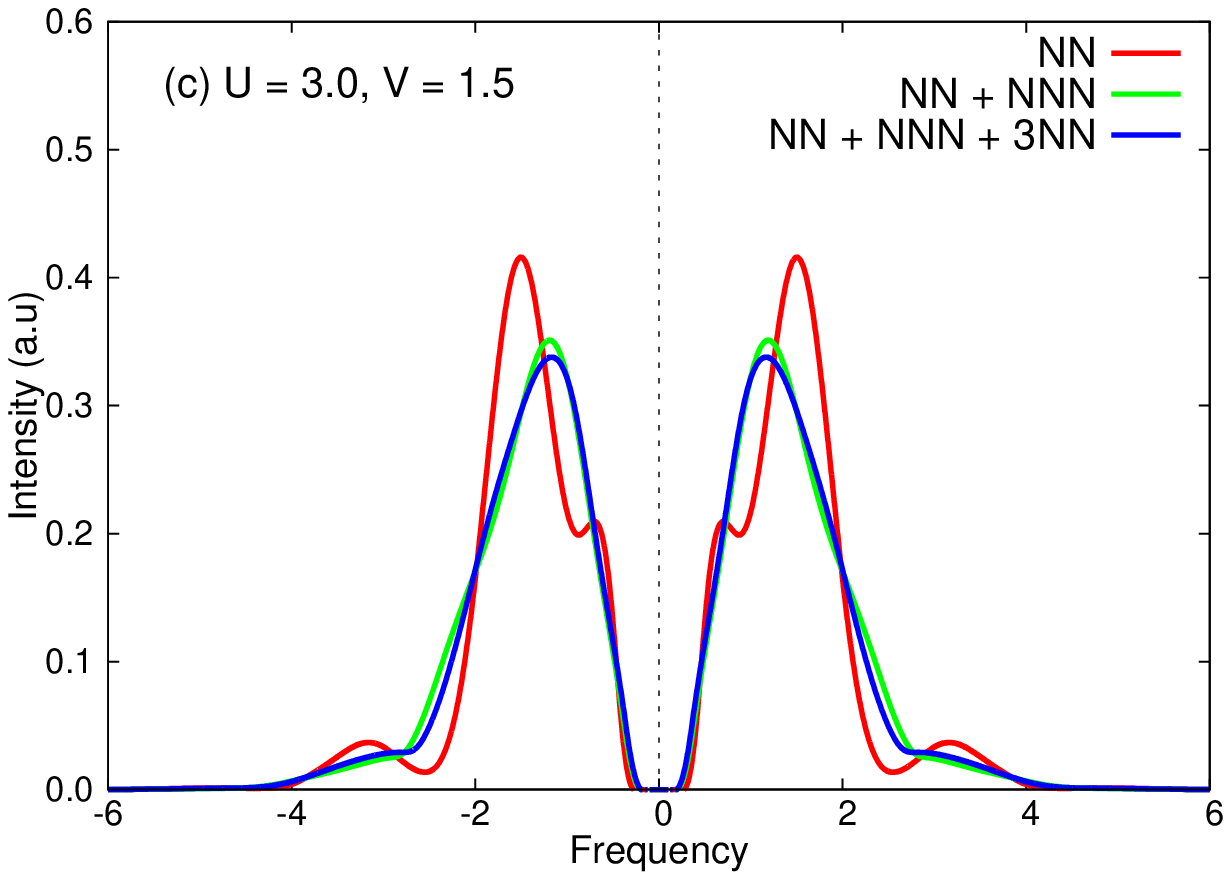}
\includegraphics[width=0.32\textwidth]{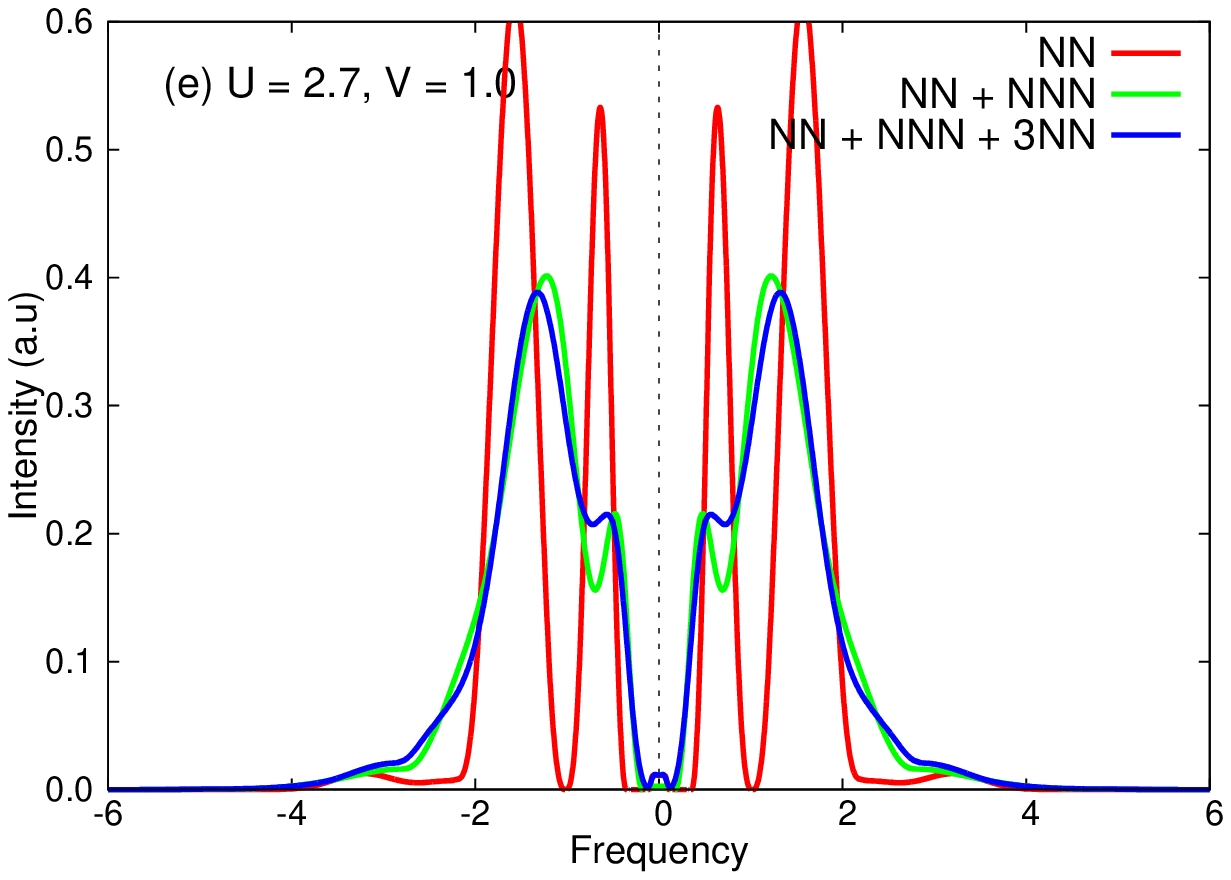}
\includegraphics[width=0.32\textwidth]{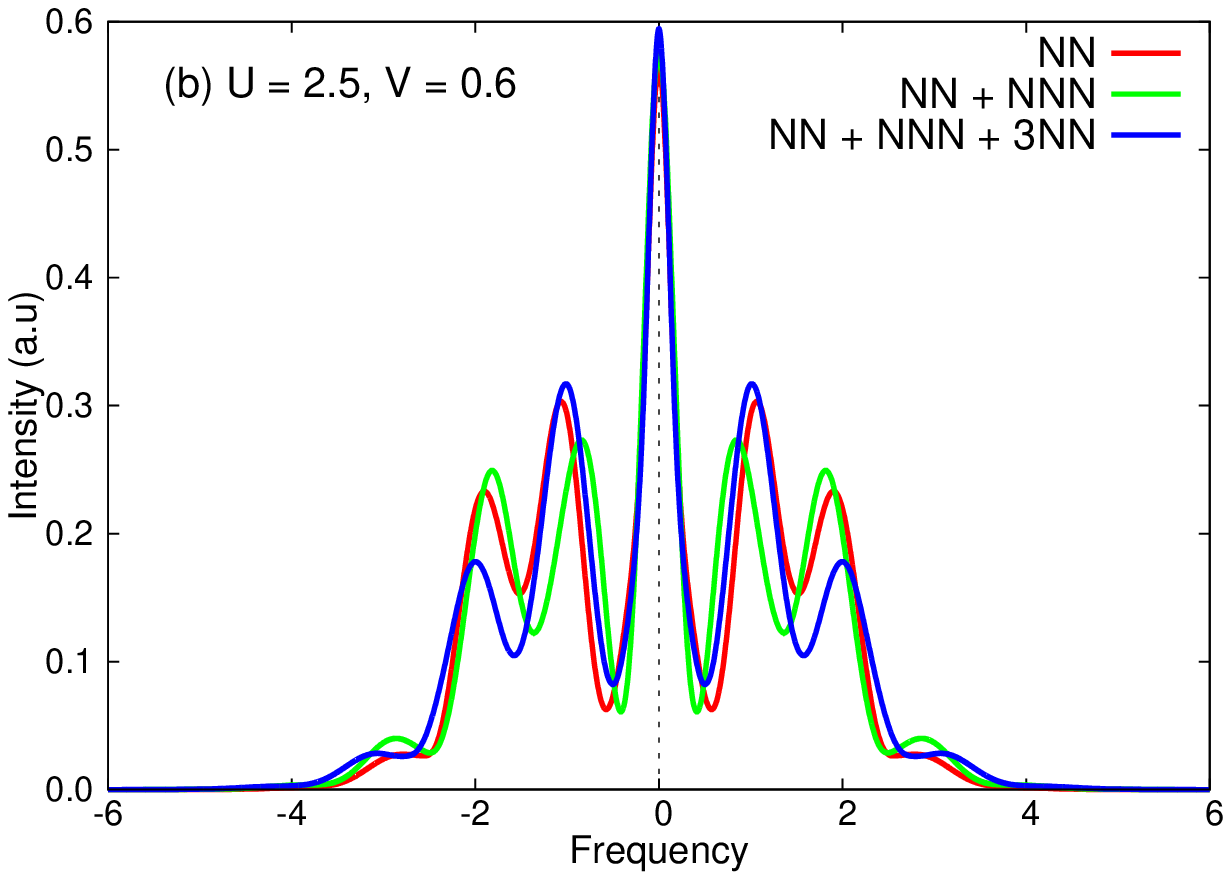}
\includegraphics[width=0.32\textwidth]{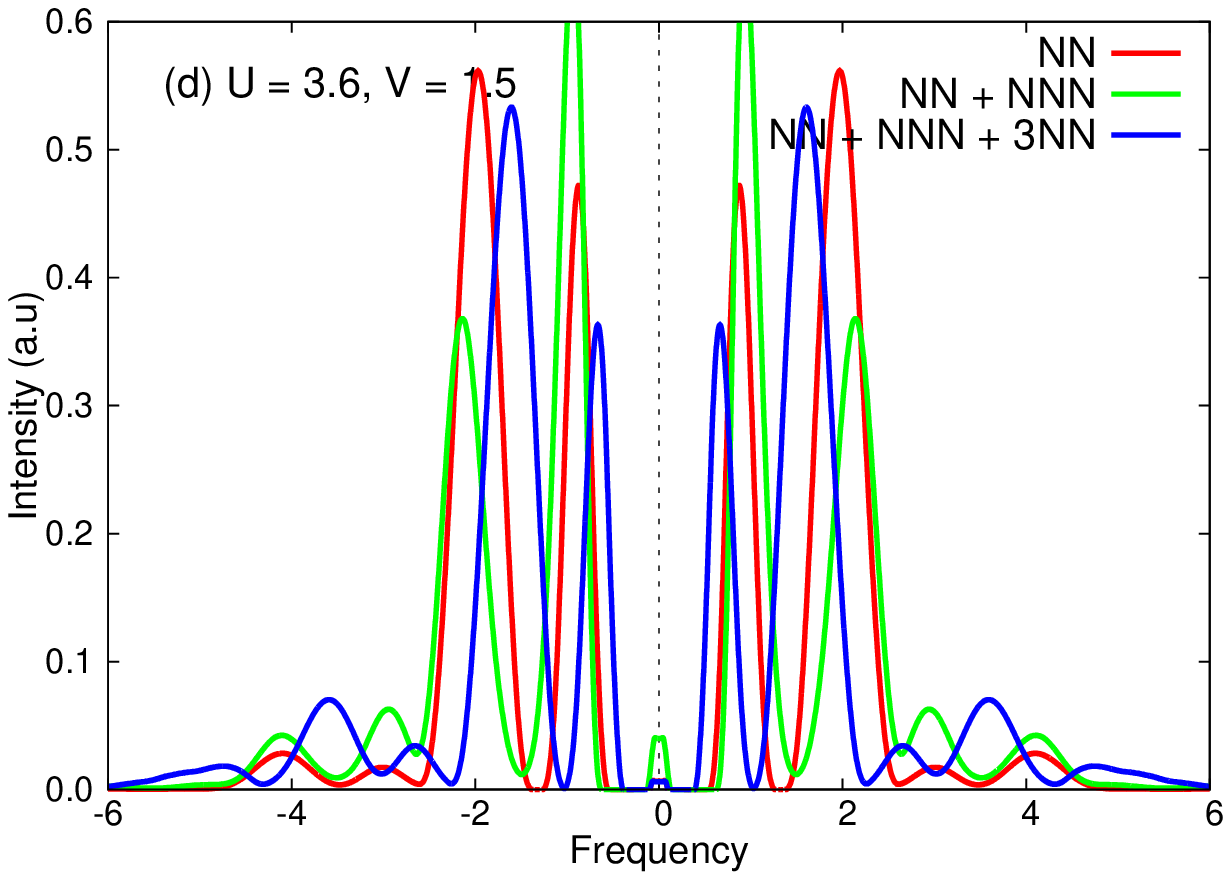}
\includegraphics[width=0.32\textwidth]{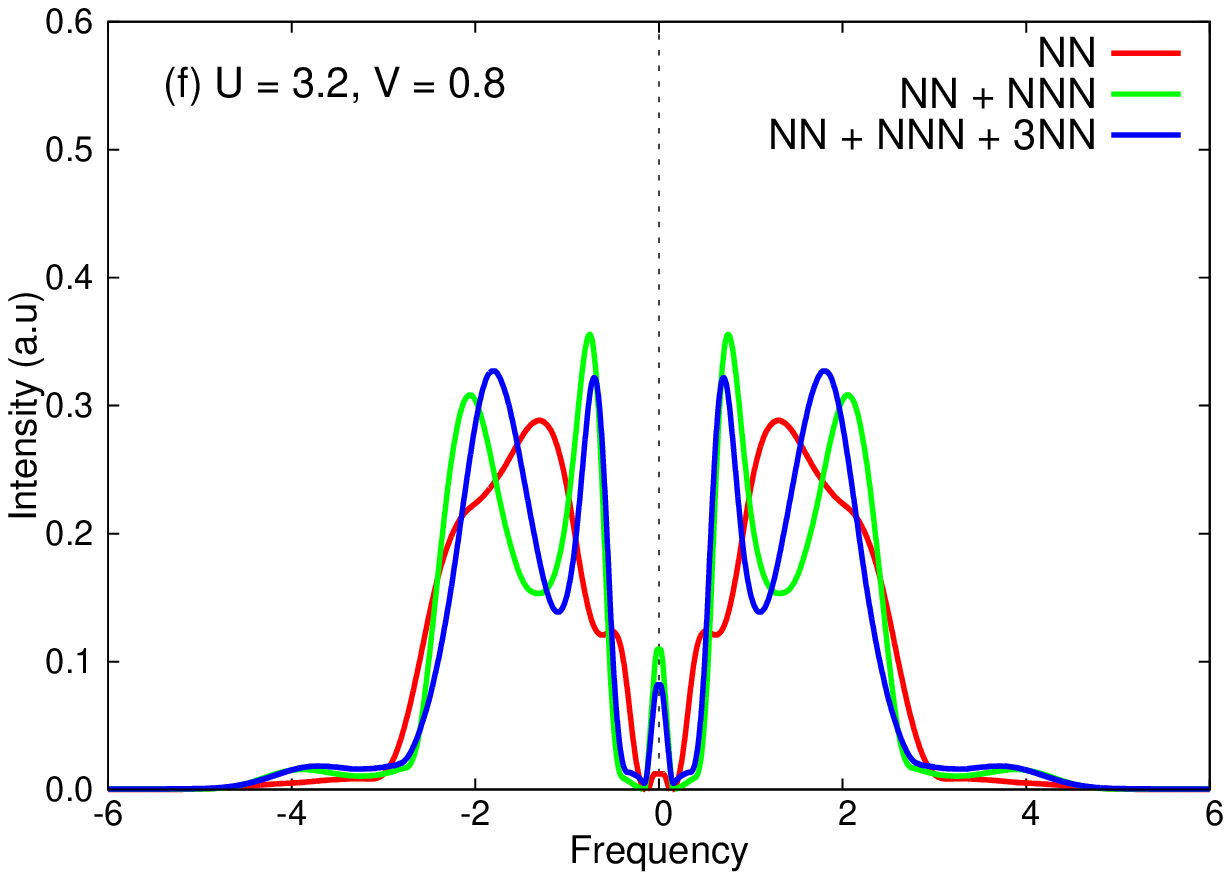}
\caption{(Color online) Spectral functions at selected points for the single-band half-filled extended Hubbard model solved by $GW$ + EDMFT. (a), (c), and (e) Results for the square lattice. (b), (d), and (e) Results for the simple cubic lattice. The parameters are as follows: (a) Metallic region, $U = 2.5$ and $V = 0.8$; (b) Metallic region, $U = 2.5$ and $V = 0.6$; (c) Mott insulating region, $U = 3.0$ and $V = 1.5$; (d) Mott insulating region, $U = 3.6$ and $V = 1.5$; (e) ``Triangle" zone, $U = 2.7$ and $V = 1.0$; (f) ``Triangle" zone, $U = 3.2$ and $V = 0.8$. The impurity spectral functions are obtained using the analytical continuation method proposed in Ref.~\onlinecite{PhysRevB.85.035115}. \label{fig:gw_aw23}}
\end{figure*}

The top panels of Fig.~\ref{fig:gw_aw23} show some typical spectral functions for the square lattice obtained by $GW$ + EDMFT. Similar results for the simple cubic lattice are shown in the bottom panels. Here we consider the FL metallic state, MI state, and the ``triangle'' zone in the $U-V$ phase diagrams. Since the parameter values are the same, one can directly compare these spectra to the EDMFT results as shown in Fig.~\ref{fig:aw23}. Consistent with the previous discussion, within the $GW$ + EDMFT scheme, the quasiparticle peak is greatly reduced, and the upper and lower Hubbard bands become more pronounced. For instance, let us focus on the ``triangle'' zone for the square lattice (parameters $U = 2.7$ and $V = 1.0$). The EDMFT local spectral function shows considerable weight at the Fermi level, i.e., the system is metallic [see Fig.~\ref{fig:aw23}(e), for the NN + NNN case]. However, the corresponding $GW$ + EDMFT spectral function has almost no weight at $\omega = 0$, which means that it is close to or even in the MI phase [see Fig.~\ref{fig:gw_aw23}(e), for the NN + NNN case]. From this fact, we conclude that there exists a small difference between the FL-MI phase boundaries calculated with EDMFT and $GW$ + EDMFT, respectively, and that the MI region in the latter case should be larger.

The influence of longer range intersite interactions on the local spectral functions $A(\omega)$ is very similar to the EDMFT case. Namely, longer range intersite interactions enhance the quasiparticle peak and shift spectral weight to high-energy satellites. The local spectral function becomes more metallic in character as a result of the additional screening. Consistent with the lower energies of the SMs in the $GW$ + EDMFT case, the satellite features appear at lower energies. For example, in Fig.~\ref{fig:gw_wloc23}(c) (with $GW$ + EDMFT) the satellites are at energy $\pm$3-3.5\ eV, while in Fig.~\ref{fig:wloc23}(c) (with EDMFT), they are at $\omega \approx \pm 4$\ eV.

\section{conclusions\label{sec:conclu}}
We studied the paramagnetic solutions of the single-band half-filled extended Hubbard model on the square and simple cubic lattices by means of the EDMFT method. Longer range intersite interactions introduce additional screening and lead to smaller effective local interactions. In the weakly correlated region, longer range intersite interactions favor the metallic phase, whereas in the strongly correlated region, they stabilize the CO phase. The obvious ``kink" in the $V_c(U)$ line near the Mott transition point in the square lattice model with NN intersite interaction becomes a smooth slope change if longer range interactions are included. At the same time, the metallic region extends to larger $U$ values, so that the transition between MI and CO phases is via an intermediate metallic phase. We showed that the slope change in the $V_c(U)$ line, which cannot be explained by a simple mean-field picture, is associated with a sudden increase in the effective screening frequency near the critical $U_c$ for the FL-MI transition.

Like DMFT, the EDMFT formalism is based on a local approximation.\cite{PhysRevLett.84.3678,PhysRevB.66.085120} To incorporate spatial correlations, we performed fully self-consistent $GW$ + EDMFT calculations for some selected $U$ and $V$ parameters. On the one hand, longer range intersite interactions enhance the screening effect, just as in the EDMFT case. The screened and retarded interactions are strongly reduced. On the other hand, within the $GW$ approximation the screening effect is weakened, which leads to a larger $U_{\text{scr}}$ [$\equiv$ Re$\mathcal{U}(i\nu = 0)$] compared to the EDMFT result. In other words, considering the nonlocal $GW$ self-energy and polarization makes the system more correlated. As a consequence, the $U_c(V)$ line (MI-FL phase boundary) will be modified slightly and shifted to smaller $U$. The results obtained from the $GW$ + EDMFT calculations confirm that the nonlocal contributions to the self-energy coming from the $GW$ diagrams are quite small in the case of the extended Hubbard model, which agrees with previous $GW$ + EDMFT studies,\cite{PhysRevB.87.125149} but is not consistent with DCA\cite{PhysRevB.80.045120,PhysRevB.80.245102} and CDMFT results.\cite{PhysRevB.73.205106,RevModPhys.77.1027} The effect of longer range intersite interactions is to enhance the nonlocal self-energy and polarization functions.

We have critically reexamined the possibility of finding simple rules of thumb for local interaction parameters incorporating screening by nonlocal interactions in an effective manner. While in the 2D case with NN interactions only, a local interaction $U$ reduced by the NN interaction $V$ provides a lower bound for such an effective interaction, in all other cases the strong charge fluctuations in the proximity of the charge-ordered phase invalidate any simple estimate. This is consistent with a growing range of charge-charge correlations close to the transition.

The single-band extended Hubbard model calculations presented in this paper can be straightforwardly extended to the general multiorbital case, paving the way for realistic first-principles materials calculations. Low dimensional $sp$-electron systems like graphene,\cite{PhysRevLett.106.236805} silicene,\cite{PhysRevLett.111.036601} aromatic molecules such as benzene,\cite{PhysRevLett.111.036601} and systems of adatoms on semiconductor surfaces such as Si(111):$X$\cite{PhysRevLett.110.166401} feature simultaneously strong local and nonlocal Coulomb interactions. Obviously, these cannot be adequately addressed in the simple DMFT framework, which cannot handle nonlocal intersite interaction $V$ beyond the Hartree level. The EDMFT and $GW$ + EDMFT approaches provide a relatively inexpensive treatment of local and (short range or long range) nonlocal interactions, making the application of them to electronic structure calculations of realistic materials worthwhile and promising.

\begin{acknowledgments}
We acknowledge fruitful discussions with Junya Otsuki and Hiroshi Shinaoka. This work was supported by SNF Grant No. 200021-140648. The calculations have been performed on the computer cluster at Fribourg University, using a code based on ALPS.\cite{alps2011} 
\end{acknowledgments}

\appendix
\section{Long range intersite interactions for extended Hubbard model\label{app:vk}}
The partition function of the single-band extended Hubbard model [see Eq.~(\ref{eq:ham})] is 
\begin{equation}
Z = \text{Tr}e^{-\beta H},
\end{equation}
with inverse temperature $\beta$. It is more convenient to express it in the path-integral form 
\begin{equation}
Z = \int \mathcal{D}[c^{*}_{i},c_{i}]e^{-S},
\end{equation}
where the effective action $S$ is
\begin{align}
\label{eq:action}
S[c^{*},c] = \int^{\beta}_{0}\text{d}\tau \Bigg\{
&\sum_{ij,\sigma}c^{*}_{i\sigma}(\tau) [ (\partial_{\tau} - \mu)\delta_{ij} - t_{ij} ] c_{j\sigma}(\tau) \nonumber \\
&+ U \sum_{i} n_{i\uparrow}(\tau) n_{i\downarrow}(\tau) \nonumber \\
&+ \frac{1}{2} \sum_{ij} V_{ij} n_{i}(\tau) n_{j}(\tau) \Bigg\}.
\end{align}
Using the identity $n_{i}n_{i} = (n_{i\uparrow} + n_{i\downarrow})^2 = n_{i} + 2n_{i\uparrow}n_{i\downarrow}$, we can rewrite the action as 
\begin{align}
S[c^{*},c] = \int^{\beta}_{0}\text{d}\tau \Bigg\{
&\sum_{ij,\sigma}c^{*}_{i\sigma}(\tau) [ (\partial_{\tau} - \tilde{\mu})\delta_{ij} - t_{ij} ] c_{j\sigma}(\tau) \nonumber \\
&+ \frac{1}{2} \sum_{ij} v_{ij} n_{i}(\tau) n_{j}(\tau) \Bigg\},
\end{align}
where $\tilde{\mu} = \mu + U/2$, and $v_{ij} = U\delta_{ij} + V_{ij}$. Thus, in reciprocal space, we have the equation: $v_{k} = U + V_{k}$. Here, $v_{k}$ is the $k$-dependent bare interaction, $U$ the static onsite interaction, $V_k$ the $k$-dependent intersite interaction.

Since both the band dispersion $\epsilon_k$ and the bare interaction $v_k$ enter the lattice Dyson equations [see Eqs.~(\ref{eq:lattice_dyson_g}) and (\ref{eq:lattice_dyson_w})], we will next give the explicit formulas for $V_k$. The formulas for $\epsilon_k$ are identical, with the interaction parameter $V_i$ replaced by the hopping parameter $-t_i$. In the present work, we only considered the following three cases (see Fig.~\ref{fig:lattice}). Unless explicitly stated otherwise, in the following the lattice constant $a_0 = 1$.

(1) The nearest neighbor (NN) case:
\begin{equation}
V_{ij} = V_{0} \delta_{\langle ij \rangle},
\end{equation}
where $\delta_{\langle ij \rangle} = 1$ if $i$ and $j$ are the nearest neighbors and 0 otherwise. The Fourier transformation of $V_{ij}$ on the square lattice is
\begin{equation}
V_k = 2V_{0} [\cos(k_x) + \cos(k_y)].
\end{equation}
On the simple cubic lattice, we obtain
\begin{equation}
V_k = 2V_{0} [\cos(k_x) + \cos(k_y) + \cos(k_z)].
\end{equation}

(2) The nearest neighbor (NN) + the next nearest neighbor (NNN) case:
\begin{equation}
V_{ij} = V_{0} \delta_{\langle ij \rangle} + V_{1} \delta_{\ll ij \gg},
\end{equation}
where $\delta_{\ll ij \gg} = 1$ if $i$ and $j$ are the next nearest neighbors and 0 otherwise. The Fourier transformation of $V_{ij}$ on the square lattice is
\begin{align}
V_k = &+2V_{0} [\cos(k_x) + \cos(k_y)] \nonumber \\
      &+2V_{1} [\cos(k_x + k_y) + \cos(k_x - k_y)].
\end{align}
On the simple cubic lattice, we obtain
\begin{align}
V_k        =& +2V_{0}[\cos(k_x) + \cos(k_y) + \cos(k_z)] \nonumber \\
            & +2V_{1}[\cos(k_x + k_y) + \cos(k_x - k_y)] \nonumber \\
            & +2V_{1}[\cos(k_y + k_z) + \cos(k_y - k_z)] \nonumber \\
            & +2V_{1}[\cos(k_z + k_x) + \cos(k_z - k_x)].
\end{align}

(3) The nearest neighbor (NN) + the next nearest neighbor (NNN) + the third nearest neighbor (3NN) case:
\begin{equation}
V_{ij} = V_{0} \delta_{\langle ij \rangle} + V_{1} \delta_{\ll ij \gg} + V_{2} \delta_{\lll ij \ggg},
\end{equation}
where $\delta_{\lll ij \ggg} = 1$ if $i$ and $j$ are the third nearest neighbors and 0 otherwise. The Fourier transformation of $V_{ij}$ on the square lattice is 
\begin{align}
V_k = &+2V_{0} [\cos(k_x) + \cos(k_y)] \nonumber \\
      &+2V_{1} [\cos(k_x + k_y) + \cos(k_x - k_y)] \nonumber \\
      &+2V_{2}[\cos(2k_x) + \cos(2k_y)].
\end{align}
On the simple cubic lattice, we obtain
\begin{align}
V_k        =& +2V_{0}[\cos(k_x) + \cos(k_y) + \cos(k_z)] \nonumber \\
            & +2V_{1}[\cos(k_x + k_y) + \cos(k_x - k_y)] \nonumber \\
            & +2V_{1}[\cos(k_y + k_z) + \cos(k_y - k_z)] \nonumber \\
            & +2V_{1}[\cos(k_z + k_x) + \cos(k_z - k_x)] \nonumber \\
            & +2V_{2}[\cos(k_x + k_y + k_z)] \nonumber \\
            & +2V_{2}[\cos(k_x + k_y - k_z)] \nonumber \\
            & +2V_{2}[\cos(k_x + k_z - k_y)] \nonumber \\
            & +2V_{2}[\cos(k_y + k_z - k_x)].
\end{align}

Now the remaining issue is how to choose reasonable $t_0$, $t_1$, $t_2$, $V_0$, $V_1$ and $V_2$ parameters. For simplicity, we only retain the hoppings between the nearest neighbours, in other words, we set $t_0 = t$, and $t_1 = t_2 = 0$. On the other hand, we assume that the intersite interaction $V_{ij}$ fulfills the following relation:
\begin{equation}
V_{ij} = \frac{V}{|\vec{r}_i - \vec{r}_j|/a},
\end{equation}
where $i \neq j$, $V$ is an adjustable parameter which controls the strength of nonlocal intersite interaction, and $a$ is the shortest distance between two neighbors. By applying this restriction, we can easily determine $V_{0}$, $V_{1}$ and $V_{2}$ for the square and simple cubic lattices.

(1) The nearest neighbor case: $|\vec{r}_i - \vec{r}_j| = a$, $V_{0} = V$.

(2) The next nearest neighbor case: $|\vec{r}_i - \vec{r}_j| = \sqrt{2}a$, $V_{1} = V/\sqrt{2}$.

(3) The third nearest neighbor case: The $V_2$ parameters for 2D and 3D lattices are different. For the square lattice, $|\vec{r}_i - \vec{r}_j| = 2a$ and $V_{2} = V/2$, while for the simple cubic lattice, $|\vec{r}_i - \vec{r}_j| = \sqrt{3}a$ and $V_{2} = V/\sqrt{3}$.

\section{Maximum entropy method for retarded interaction $\mathcal{U}(i\nu)$ and fully screened interaction $W(i\nu)$\label{app:mem}}
\begin{figure}[tp]
\centering
\includegraphics[width=\columnwidth]{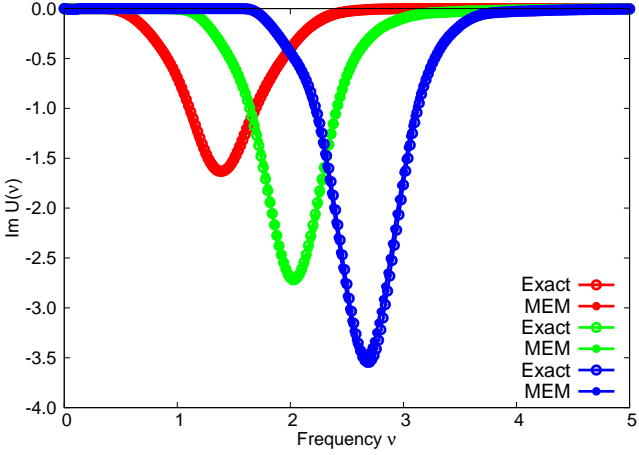}  
\caption{(Color online) Benchmarks for the maximum entropy method for retarded interaction $\mathcal{U}(i\nu)$. The exact spectra for $\text{Im}\mathcal{U}(\nu)/\nu$ are generated using classic Gaussian model. They are converted into $\mathcal{U}(\tau)$, and then processed by the proposed maximum entropy method. In the simulations, we assume $\beta = 100$ and $U - U_{\text{scr}} = 2.0/\pi$. \label{fig:imu_n1}}
\end{figure}

\begin{figure}[tp]
\centering
\includegraphics[width=0.48\columnwidth]{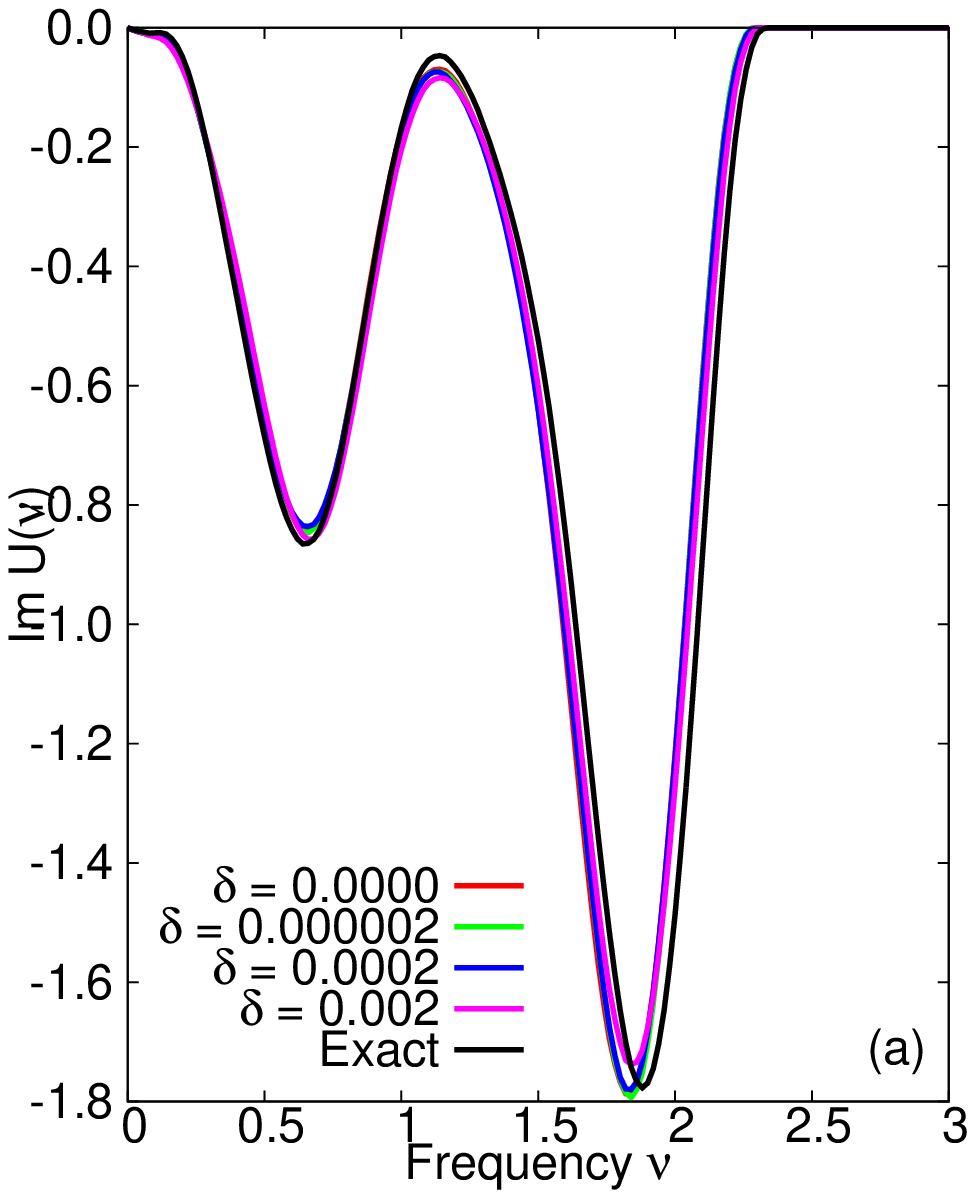}  
\includegraphics[width=0.48\columnwidth]{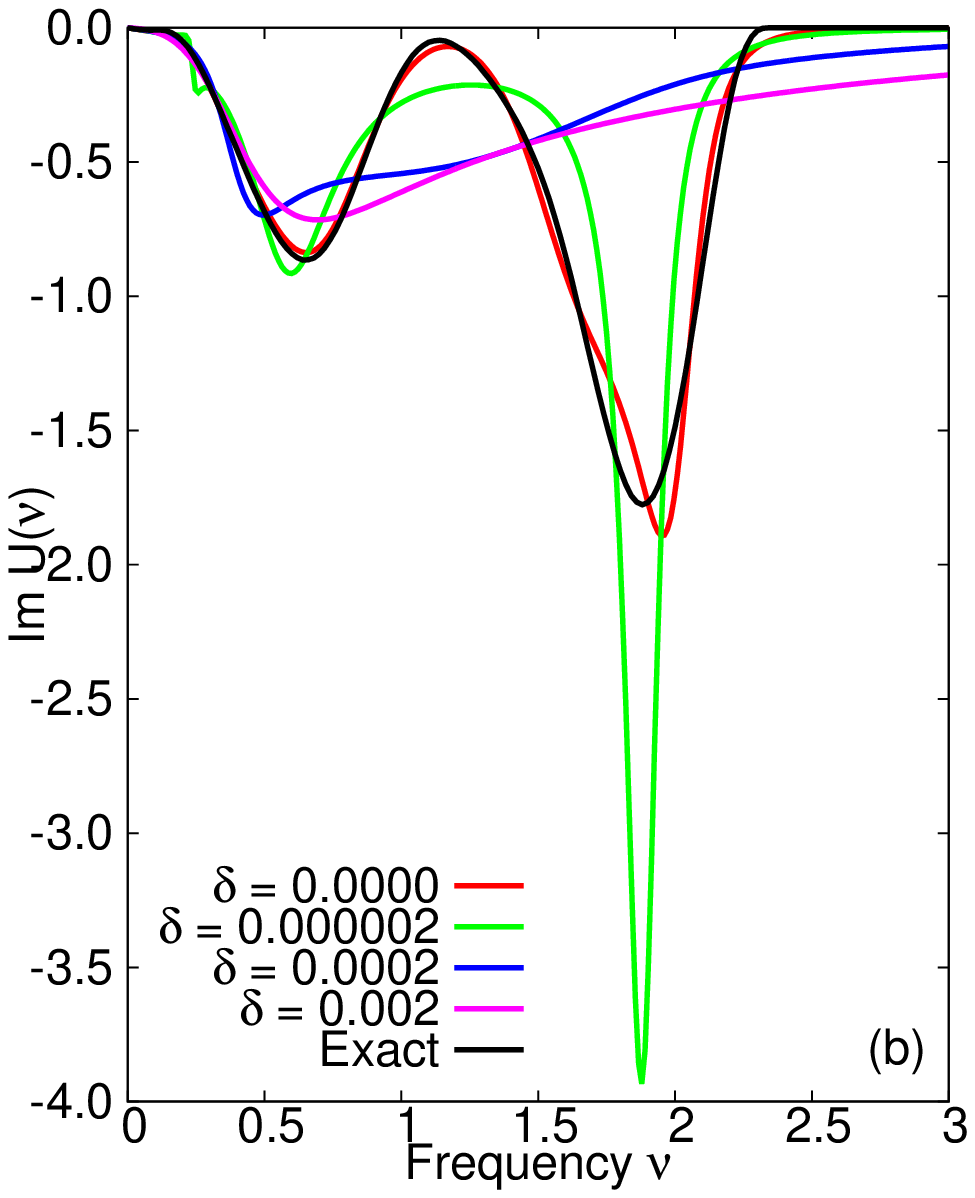}  
\caption{(Color online) Benchmarks for the maximum entropy method and Pad\'{e} approximation against numerical noise. (a) Results obtained by maximum entropy method. (b) Results obtained by Pad\'{e} approximation. In the calculations, we set $\beta = 100$ and $\alpha = 1.288$. The $\delta$ parameter is used to control the strength of data noise. Please see the text for the details. \label{fig:imu_noise}}
\end{figure}

In the self-consistent EDMFT and $GW$ + EDMFT calculations, the frequency-dependent retarded interaction $\mathcal{U}(i\nu)$ can be calculated via the local Dyson equation [see Eq.~(\ref{eq:local_dyson_u})]. In order to determine the effective screening frequency $\nu_0$ and reveal the high-energy plasmonic peaks in the local spectral function $A(\omega)$, we need $\mathcal{U}(\nu)$ [in fact, Im$\mathcal{U}(\nu)$]. However, the analytical continuation of $\mathcal{U}(i\nu)$ is not a trivial task due to the unavoidable numerical noise. In that case, the commonly used Pad\'{e} procedure\cite{vid.29.179} is questionable, and is not the first choice any more. The maximum entropy method is widely used in the Monte Carlo community to extract real frequency data from imaginary time correlation functions.\cite{mem:1996} In this appendix, we will extend it to support the analytical continuation of retarded interaction function $\mathcal{U}(i\nu)$.

First of all, the retarded interaction $\mathcal{U}(i\nu)$ obeys the following relation:\cite{PhysRevLett.104.146401,Werner2012} 
\begin{equation}
U_{\text{scr}} = U + 2\int^{\infty}_{0} \frac{\text{d}\nu}{\pi} \frac{\text{Im}\mathcal{U}(\nu)}{\nu},
\end{equation}
with $U_{\text{scr}} = \text{Re}\mathcal{U}(i\nu = 0)$ and $U$ is the static onsite interaction. This equation can be rewritten as
\begin{equation}
\label{eq:normal}
\int^{\infty}_{0} \tilde{\mathcal{U}}(\nu) \text{d}\nu = 1,
\end{equation}
where
\begin{equation}
\label{eq:tilde_u}
\tilde{\mathcal{U}}(\nu) = -\frac{\text{Im}\mathcal{U}(\nu)}{\pi} \frac{2}{\nu(U - U_{\text{scr}})}.
\end{equation}
Eqs.~(\ref{eq:normal}) and (\ref{eq:tilde_u}) can be viewed as the sum-rule for $\mathcal{U}(\nu)$, which is important for the maximum entropy algorithm. On the other hand, the kernel equation for the  maximum entropy method is\cite{mem:1996}
\begin{equation}
\label{eq:mem_u}
\mathcal{U}(\tau) = \int^{\infty}_{0} \text{d}\nu \frac{e^{-\tau\nu}}{1-e^{-\beta\nu}} \left[\frac{-\text{Im}\mathcal{U}(\nu)}{\pi} \right].
\end{equation}
Using Eq.~(\ref{eq:tilde_u}), it is easy to rewrite Eq.~(\ref{eq:mem_u}) as
\begin{equation}
\label{eq:core_mem}
\mathcal{U}(\tau) = \int^{\infty}_{0} \text{d}\nu K(\nu,\tau) \tilde{\mathcal{U}}(\nu),
\end{equation}
where $K(\nu,\tau)$ is the so-called bosonic Kernel function. The explicit definition of $K(\nu,\tau)$ is
\begin{equation}
K(\nu,\tau) = \frac{e^{-\tau\nu}}{1-e^{-\beta\nu}} \frac{\nu(U - U_{\text{scr}})}{2}.
\label{eq:kernel}
\end{equation}
Note that $U - U_\text{scr}$ = $U - \text{Re}\mathcal{U}(i\nu = 0)$ parameter is determined by the self-consistency equation [see Eq.~(\ref{eq:local_dyson_u})]. Now we can apply the standard maximum entropy algorithm\cite{mem:1996} to solve Eqs.~(\ref{eq:normal}), (\ref{eq:tilde_u}), (\ref{eq:core_mem}), and (\ref{eq:kernel}) to obtain the solutions $\tilde{\mathcal{U}}(\nu)$ and $\text{Im}\mathcal{U}(\nu)$.

Once we have determined $\text{Im}\mathcal{U}(\nu)$, the following equation can be used to verify its correctness:\cite{PhysRevLett.104.146401,PhysRevLett.109.126408}
\begin{equation}
\label{eq:verify}
\mathcal{U}(\tau) = \int^{\infty}_{0} \frac{\text{d}\nu}{\pi} \text{Im}\mathcal{U}(\nu) B(\nu,\tau),
\end{equation}
with $B(\nu,\tau) = \text{cosh}[(\tau - \frac{\beta}{2})\nu] / \text{sinh}[\frac{\nu\beta}{2}]$ for $0 \leq \tau \leq \beta$. Additionally, with $\text{Im}\mathcal{U}(\nu)$, the corresponding real part of retarded interaction $\text{Re}\mathcal{U}(\nu)$ can be easily calculated via the Kramers-Kronig relation 
\begin{equation}
\label{eq:kk}
\text{Re}\mathcal{U}(\nu) = \frac{1}{\pi} \mathcal{P} \int^{\infty}_{-\infty} \frac{\text{Im}\mathcal{U}(\nu')}{\nu'-\nu} \text{d}\nu',
\end{equation}
where $\mathcal{P}$ denotes the Cauchy principal value. 

In summary, the procedure to apply the maximum entropy method for the analytical continuation of retarded interaction $\mathcal{U}(i\nu)$ is as follows: 

(i) Calculate $\mathcal{U}(\tau)$ from $\mathcal{U}(i\nu)$ by using the invert Fourier transformation:
\begin{equation}
\label{eq:fourier}
\mathcal{U}(\tau) = \frac{1}{\beta} \sum^{\infty}_{n=-\infty} e^{-i\nu_n\tau} \mathcal{U}(i\nu_n).
\end{equation}

(ii) Use the classic maximum entropy algorithm\cite{mem:1996} to solve Eq.~(\ref{eq:core_mem}). The normalization condition is Eq.~(\ref{eq:normal}). In general, we have to specify the default model in the maximum entropy algorithm. According to our experience, the flat default model is sufficient.

(iii) With $\tilde{\mathcal{U}}(\nu)$, the $\text{Im}\mathcal{U}(\nu)$ can be determined by using Eq.~(\ref{eq:tilde_u}).

(iv) Apply Eq.~(\ref{eq:verify}) to check the correctness of the spectral function $\text{Im}\mathcal{U}(\nu)$ if need.

(v) Apply Kramers-Kronig relation Eq.~(\ref{eq:kk}) to evaluate $\text{Re}\mathcal{U}(\nu)$ if necessary.

Next, we will benchmark this modified maximum entropy method. At first, we will generate some exact spectra with a Gaussian distribution. Starting from an initial $\text{Im}\mathcal{U}(\nu)$, we calculate $\mathcal{U}(\tau)$ via Eq.~(\ref{eq:verify}). Then, applying the maximum entropy method as introduced above to it, we can obtain a new spectrum for $\text{Im}\mathcal{U}(\nu)$. At last, we should verify whether the new spectrum coincides with the exact one. Figure~\ref{fig:imu_n1} shows some representative results. It is apparent that the extended maximum entropy method works well, and allows to reproduce the initial spectra accurately.

Finally, we will test the robustness of this maximum entropy method, i.e., benchmark its stability and ability to deal with the numerical noises contained in realistic $\mathcal{U}(i\nu)$ data. Let's start from an exact spectrum again. Here we consider a more complicated two-hump spectrum. As usual, we first convert it to $\mathcal{U}(\tau)$ and then calculate $\mathcal{U}(i\nu)$ by the Fourier transformation  
\begin{equation}
\mathcal{U}(i\nu_n) = \int^{\beta}_{0} \text{d}\tau e^{i\nu_n\tau} \mathcal{U}(\tau).
\end{equation}
Next, we use the following algorithm to introduce some random noises to the real part of $\mathcal{U}(i\nu)$. The strength of the numerical noise is controlled by a $\delta$ parameter:
\begin{equation}
\mathcal{U}(i\nu) \to 
\begin{cases}
\mathcal{U}(i\nu) + \xi_1\delta/2, \quad \xi_2 < 0.5, \\
\mathcal{U}(i\nu) - \xi_1\delta/2, \quad \xi_2 \geq 0.5,
\end{cases}
\end{equation}
where $\xi_1$ and $\xi_2$ are two random numbers in the interval [0,1]. Then we transform it back to $\mathcal{U}(\tau)$ again using Eq.~(\ref{eq:fourier}), and apply the maximum entropy method to obtain the spectral function $\text{Im}\mathcal{U}(\nu)$. Through this benchmark, we can assess the influence of numerical noise on the maximum entropy method. The benchmark results are shown in Fig.~\ref{fig:imu_noise}(a). In principle, the Pad\'{e} approximation can also be used to extract $\text{Im}\mathcal{U}(\nu)$ from $\mathcal{U}(i\nu)$ directly.\cite{vid.29.179} The results from the Pad\'{e} analytical continuation are shown in Fig.~\ref{fig:imu_noise}(b), to enable a direct comparison. We see that when $\delta$ is small, the two-hump structure can be roughly reproduced by the Pad\'{e} approximation. But when $\delta$ is large, the Pad\'{e} approximation fails -- it gives a wrong single-peak spectrum with a very broad tail. On the other hand, it seems that the maximum entropy method is not sensitive to this level of numerical noise. The maximum entropy spectra agree well with the exact spectra, irrespective of the details of the numerical noise. For this reason, we believe that the maximum entropy method is superior to the Pad\'{e} approximation for the analytical continuation of the retarded interaction computed within EDMFT or $GW$ + EDMFT schemes. 

In this appendix, so far we have focused on the analytical continuation of the retarded interaction $\mathcal{U}(i\nu)$. However, it should be emphasized that the method is general and can be applied to the analytical continuation of the fully screened interaction $W(i\nu)$ as well. We merely need to replace $\mathcal{U}$ with $W$ in the above equations. 

\bibliography{test}
\end{document}